\documentclass[twocolumn]{aastex62}
\usepackage{amsmath,apjfonts,amsfonts,amssymb,enumerate,graphicx, longtable, float, times,textcomp,verbatim,url,nicefrac}                                   
\def\apjl{ApJ}%
\def\apjs{ApJS}%
\def\ao{Appl.~Opt.}%
\def\aap{A\&A}%
\newcommand{\rsun}{R$_\odot$}

\tabletypesize{\footnotesize}                                                             
\shorttitle{TESS Supernovae}    
 \shortauthors{Fausnaugh et al.}    

\begin{document}

\title{Early Time Light Curves of Type Ia Supernovae Observed with TESS}

\author[0000-0002-9113-7162]{M.~M.~Fausnaugh}
\affil{Department of Physics and Kavli Institute for Astrophysics and Space Research, Massachusetts Institute of Technology, Cambridge, MA 02139, USA}

\author{P.~J.~Vallely}
\affil{Department of Astronomy, The Ohio State University, 140 West 18th Avenue, Columbus, OH 43210, USA}
\author{C.~S.~Kochanek}
\affil{Department of Astronomy, The Ohio State University, 140 West 18th Avenue, Columbus, OH 43210, USA}
\affil{Center for Cosmology and AstroParticle Physics, The Ohio State University, 191 W. Woodruff Ave., Columbus, OH 43210, USA}
\author[0000-0003-4631-1149]{B.~J.~Shappee}
\affil{Institute for Astronomy, University of Hawai'i, 2680 Woodlawn Drive, Honolulu, HI 96822, USA}
\author{K.~Z.~Stanek}
\affil{Department of Astronomy, The Ohio State University, 140 West 18th Avenue, Columbus, OH 43210, USA}
\affil{Center for Cosmology and AstroParticle Physics, The Ohio State University, 191 W. Woodruff Ave., Columbus, OH 43210, USA}
\author[0000-0002-2471-8442]{M.~A.~Tucker}
\affil{Institute for Astronomy, University of Hawai'i, 2680 Woodlawn Drive, Honolulu, HI 96822, USA}
\affil{DOE CSGF Fellow}

\author{George R.\ Ricker}
\affil{Department of Physics and Kavli Institute for Astrophysics and Space Research, Massachusetts Institute of Technology, Cambridge, MA 02139, USA}

\author{Roland\ Vanderspek}
\affil{Department of Physics and Kavli Institute for Astrophysics and Space Research, Massachusetts Institute of Technology, Cambridge, MA 02139, USA}

\author{David W.\ Latham}
\affil{Center for Astrophysics | Harvard \& Smithsonian, 60 Garden Street, Cambridge, MA 02138}

\author[0000-0002-6892-6948]{S.~Seager}
\affil{Department of Physics and Kavli Institute for Astrophysics and Space Research, Massachusetts Institute of Technology, Cambridge, MA 02139, USA}
\affil{Department of Earth, Atmospheric, and Planetary Sciences, Massachusetts Institute of Technology, Cambridge, MA 02139, USA}
\affiliation{Department of Aeronautics and Astronautics, MIT, 77 Massachusetts Avenue, Cambridge, MA 02139, USA}

\author[0000-0002-4265-047X]{Joshua~N.~Winn}
\affil{Department of Astrophysical Sciences, Princeton University, 4 Ivy Lane, Princeton, NJ 08544, USA}

\author{Jon M.\ Jenkins}
\affil{NASA Ames Research Center, Moffett Field, CA, 94035, USA}

\author[0000-0002-3321-4924]{Zachory~K.~Berta-Thompson}
\affiliation{University of Colorado Boulder, Boulder, CO 80309, USA}

\author[0000-0002-6939-9211]{Tansu~Daylan}

\affiliation{Department of Physics and Kavli Institute for Astrophysics and Space Research, Massachusetts Institute of Technology, Cambridge, MA 02139, USA}

\affiliation{Kavli Fellow}

\author{John~P.~Doty}
\affiliation{Noqsi Aerospace Ltd, 15 Blanchard Avenue, Billerica, MA 01821, USA}

\author{G{\'a}bor F{\H u}r{\' e}sz}
\affiliation{Department of Physics and Kavli Institute for Astrophysics and Space Research, Massachusetts Institute of Technology, Cambridge, MA 02139, USA}

\author{Alan~M.~Levine}
\affiliation{Department of Physics and Kavli Institute for Astrophysics and Space Research, Massachusetts Institute of Technology, Cambridge, MA 02139, USA}

\author{Robert~Morris}
\affiliation{NASA Ames Research Center, Moffett Field, CA, 94035, USA}
\affiliation{SETI Institute, Mountain View, CA 94043, USA}

\author{Andr\'as P\'al}
\affiliation{Konkoly Observatory, Research Centre for Astronomy and
Earth Sciences, Hungarian Academy of Sciences, Konkoly Thege
Mikl\'os \'ut 15-17, H-1121 Budapest, Hungary}
\affiliation{Department of Physics and Kavli Institute for Astrophysics and Space Research, Massachusetts Institute of Technology, Cambridge, MA 02139, USA}
\affiliation{Department of Astronomy, Lor\'and E{\H o}tv{\H o}s University,
P\'azm\'any P. stny. 1/A, Budapest H-1117, Hungary}

\author[0000-0001-5401-8079]{Lizhou~Sha}
\affiliation{Department of Physics and Kavli Institute for Astrophysics and Space Research, Massachusetts Institute of Technology, Cambridge, MA 02139, USA}

\author{Eric~B.~Ting}
\affiliation{NASA Ames Research Center, Moffett Field, CA, 94035, USA}

\author[0000-0002-5402-9613]{Bill~Wohler}
\affiliation{NASA Ames Research Center, Moffett Field, CA, 94035, USA}
\affiliation{SETI Institute, Mountain View, CA 94043, USA}

\begin{abstract}
We present early time light curves of Type Ia supernovae observed in the first six sectors of TESS data.  Ten of these supernovae were discovered by ASAS-SN, seven by ATLAS, six by ZTF, and one by \textit{Gaia}.  For nine SNe with sufficient dynamic range ($>$3.0 mag from detection to peak), we fit power law models and search for signatures of companion stars. We find a diversity of early time light curve shapes, although most of our sources are consistent with fireball models where the flux increases $\propto t^2$.  Three SN display a flatter rise with flux $\propto t$.  We do not find any evidence for additional structure such as multiple power law components in the early rising light curves.   For assumptions about the SN properties and the observer viewing angle, and further assuming that companion stars would be in Roche-lobe overflow, we place limits on the radii of companions for six SNe with complete coverage of the early time light curves.  The upper limits are $\lesssim$\,32 R$_\odot$ for these six supernovae, $\lesssim$\,20 R$_\odot$ for five of these six, and $\lesssim$\,4 R$_\odot$ for two of these six.  The small sample size does not constrain occurrence rates of single degenerate Type Ia SN progenitors, but we expect that TESS observed enough SNe in its primary mission (26 sectors) to inform this measurement. We also show that TESS is capable of detecting emission from a 1 \rsun\  companion for a Type Ia SN within 50 Mpc, and may do so after about six years.

\end{abstract}
\keywords{supernovae:general}

\section{Introduction \label{sec:intro}}

A key observation for studying the progenitors of supernovae (SNe) is the early time light curve.   The shape and duration of the rising light curve just after the explosion contain information about the initial shock breakout and cooling of the SN ejecta, as well as the distribution of circumstellar material near the SN, the density/composition profile of the progenitor star, and the properties of any companion stars  (e.g., \citealt*{Piro2010, Kasen2010, Rabinak2011,  Piro2016, Kochanek2019}).

However, catching SN light curves significantly before peak brightness from the ground is difficult.  Type Ia SNe dominate the SN yield of flux-limited surveys, and about twenty of these objects have been serendipitously observed within $\sim$3 days of the explosions (see \citealt{Stritzinger2018}, and references therein). The case of SN2011fe was particularly favorable, with observations obtained just 2--3 hours after the explosion \citep{Nugent2011,Bloom2012}.  Ground-based transient surveys are reducing the delay from several days to 10--20 hours, using high-cadence observations from networks of telescopes (ASAS-SN, \citealt{Shappee2014}), wide-field instruments (ZTF, \citealt{Bellm2019}; ATLAS, \citealt{Tonry2018}), or high-cadence studies of specific targets (1M2H, \citealt{Coulter2017}; DLT40, \citealt{Valenti2017, Tartaglia2018}).  Recent Type Ia SNe with early time observations include  2019ein (observed about two days after explosion, \citealt{Kawabata2020, Pellegrino2020}) and SN2019yvq (estimated to have been observed within 0.1 to 1.3 days of the explosion, \citealt{Miller2020a}).  ZTF has also published early time observations of 127 Type Ia SNe from 2018 with a cadence of 1 day \citep{Yao2019, Miller2020b}.

The \textit{Kepler} spacecraft opened a new window on the early time light curves of SNe by continuously monitoring several hundred galaxies over the four years of its primary mission, and more than 9000 galaxies in the \textit{K2} Campaign 16 Supernova Experiment.  These programs yielded light curves of six SNe from before the explosion through the early rise, including ASASSN-18bt, which has the highest photometric precision light curve of any SN to date \citep{Olling2015, Garnavich2016, Shappee2019, Dimitriadis2019}.

The Transiting Exoplanet Survey Satellite (TESS, \citealt{Ricker2015}) has the potential to significantly expand the sample of early time SN light curves.  TESS combines an ability to perform nearly continuous monitoring from a stable space-based platform over time intervals as long as one month to one year, with an extremely wide field-of-view (24$\times$96 degrees).  The continuous monitoring allows TESS to observe a SN at the moments just after the explosion, while the wide field-of-view greatly increases the probability of observing bright SNe.   Despite the small apertures of its cameras, TESS can achieve a $3\sigma$ limiting magnitude in 8 hours of $\sim$20 mag, and thereby make useful photometric measurements of SNe and other extragalactic transients. For example, see \citet{Holoien2019} for an analysis of the TESS light curve of a tidal disruption event, ASASSN-19bt.

In this work, we present early time light curves of Type Ia SNe for the first six sectors of TESS observations, which we use to constrain the explosion physics and the properties of possible companion stars.  Ten of these SNe were discovered by ASAS-SN \citep{Shappee2014}, seven by ATLAS \citep{Tonry2018}, six by ZTF \citep{Bellm2019}, and one by \textit{Gaia} \citep{gaia2016}.  ASAS-SN recovered four of the SNe found by the other three projects, and has increased the cadence with which it observes the TESS fields in order to discover interesting transients that will benefit from continuous TESS data.  For the TESS survey of the northern ecliptic hemisphere, ZTF followed suit and observed the second-year TESS fields nightly \citep{vanRoestel2019}. This is an important point, because TESS data are downloaded and released several weeks after the observations, but the transients must be identified earlier in order to obtain timely multi-wavelength observations and spectroscopy.  One of these SNe, SN2018fhw (ASASSN-18tb), is discussed in detail by \citet{Vallely2019}. 

In \S2 we review the TESS observations and in \S3 we describe our data reduction.  In \S4 we present our analysis of the early time light curves, and in \S5 we compare our light curves to models of companion stars interacting with the SNe ejecta. Finally, in \S6 we assess the impact that TESS will have on the sample of early time SNe light curves and summarize our conclusions.  Throughout, we assume a consensus cosmology with $H_0 = 70$\, km s$^{-1}$, $\Omega_m = 0.3$, and $\Omega_\Lambda = 0.7$. We correct for Galactic extinction estimated by \citet{Schlafly2011} with a \citet*{Cardelli1989} extinction law and $R_V = 3.1$.

\section{Observations\label{sec:obs}}

TESS \citep{Ricker2015} began its survey of the southern ecliptic hemisphere in July of 2018.  Every 27 days, TESS slews 14 degrees eastward of the antisolar direction while keeping the center of the field of Camera 4 fixed at the ecliptic pole.  For the first year of the mission the fields were in the southern ecliptic hemisphere, and for the second year the fields were in the northern ecliptic hemisphere.   During each 27 day pointing, the fields of the four wide-field cameras ($24^{\circ}\times 24^{\circ}$ per camera) define a "sector" that covers approximately 1/18th of the sky stretching from $6\arcdeg$ from the ecliptic plane to $12\arcdeg$ beyond the ecliptic pole\footnote{Sectors~14, 15, 16, 24, 25, and 26 are exceptions; see \url{https://tess.mit.edu/observations}.}.  The first six sectors of TESS observations swept over nearly a quarter of the sky from 2018 July 25 through 2019 January 06.  In each sector, full-frame images (FFIs) are continuously collected at a 30-minute cadence.  Cosmic rays are corrected on-board by the flight software, resulting in an effective exposure time of 1440 seconds per FFI.  TESS observes in a single broad-band filter, ranging from about 600--1000\,nm with an effective wavelength of 800 nm.

\begin{figure*}
    \centering
    \includegraphics[width=0.95\textwidth]{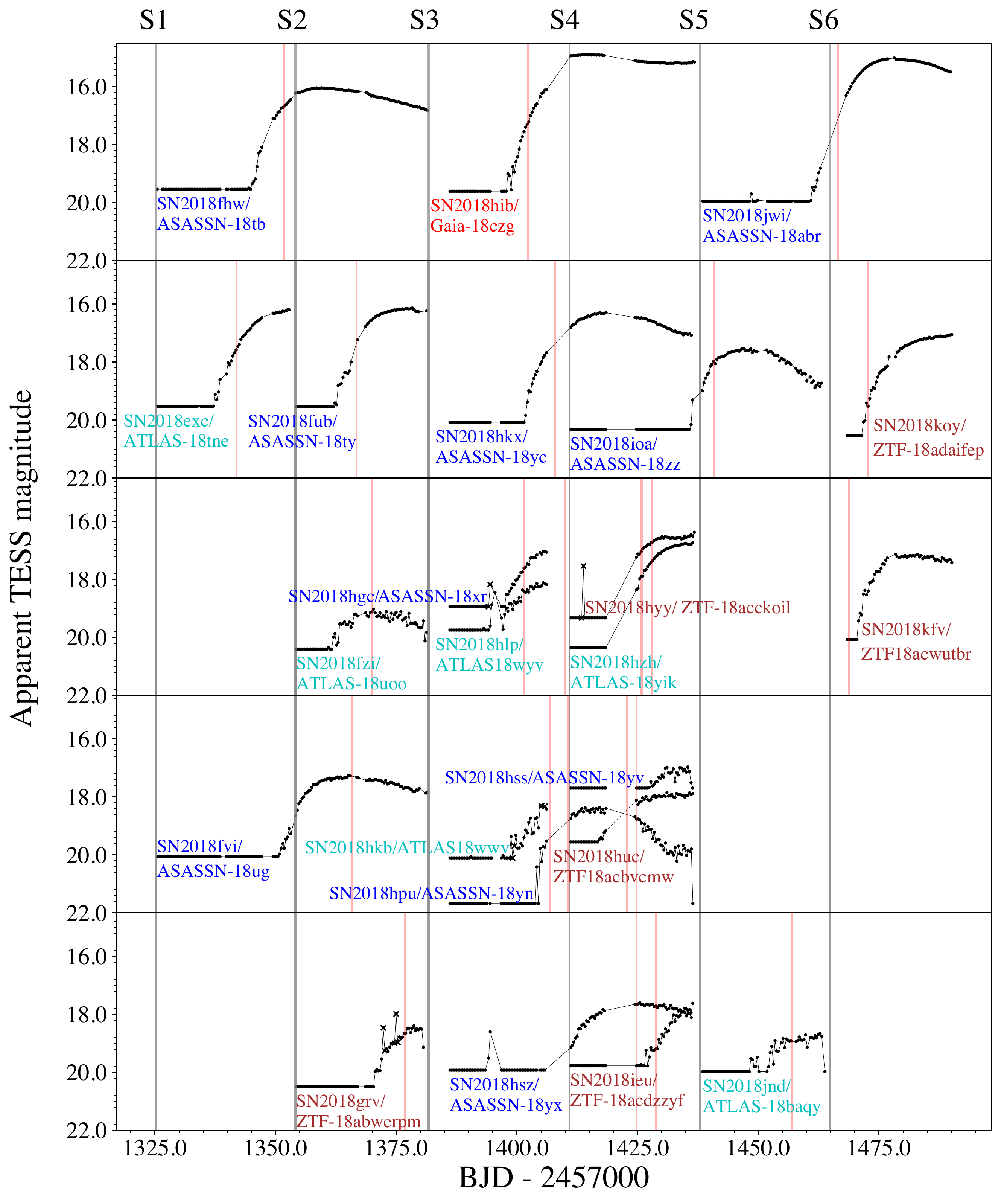}
    \caption{Light curves of the Type Ia SNe observed in the first six sectors of TESS data.  The data are binned to 8 hours and flux-calibrated as described in \S\ref{sec:data_analysis}.  Pre-explosion magnitudes are 3$\sigma$ upper limits.  The SNe are roughly ordered from top to bottom by apparent magnitude at peak.  The vertical red lines mark the times of discovery, and the Xs mark outliers caused by asteroids moving through the photometric aperture.}
    \label{fig:light_curves}
\end{figure*}

\floattable
\movetabledown=2.55in\begin{rotatetable}\begin{deluxetable*}{llrrrrrrrrcccccc}\tablewidth{0pt}\tablecaption{Properties of the Supernova Sample\label{tab:physical_data}}\tablehead{\multicolumn{2}{c}{Name}& \colhead{$T_{\rm peak}$}&\colhead{$\Delta T$}&\colhead{$M_T$}&\colhead{$\nu L_{\nu}$}&\colhead{Redshift}&\colhead{$E(B-V)$}&\colhead{Discovery}&\colhead{Classification}&\multicolumn{6}{c}{Analysis Notes}\\ &&\colhead{(mag)}&\colhead{(mag)}&\colhead{(mag)\tablenotemark{a}}&\colhead{(erg s$^{-1}$)}&&\colhead{(mag)} & & &\colhead{\S\ref{sec:analysis}\tablenotemark{b}} & \colhead{\S\ref{sec:models}\tablenotemark{c}} & \colhead{Flux Cal\tablenotemark{d}} & \colhead{Bkg\tablenotemark{e}}&\colhead{Strap\tablenotemark{f}}&\colhead{Detrend\tablenotemark{g}}   }\startdata
SN2018exc & ATLAS-18tne &16.21 & 3.32 & $-20.84\pm 0.14$ & 3.30$\times~10^{47}$ & 0.0570 & 0.034 &   \citet{2018TNSTR1154}&\citet{2018ATel11947} & X & X & -- & -- & -- & --\\
SN2018fhw & ASASSN-18tb &16.05 & 3.49 & $-18.31\pm 0.43$ & 2.96$\times~10^{46}$ & 0.0170 & 0.027 &  \citet{2018ATel11976} & \citet{2018ATel11980} & X & X & -- & -- & -- & X\\ & & & & & & & & & \citet{Kollmeier2019}\\& & & & & & & & & \citet{Vallely2019}\\
SN2018fub & ASASSN-18ty &16.16 & 3.38 & $-19.35\pm 0.26$ & 8.07$\times~10^{46}$ & 0.0288 & 0.011 &   \citet{2018ATel12011}&\citet{2018TNSCR1322} & X & X & -- & -- & -- & --\\
SN2018fvi & ASASSN-18ug &17.34 & 2.72 & $-20.06\pm 1.14$ & 5.54$\times~10^{46}$ & 0.0404 & 0.025 &   \citet{2018ATel12015}&\citet{2018TNSCR1322} & -- & -- & X & -- & -- & X\\
SN2018fzi & ATLAS-18uoo &19.18 & 1.22 & $-18.70\pm 0.10$ & 4.29$\times~10^{46}$ & 0.0800 & 0.174 &   \citet{2018TNSTR1335}&\citet{2018TNSCR1567} & -- & -- & -- & -- & -- & --\\
SN2018grv & ZTF-18abwerpm &18.51 & 1.98 & $-19.00\pm 0.11$ & 6.21$\times~10^{46}$ & 0.0700 & 0.032 &   \citet{2018TNSTR1426}&\citet{2018TNSCR1846} & -- & -- & -- & -- & -- & --\\
SN2018hgc & ASASSN-18xr &17.02 & 1.91 & $-19.82\pm 0.15$ & 1.28$\times~10^{47}$ & 0.0520 & 0.026 &   \citet{2018ATel12104}&\citet{2018TNSCR1578} & -- & -- & -- & -- & X & --\\
SN2018hib & Gaia-18czg &14.90 & 4.70 & $-19.36\pm 0.45$ & 7.85$\times~10^{46}$ & 0.0163 & 0.019 &   \citet{2018TNSTR1575}&\citet{2018ATel12131} & X & X & -- & -- & -- & X\\
SN2018hka & ATLAS-18wwt &  -- & -- & -- & -- & 0.0370 & 0.029 & \citet{2018TNSTR1582} & \citet{2018TNSCR1667}& -- & -- & -- & -- & -- & --\\
SN2018hkb & ATLAS-18wwv &18.20 & 1.90 & $-18.30\pm 0.17$ & 2.89$\times~10^{46}$ & 0.0440 & 0.105 &   \citet{2018TNSTR1582}&\citet{2018Atel12153} & -- & -- & -- & -- & -- & --\\
SN2018hkx & ASASSN-18yc &16.32 & 3.75 & $-18.88\pm 0.30$ & 5.11$\times~10^{46}$ & 0.0250 & 0.036 &   \citet{2018ATel12119}&\citet{2018ATel12122} & X & X & -- & -- & -- & --\\
SN2018hlp & ATLAS-18wyv &18.15 & 1.59 & $-19.18\pm 0.12$ & 7.37$\times~10^{46}$ & 0.0650 & 0.021 &   \citet{2018TNSTR1598}&\citet{2018ATel12143} & -- & -- & -- & -- & -- & --\\
SN2018hpu & ASASSN-18yn &18.43 & 3.25 & $-18.93\pm 0.32$ & 5.94$\times~10^{46}$ & 0.0660 & 0.013 &   \citet{2018ATel12149}&\citet{2018TNSCR1666} & X & -- & -- & -- & X & --\\
SN2018hss & ASASSN-18yv &17.10 & 0.59 & $-19.30\pm 0.18$ & 7.92$\times~10^{46}$ & 0.0430 & 0.009 &   \citet{2018ATel12171}&\citet{2018TNSCR1704} & -- & -- & -- & -- & -- & X\\
SN2018hsz & ASASSN-18yx &17.55 & 2.38 & $-20.52\pm 1.16$ & 1.16$\times~10^{47}$ & 0.0617 & 0.013 &   \citet{2018ATel12172}&\citet{2018TNSCR1719} & -- & -- & X & -- & -- & --\\
SN2018huc & ZTF-18acbvcmw &17.92 & 1.63 & $-19.68\pm 0.11$ & 1.19$\times~10^{47}$ & 0.0730 & 0.019 &   \citet{2018TNSTR1690}&\citet{2019TNSCR892} & -- & -- & -- & -- & -- & --\\
SN2018hyy & ZTF-18acckoil &16.53 & 2.79 & $-19.00\pm 0.25$ & 5.77$\times~10^{46}$ & 0.0290 & 0.025 &   \citet{2018TNSTR1718}&\citet{2018TNSCR1720} & -- & -- & -- & -- & -- & --\\
SN2018hzh & ATLAS-18yik &16.77 & 3.59 & $-19.17\pm 0.21$ & 6.99$\times~10^{46}$ & 0.0350 & 0.005 &   \citet{2018TNSTR1713}&\citet{2018ATel12246} & X & -- & -- & -- & -- & --\\
SN2018ieu & ZTF-18acdzzyf &17.76 & 2.01 & $-18.51\pm 0.19$ & 3.79$\times~10^{46}$ & 0.0405 & 0.019 &   \citet{2018TNSTR1744}&\citet{2018TNSCR1760} & -- & -- & -- & -- & -- & --\\
SN2018ioa & ASASSN-18zz &17.59 & 2.72 & $-19.38\pm 1.10$ & 4.59$\times~10^{46}$ & 0.0412 & 0.021 &   \citet{2018ATel12228}&\citet{2018ATel12246} & -- & -- & X & X & -- & X\\
SN2018jnd & ATLAS-18baqy &18.81 & 1.16 & $-19.30\pm 0.09$ & 8.13$\times~10^{46}$ & 0.0900 & 0.088 &   \citet{2018TNSTR1887}&\citet{2018TNSCR1987} & -- & -- & -- & -- & X & --\\
SN2018jwi & ASASSN-18abr &15.05 & 4.89 & $-19.29\pm 0.43$ & 7.25$\times~10^{46}$ & 0.0170 & 0.064 &   \citet{2018ATel12302}&\citet{2018TNSCR1936} & X & -- & X & X & -- & --\\
SN2018kfv & ZTF-18acwutbr &  17.17 & 2.90 & $-16.36\pm0.66$ & 3.72$\times~10^{45}$ &0.0110& 0.299 & \citet{2018TNSTR1950} & \citet{2019TNSCR89}& -- & -- & -- & -- & -- & X\\
SN2018koy & ZTF-18adaifep &17.09 & 3.44 & $-18.65\pm 0.24$ & 3.79$\times~10^{46}$ & 0.0310 & 0.122 &   \citet{2018TNSTR1975}&\citet{2019TNSCR-2019-8} & X & X & -- & -- & -- & --\\
\enddata\tablecomments{Absolute magnitudes $M_T$ are calculated from the classification redshift and corrected for Galactic extinction using the \citet{Cardelli1989} extinction law with $R_V=3.1$ and the given value of $E(B-V)$ from \citet{Schlafly2011}.  We assume an error of 1000 km s$^{-1}$ on the redshifts, due to the difficulty of separating blue-shifted SN emission in the line profile from the hosts cosmological redshift. No $K$ correction is applied.  Apparent peak magnitudes $T_{\rm peak}$ are observed values, uncorrected for Galactic extinction. \tablenotetext{a}{The absolute magnitude for SN2018kfv is implausibly low for the redshift reported to TNS.  The classification spectrum is low-resolution and consistent with redshifts up to 0.03.  At $z=0.03$, the absolute magnitude of SN2018kfv would be $-18.9$.}\tablenotetext{b}{Sources marked with an "X"  were fit with a power law to characterize the early light curve, as described in \S\ref{sec:analysis}.  They all have $\Delta T > 3.0$ (column 4).}\tablenotetext{c}{Sources marked with an "X"  were searched for signatures of companion stars, as described in \S\ref{sec:models}.  SN2018hpu, SN2018hzh, and SN2018jwi were not analyzed because of large gaps in the light curves near the times of first light.  }\tablenotetext{d}{Sources marked with an "X"  were flux calibrated across two sectors using a joint fit to both data sets, as described in \S\ref{sec:data_analysis}.}\tablenotetext{e}{Sources marked with an "X"  were near strong scattered light backgrounds at the time of first light, and these times were manually excluded as described in \S\ref{sec:systematics}.}\tablenotetext{f}{Sources marked with an "X" are affected by pixels with enhanced sensitivity from "straps," as described in \S\ref{sec:systematics}.}\tablenotetext{g}{Sources marked with an "X"  were near bright stars and required detrending, as described in \S\ref{sec:systematics}.}}\end{deluxetable*}\end{rotatetable}

\clearpage

Several events occurred in the first six sectors that affect the quality of the data, as documented in the TESS data release notes\footnote{\url{https://archive.stsci.edu/tess/tess_drn.html}}.   In Sector~3, time was taken to perform additional tests of the attitude control system.  In Sector~4, the cameras were turned off for several days in response to an instrument data-system anomaly, which led to a significant change in the thermal state of the instrument.  Finally, at the end of Sector~5, a high level of scattered light from the Earth affected the spacecraft pointing.  In our analyses, we ignore all data from these suspect epochs.
 
\section{Data Reduction \label{sec:data_analysis}}
We extracted light curves using our custom TESS transient pipeline.  The pipeline interfaces directly with the Transient Name Server\footnote{\url{https://wis-tns.weizmann.ac.il/}} (TNS) to produce differential light curves of reported astrophysical transients that land in the TESS field-of-view.  Details of the pipeline are provided in this section.

We processed TESS FFIs from each sector using the difference imaging technique described by \citet{Alard1998} and \citet{Alard2000}, implemented in the software package {\tt ISIS} .  In contrast to other detrending techniques that work at the light curve level, image subtraction removes systematic errors based on pixel data.  For each sector, we first construct a reference image by median stacking 20 FFIs with low backgrounds. We then fit for a spatially variable smoothing kernel that transforms the reference image to match an individual FFI.  This transform corrects systematic errors caused by pointing shifts, pointing jitter, and slight variations in the pixel-response function (PRF) due to thermal changes.  The TESS PRF is slightly undersampled, and we found that smoothing the images with a narrow Gaussian ($\sigma = 0.9$\ pixels) before running {\tt ISIS} improved the kernel fits.  Lastly, the reference image is convolved with the optimized kernel and subtracted from the FFI, leaving only the variable flux in each pixel.   This step eliminates sources of constant contaminating flux, such as nearby faint stars and the supernovae host galaxies.  During the final step, {\tt ISIS} also fits and removes the sky background with a two-dimensional polynomial.

We made photometric measurements for each transient source using forced photometry and models of the instrument PRF\footnote{The PRF models are available at \url{https://archive.stsci.edu/missions/tess/models/prf_fitsfiles/}}.   First, we collected the coordinates of each SN reported on TNS and used a model of the TESS focal plane geometry to identify the locations of the transients in the reference images.  The focal plane geometry model converts sky coordinates (in RA and Dec) to TESS FFI pixel coordinates, and accounts for the effect of velocity aberration due to motion of the spacecraft around the solar system barycenter.  The focal plane geometry model is used to predict the locations of guide stars for the TESS mission, and is known to be accurate to a few hundredths of a pixel.  Using the predicted image coordinates, we then fit PRF models (also smoothed by the narrow Gaussian) to the difference images at the location of each transient, and integrated the fitted PRF to estimate differential flux measurements.  We also applied a local background correction to each flux measurement, estimated with the median of the difference image pixels in a circular annulus centered on the transient with inner/outer radius of 4/8 pixels. The local background correction accounts for any residuals from the global sky background polynomial fits.

In total, there were 42 SNe brighter than 20th magnitude at discovery reported to TNS and observed by TESS in the first six sectors (2018 July 25 to 2019 January 6 UTC).  Of these, 34 were Type Ia SNe, which we focus on here.  Twenty-four of the Type Ia SNe unambiguously show a rising light curve after a time where the SNe flux is either zero or well below the TESS detection limit.  For nine other SNe, we can clearly detect a SNe signal, but no pre-explosion measurements were included in the TESS observations (SN2018eod, SN2018evo, SN2018fqn, SN2018hdo, SN2018hrs, SN2018itr, SN2018iyx, SN2018jeb, and SN2018lla).  For SN2018fwi, we did not detect any SN signal in the TESS FFIs.  Finally, for SN2018hss, although the supernova signal is clearly present, the early time light curve suffers from strong systematic issues related to a nearby bright star (see \S\ref{sec:systematics}).  For completeness, we present the light curve of SN2018hss here, although we do not use it in any analysis.  

For the 24 Type Ia SNe with pre-explosion observations, we flux-calibrated the light curves by shifting the differential light curve so that the pre-explosion flux was equal to zero. If the SN was observed across two sectors, we solved for a sector-to-sector offset that best aligns the first 0.5 days of the second sector to an extrapolated rising power law fit of the light curve in the first sector (see \S\ref{sec:analysis}). There are four exceptions: SN2018fvi, SN2018hsz, SN2018ioa, and SN2018jwi.  In each of these cases, we detect the initial onset of the SN at the very end of the sector in which it was discovered, but most of the rise is observed in next sector.  For these SNe, we jointly fit the power law rise to both sectors, which better constrains the power law slope and calibration offset.  We then converted the units of our light curves from electrons per second ($r_e$) to TESS magnitudes ($T$) with $T = -2.5\log_{10}\left( r_e \right) + Z_T$, where $Z_T$ is the TESS image zero-point.  The average image zero-point  across the four cameras is $Z_T = 20.44\pm 0.05$ mag, measured during commissioning from observations of bright isolated stars \citep{TIH}.  We calculated the eight-hour limiting magnitude $T_{\rm limit}$ by binning the light curve of each source to 8 hours, measuring the root-mean-square scatter $\sigma$ at early times (before explosion), and setting $T_{\rm limit} = -2.5\log_{10}\left(3\sigma\right) + 20.44$.  The limiting magnitudes range from 18.93 to 21.68, depending on the quality of the subtractions, the proximity of the SNe to bright stars, and time-variable structure in the image backgrounds.  We also rescaled the default pipeline uncertainties, which only account for photon noise in the images, to match the observed pre-explosion root-mean-square scatter.  The rescaling factor is between 1.0 and 2.2.

Figure~\ref{fig:light_curves} shows the final flux-calibrated light curves binned to 8 hours for 23 out of 24 of the SNe listed in Table~\ref{tab:physical_data}; 2018hka is not shown because the TESS light curve does not capture the peak of the SN (the TESS light curve of this object is shown in Appendix~A).  Table~\ref{tab:lc_data} provides the light curves.  Of the physical properties given in Table~\ref{tab:physical_data}, the peak magnitude $T_{\rm peak}$ and the difference between the peak and limiting magnitudes $\Delta T = T_{\rm limit} - T_{\rm peak}$ are particularly important, because they determine the signal-to-noise ratio and dynamic range of the early time light curves.  Absolute magnitudes $M_T$ were determined based on the redshifts reported in the classification references.  We assume uncertainties on the redshifts of 1000~km~s$^{-1}$, due to systematic
issues in interpreting SN redshifts.  For one source, SN2018kfv, the inferred absolute magnitude is too faint to be a Type Ia SNe  ($M_T =- 16.36 $).  The classification spectrum was taken at low resolution and is consistent with a range of redshifts from $z\sim0.01$ to $z\sim0.03$.  At the upper end of this redshift range, the absolute magnitude of SN2018kfv would be about $-$18.9, and so we regard the redshift reported to TNS as erroneous.  SN2018kfv is also at low Galactic latitude, and so errors in the reddening may contribute additional errors to the estimate of the absolute magnitude.

\begin{deluxetable}{rrrrr} \tablewidth{0pt}\tablecaption{TESS Supernova Light Curves\label{tab:lc_data}}\tablehead{ \multicolumn{2}{c}{Name}&\colhead{BJD$-$2457000.0}&\colhead{Flux}  &\colhead{Asteroid}\\  & &\colhead{(days)}&\colhead{(e s$^{-1}$)}  &    }\startdata
SN2018exc  &    ATLAS$-$18tne  &   1325.325680  &         $-$1.87 $\pm$         2.02  &               0 \\
SN2018exc  &    ATLAS$-$18tne  &   1325.346510  &          3.12 $\pm$         2.03  &               0 \\
SN2018exc  &    ATLAS$-$18tne  &   1325.367340  &         $-$0.44 $\pm$         2.06  &               0 \\
\ldots &\ldots &\ldots &\ldots  \\ \enddata\tablecomments{Light curves for all SNe are presented in a single table.  The "Asteroid" column is a boolean value that marks epochs for which we identified asteroids moving through the photometric apertures. A machine readable version of this table is available in the online version of this article.}\end{deluxetable}

For six objects (SN2018fhw, SN2018fvi, SN2018hib, SN2018hkx, SN2018ioa, and SN2018jwi), we were also able to measure  $\Delta m_{15}$ (the decline in the light curve 15 days after peak, \citealt{Phillips1993}) and thereby estimate the absolute magnitudes independently of the redshifts. To determine $\Delta m_{15}$, we used SNooPy \citep{Burns2011} to fit \textit{i}-band Type Ia SN templates to the TESS light curves. The templates were adapted to the TESS instrument response neglecting any internal host-galaxy extinction and 
using the template Type Ia spectral energy distributions (SED) of \citet{Hsiao2007} to calculate K-corrections. The differences between the absolute magnitudes determined using the redshift-based distances and those determined using $\Delta m_{15}$ range from $-$0.29 to 0.35 mag.  The mean difference is 0.03 mag in the sense that the $\Delta m_{15}$ distances are slightly larger than the redshift-based distances.  The values of $\Delta m_{15}$ for SN2018fhw, SN2018fvi, SN2018hib, SN2018hkx,  SN2018ioa, and SN2018jwi are 1.80, 0.84, 0.18, 1.68, 0.60, and 1.26 mag, respectively, all with uncertainties of 0.06 mag.    Our estimate of $\Delta m_{15}$ for SN2018fhw is 0.2 magnitudes smaller than the estimates in \citet{Kollmeier2019} and \citet{Vallely2019}, who used ground-based \textit{BVri}-band photometry and \textit{Swift} UV/optical light curves to derive $\Delta m_{15}$.  Part of this discrepancy is because the value of $\Delta m_{15}$ reported here is for the \textit{i}-band, while \citet{Kollmeier2019} and \citet{Vallely2019} report $\Delta m_{15}$ for the \textit{B}-band.  Those studies also used color information to constrains the intrinsic SN spectrum and K-corrections, and so we expect their estimates of $\Delta m_{15}(B)$ to be more accurate. The systematic errors in the redshift-based distances of local SNe primarily arise from the difficulty of separating blue-shifted SN emission in the line profile from the host's cosmological redshift. Since we do not have late-time light curves for all of the SNe and therefore do not have a complete set of $\Delta m_{15}$ estimates, we use the redshift-based distances for consistency.  Future work may leverage additional multiwavelength observations of these SNe to tightly constrain $\Delta m_{15}$ and provide more precise distances.  

\floattable
\begin{deluxetable}{llr}\tablewidth{0pt}\tablecaption{Systematic Errors in TESS Supernova Light Curves\label{tab:systematics_description}}\tablehead{\colhead{Issue}&\colhead{Solution}&\colhead{Number of}\\ & &\colhead{affected SNe} }\startdata
Pointing Jitter & Three rounds of 5$\sigma$\,clipping on & \\
   &  (a) 30-minute binned guiding offset quaternions and   & \\
   &  (b) standard deviations of quaternions within each bin &  24\\
Scattered light/glints & Three rounds of 5$\sigma$\,clipping on the local background estimates & 24\\
 & Removed 0.8 days of data at the ends of Sector~4 and Sector~5 & 2\\
Enhanced background  &   & \\
from "straps" & Median filter image columns and interpolate along rows & 3\\
Asteroids & Visual inspection of difference images & 5\\
 Blended bright star & Fit a scaled version of the star light curve & 6\\
\enddata\tablecomments{Summary of systematic errors in TESS supernova light curves and steps taken to flag/remove problematic data.  A detailed discussion is given in \S\ref{sec:systematics}.  Appendix A shows the full supernova light curves and the timeseries of binned quaternions, standard deviations within each bin, and local background estimates used for sigma-clipping.}\end{deluxetable}

\subsection{TESS Systematic Errors \label{sec:systematics}}
Even after image subtraction, TESS FFIs are subject to several systematic errors.  We describe each effect and our strategy to mitigate these issues below.  Table~\ref{tab:systematics_description} summarizes the steps that we apply to flag and remove problematic data.   Appendix~A shows the detailed light curves for each SNe and the auxiliary data used to identify and remove problematic epochs.

\begin{itemize}

\item Pointing jitter during an exposure increases the observed noise due to flux leaving the photometric aperture and intrapixel variations. Although the difference imaging method helps correct some of these losses, periods of large jitter are imperfectly corrected.  We consider time periods with high pointing jitter unreliable, and remove them from further analysis.  Such time periods are identified by binning the mission-supplied guiding offsets in quaternion form\footnote{The guiding quaternions are available at \url{https://archive.stsci.edu/missions/tess/engineering/}} to the 30-minute intervals of each FFI, and performing three rounds of $5 \sigma$ clipping using both the binned quaternions and the standard deviation within each bin.  Epochs removed for each light curve using this method are shown in the figures in Appendix~A.
 
\item Scattered light from the Earth and Moon is the main source of systematic errors in our light curves and set the limit on the accuracy of our photometry.  Scattered light is apparent in the FFIs when the Earth or Moon moves within 37 degrees of a camera boresight.  At these times, the image backgrounds become quite high, and strong, rapidly changing glints can appear and move through the images.  These issues are described in detail in \citet[see especially their Figures~7.2, 7.4, and 7.7]{TIH}.  Times affected by scattered light are also recorded in the data release notes for each sector.  Although the local background corrections described in \S\ref{sec:data_analysis} account for most of these issues, strong gradients and high-frequency spatial features in the scattered light pattern still cause significant systematic errors for faint sources such as SNe.  
We estimate that the background corrections during these problematic times are typically accurate to about 0.2--0.6\% of the mean background level.  However, larger deviations have been observed, and even $\sim$0.1\% errors in the background corrections are significant for faint SN light curves.
For time periods with enhanced background levels or strong scattered light signals, we do not consider the photometry reliable and remove such data from further analysis.  We identified these time periods using three rounds of 5$\sigma$ clipping on the local background estimates.  This procedure removes outliers in the distribution of background estimates, and catches most rapidly moving glints.  All epochs removed by sigma-clipping of the backgrounds are shown in Appendix~A.

Sigma-clipping does not perform very well as the Earth or Moon rises above or sets below the spacecraft sunshade.  We therefore flagged additional images based on visual inspection of the difference images.  The problem is most acute at the ends of Sector~4 and Sector~5, for which we manually excluded data from the light curves of SN2018ioa (Sector~4, BJD\,$-$\,2457000~$=$~1436.020 to 1436.812)  and SN2018jwi  (Sector~5, BJD\,$-$\,2457000~$=$~1463.605 to 1464.251).  The TESS mission also excluded these time intervals from their transiting planet searches because of strong and rapidly changing background features.  SN2018ioa and SN2018jwi are the only sources for which these features occur during the early time light curves---other sources are clearly affected at these times, but the scattered light does not affect the initial rise and so the data are shown in Figure~\ref{fig:light_curves} for completeness.  We also found some evidence of residual background errors in the light curves from Sectors~1 and 2 of SN2018exc, SN2018fhw, and SN2018fub based on visual inspection.  In these cases, there is no clear justification for excluding the data, but it is likely that the early time light curves are affected.  We therefore include the data in further analysis, but regard the results with caution (see \S\ref{sec:analysis} below).

\item Metallic "straps" on the bottom of the CCDs reflect long wavelength photons and result in heightened pixel sensitivity for certain columns in the imaging array \citep[\S6.6.1]{TIH}.  Although the straps do not affect differential photometry of the SN flux, they do affect the local background during periods of strong scattered light.  We have found that it is possible to correct for the effects of the straps using a technique similar to illumination corrections developed for long-slit spectroscopy.  We median filter the difference images along each affected column to make a smooth estimate of the enhanced background, subtract the result from the column in question, and interpolate the local background along rows.  Although the correction is not perfect, in most cases it greatly improves the photometry.  Only SN2018hgc, SN2018hpu, and SN2018jnd are near enough to CCD straps to be affected by this procedure.

\item In some cases, bright (T $<$ 18) asteroids move through the photometric aperture and cause a small bump in the light curve over several hours.  The simplest way to identify these cases is inspection of the difference images.  Given the short period of time over which such events occur, we flag and ignore the affected parts of the light curve. The probability of an asteroid passing through the photometric aperture is highest for transients in Camera~1, which was pointing nearest the ecliptic plane.  SN2018grv, SN2018hgc, SN2018hka, SN2018hkb, and SN2018hyy all suffer from this effect.  The epochs affected by asteroids are flagged in the light curves given in Table~\ref{tab:lc_data}, and explicitly highlighted in the figures in Appendix~A.

\item The final class of systematic errors is for SN that occur near a bright star in the TESS images.  Imperfect residuals in the difference images can then contaminate the photometric aperture.  Given the large plate scale (21$\farcs$19 per pixel), this is a common effect:  SN2018fhw, SN2018fvi, SN2018hib, SN2018hss, SN2018ioa, and 2018kfv are all affected.  For all of these sources, there is a very bright star within 2 or 3 pixels with obvious residuals in the difference images.  We were able to remove the main effects of these residuals by (1) extracting light curves of the contaminating stars, (2) smoothing with a median filter, and (3) fitting a shift and scale factor to the affected parts of the SNe light curves.  We then subtracted the scaled and smoothed star light curve from the SN light curve.  The light curves of the contaminating star, the smoothed model, the fit to the SN light curve, and the corrected SN light curve are shown in Appendix~B for all affected sources.  For SN2018fhw, we found better results using FFIs without Gaussian smoothing in the image subtraction (see \S\ref{sec:data_analysis}).  A detailed discussion of detrending for SN2018fhw is given by by \citet[][see their Figures 3 and 5]{Vallely2019}.  For SN2018hss, a bright star is extremely close to the SN in the image and we were not able to improve the light curve using this method.  We present our differential measurements in Appendix~A for completeness, but we do not regard the light curve as reliable.  A more sophisticated detrending algorithm might help with this problem in the future.  

\end{itemize}

\floattable
\begin{deluxetable}{llrrrrr} \tablewidth{0pt}\tablecaption{Power law Fit Parameters\label{tab:fit_params}}\tablehead{\multicolumn{2}{c}{Name}&\colhead{$t_0$}&\colhead{$\beta$}&\colhead{$dof$}&\colhead{$\chi^2/dof$}&\colhead{Comments\tablenotemark{a}}\\  & &\colhead{(BJD$-$2457000)}& & &   }\startdata
SN2018exc & ATLAS-18tne & $1329.09\pm1.26$ & $4.00\pm0.41$ & 754 & $0.83\pm0.05$ & Complicated background\\
SN2018fhw & ASASSN-18tb & $1343.12\pm0.84$ & $1.63\pm0.33$ & 984 & $1.34\pm0.05$ & Complicated background; Fit up to 60\% of peak\\
SN2018fub & ASASSN-18ty & $1356.13\pm1.11$ & $3.67\pm0.55$ & 530 & $1.19\pm0.06$ & Complicated background\\
SN2018hib & Gaia-18czg & $1395.70\pm0.36$ & $2.40\pm0.12$ & 734 & $1.22\pm0.05$ & \\
SN2018hkx & ASASSN-18yc & $1400.66\pm0.19$ & $1.26\pm0.09$ & 725 & $0.90\pm0.05$ & \\
SN2018hpu & ASASSN-18yn & $1401.28\pm0.87$ & $1.00\pm0.39$ & 721 & $1.55\pm0.05$ & \\
SN2018hzh & ATLAS-18yik & $1423.88\pm2.11$ & $0.58\pm0.83$ & 406 & $1.04\pm0.07$ & Gaps near first light\\
SN2018jwi & ASASSN-18abr & $1459.07\pm1.17$ & $2.85\pm0.57$ & 325 & $1.55\pm0.08$ & Gaps near first light; Fit up to 60\% of peak\\
SN2018koy & ZTF-18adaifep & $1470.32\pm0.35$ & $1.28\pm0.14$ & 359 & $1.12\pm0.07$ & \\
\enddata\tablecomments{Uncertainties are the central 68\% confidence intervals for fits to 1000 iterations of bootstrap resampling. \tablenotetext{a}{This column lists various issues associated with each SN light curve and the resulting fits.  Details are discussed in \S\ref{sec:analysis}.}}\end{deluxetable}

\begin{figure}
    \centering
    \includegraphics[width= 0.49\textwidth]{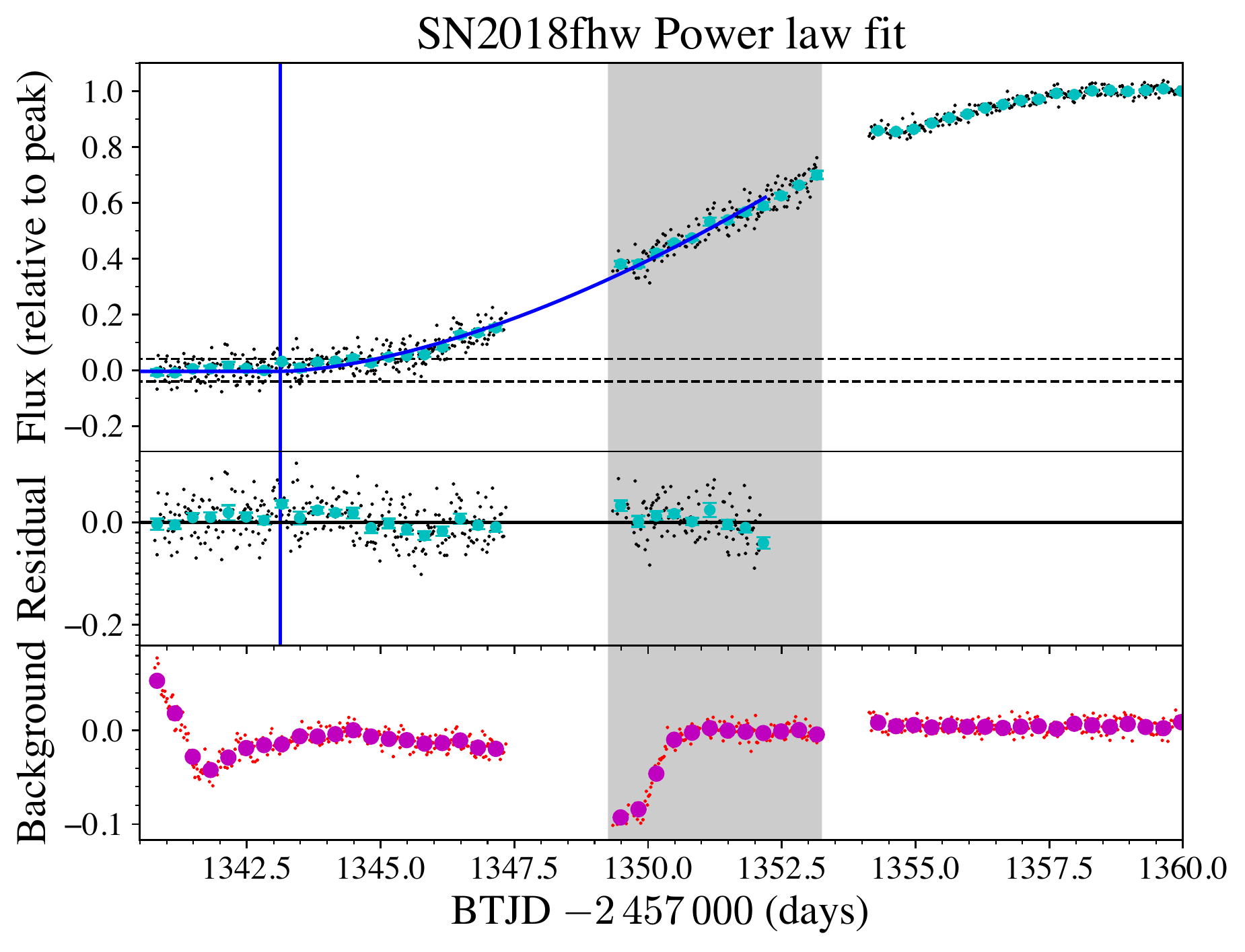}    
    \caption{Power law fit, residuals, and local background estimates for SN2018fhw near the time of first light.  The small points are the individual 30 minute FFI exposures, and the larger circles are averages over eight hours.  All units are flux relative to peak.  The vertical blue line is the fitted time of first light, while the solid blue line shows the best fit power law model.  The Earth sets below the spacecraft sunshade at approximately BJD $- 2\,457\,000 \sim 1350.4$ days, which is responsible for the sudden change in the local background at this time.  The gray shaded region highlights times over which there is a visual change in curvature in the light curve, discussed in \S\ref{sec:analysis}.  See Appendix~A for the full light curve.} 
    \label{fig:2018fhw_fit}
\end{figure}

\begin{figure}
    \centering
   \includegraphics[width= 0.49\textwidth]{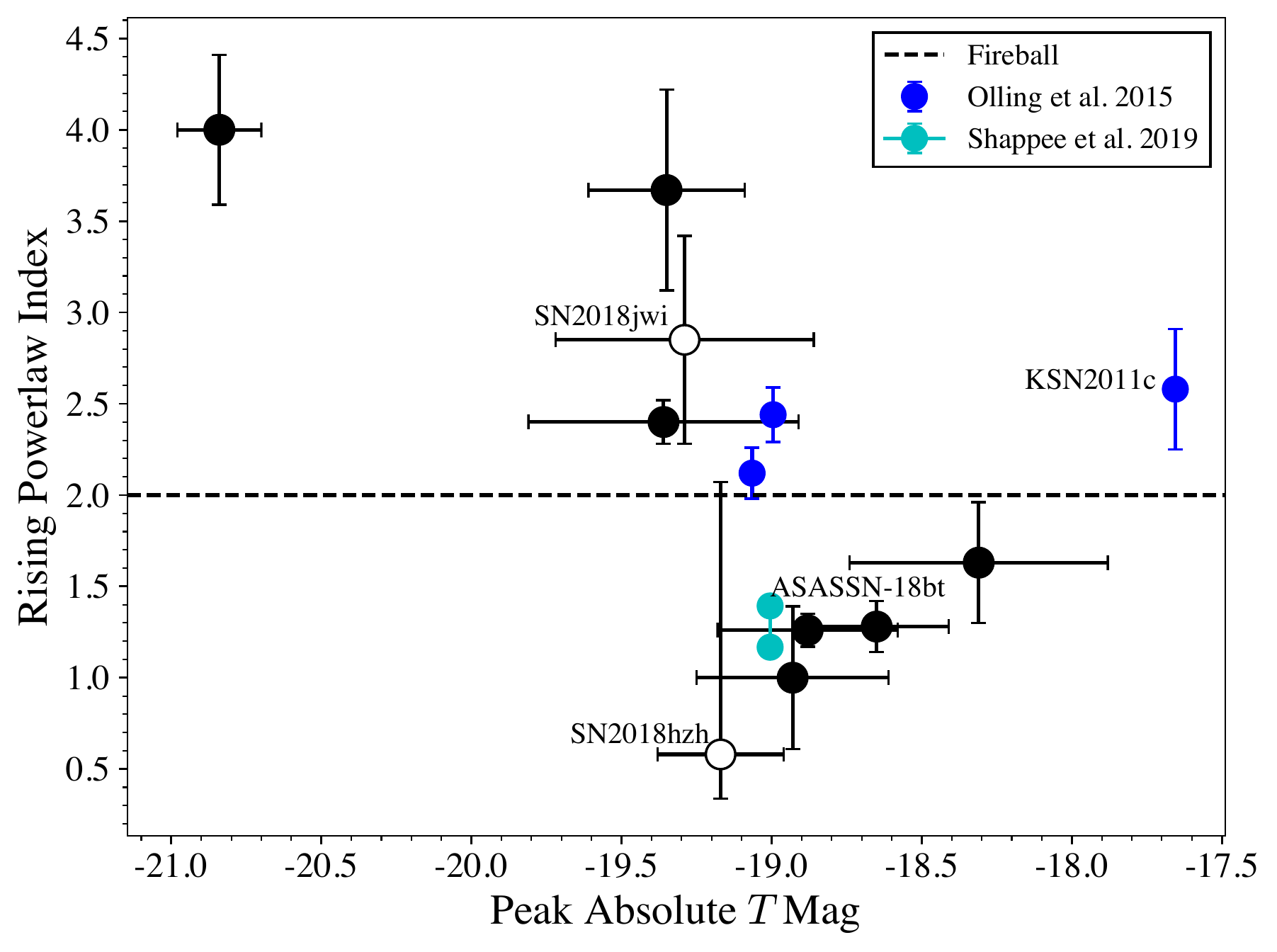}    
    \caption{Rising power law index $\beta$ as a function of absolute magnitude for the nine SNe observed by TESS with $\Delta T > 3.0$ mag (black points).  Objects that require extrapolation to $t_0$ are shown with open circles.   
    For comparison, results from \textit{Kepler} \citep{Olling2015, Shappee2019} are also shown (including both power law indices of the double component fit for ASASSN-18bt).}
    \label{fig:index_vs_abs_mag}
\end{figure}

\section{Analysis\label{sec:analysis}}

For light curves with sufficient dynamic range between the initial detection and the peak, it is possible to gain insight into the explosion physics and place limits on the size of any companions.  We begin by parameterizing the light curves with a rising power law fit of the form
\begin{align}
\label{equ:model}
    F(t) = H(t-t_0)A(t-t_0)^{\beta} + B
\end{align}
where $t_0$ is the time of the first detection, $A$, $B$ and $\beta$ are constants, and $H(t)$ is the Heaviside function.  The parameter $B$ accounts for any residual background flux at early times, and must be included to properly estimate the uncertainties in $\beta$.  We constrain $\beta$ to be between 0.5 and 4.0, and we require that $A$ be greater than 0.  Following \citet{Olling2015}, we only fit the light curves up to 40\% of the peak flux.  We estimate the uncertainties on fitted parameters with 1000 iterations of bootstrap resampling of the light curves, and report the widths of the central 68\% confidence intervals of the resulting parameter distributions.

We focus on the nine SNe with $\Delta T > 3.0$ mag, namely SN2018exc, SN2018fhw, SN2018fub, SN2018hib, SN2018hkx, SN2018hpu, SN2018hzh, SN2018jwi, and SN2018koy.  For SN2018hzh, the onset of the light curve rise is missing due to data gaps.  A large portion of the early rise of SN2018jwi is also missing, although we detect the initial onset of its rise.  For these cases, our fitting procedure is still able to constrain $t_0$ and $\beta$ if we extrapolate the observed power law through the gaps.  However, we must fit SN2018jwi up to 60\% of the peak flux to meaningfully constrain the power law index after the gap.  We also fit SN2018fhw up to 60\% of the peak of flux, because the data are affected by scattered light at the time that the light curve reaches 40\% of the peak flux.  There is a visual change in the curvature of the light curve at this time, which is related to the Earth setting below the TESS sunshade at BJD \,$-$\,2457000~$\sim$1350.4.  Fitting over the longer baseline averages over any errors in the background subtraction and should lead to a more robust result.  A detailed view of this part of the SN2018fhw light curve is shown in Figure~\ref{fig:2018fhw_fit}.

Table~\ref{tab:fit_params} gives the best-fit values of $t_0$ and $\beta$ from the rising power law fits.  For reference, a fiducial "fireball" model, with a constant temperature photosphere expanding at a constant velocity \citep{Riess1999}, has a rising power law index of $\beta = 2.0$ at wavelengths on the Rayleigh-Jeans tail of the SED. Five of the nine SNe  have $\beta > 1.6$, and all but one of these (SN2018exc) are consistent with the fireball model at or very near the $3\sigma$ level.  The best-fit index for two sources is larger than $\beta = 3.0$ (SN2018exc and SN2018fub).  Of the remaining four SNe, three have indices $\beta \approx 1.0$ (SN2018hkx, SN2018hpu, and SN2018koy), although SN2018hpu is consistent with 2.0 at the 3$\sigma$ level.  Finally, the index for SN2018hzh (2018hzh) is too uncertain to conclude that it is either low or high ($\beta = 0.58\pm 0.83$). 

We do not find any correlation between the power law indices and $T_{\rm peak}$ or $\Delta T$, so we do not attribute these results to observational effects such as signal-to-noise ratio or our fitting procedure.  However, there are reasons to be cautious in regard to three of the fits.  The light curves for the SNe with the power law indices $\beta > 3.0$, SN2018exc and SN2018fub, were obtained from images with a complicated background that may affect the light curve (see \S\ref{sec:systematics}).  A similar worry exists for SN2018fhw, although our fits for $t_0$ and $\beta$ are in good agreement with those reported by \citet[$t_0 = 1341.68\pm0.16$ in our units and $\beta=1.69 \pm 0.04 $]{Vallely2019}.\footnote{If we instead fit this object up to 40\% of the peak flux, we find $t_0= 1339\pm1.68$ and $\beta=3.17\pm 0.66$.  This fit has larger uncertainties than fitting up to 60\% of the flux but is consistent with the values in Table~\ref{tab:fit_params}.  However, the fit up to 40\% of the flux puts $t_0$ just at the end of a large data gap, which is unlikely to be an accurate estimate of the time of explosion.}  Meanwhile, the light curve of the SN with the smallest power law index, SN2018hzh, has a large gap around the onset of the rise in flux.  For this SN, $t_0$ is determined by extrapolating from the power law slope observed at significantly later times, which may not reflect the properties of the rise at the time of explosion.   

The values of $\chi^2$ per degree of freedom ($\chi^2/dof$) are consistent with unity for all sources except SN2018fhw, SN2018hpu and SN2018jwi.  For SN2018fhw and SN2018hpu, the discrepancy results from outliers in the pre-explosion light curves, which are caused by errors in the local background corrections (see Appendix~A).  For SN2018jwi, the issue is probably related to the gap between Sectors~5 and 6 during the early part of the light curve's rise, and the difficulty of simultaneously fitting the power law index and the calibration offset for the second sector of data.

In an analysis of a large sample of ground based early time Type Ia SNe light curves, \citet{Firth2015} found a range of rising power law indices, with a mean value of $\beta = 2.44 \pm 0.13$.   In a study of 127 Type Ia SNe using ZTF data, \citet{Miller2020b} found an average rising power law index of $\sim$2 in the $r$-band with a root-mean-square scatter of 0.5 and a considerable diversity among SNe.  Our findings are consistent with both of these results (although see \citealt{Miller2020b} \S9.2 for a discussion of bias in the estimate of the power law index from \citealt{Firth2015}). We find three objects out of nine with a power law index significantly less than 2: SN2018hkx, SN2018hpu, and SN2018koy. Departures from a simple fireball model have been seen in other objects, for example in SN2012fr, SN2013dy, SN2014J, iPTF16abc, and ASASSN-18bt \citep{Contreras2018,Zheng2013, Zheng2014, Siverd2015, Goobar2015, Miller2018, Shappee2019}.  These sources exhibited more nearly linear rises in the optical for a few days, followed by sharper rises with power law indices closer to 2.  \citet{Contreras2018} show that the light curve of SN2012fr is consistent with a moderately mixed initial nickel profile, as modeled by \citet{Piro2016}. \citet{Shappee2019} find that the early light curve of ASASSN-18bt is also broadly consistent with the models of \citet{Piro2016}, although \citet{Magee2020b} argue that the color and spectral evolution of the transient are not consistent with an inhomogeneous $^{56}$Ni distribution.  In our sample, the light curves of SN2018hkx and SN2018hpu could be consistent with this picture, although there is a 4-day gap in observations at the end of Sector~3 for both sources when the non-linear phase would be expected to begin. On the other hand, the behavior of SN2018koy is difficult to reconcile with the \citet{Piro2016} models.  There is no strong evidence for an increase in the slope during the rise of this source; if anything, the slope may be diminishing after seven to ten days post-explosion.  

Interactions between the expanding SN ejecta and a companion star also affect the observed behavior of early time Type Ia SN light curves.   Based on its blue color at early times, \citet{Dimitriadis2019} interpret the double power law in ASASSN-18bt as the result of a flux excess from a companion interaction superimposed on a fireball light curve.  We investigate the effects of companion stars on the TESS light curves in \S\ref{sec:models}, to determine if such models can account for the behavior observed in SN2018hkx, SN2018hpu, and SN2018koy described above.  We also discuss the interpretation of Type Ia SNe with blue flux excesses in \S\ref{sec:progenitor_discussion} (SN2012cg, iPTF14atg, SN2017cbv, HSC17bmhk, and SN2019yvq).

Figure~\ref{fig:index_vs_abs_mag} shows the power law indices as a function of peak SN absolute magnitude, with the sources requiring interpolation near $t_0$ flagged with open circles.  The SNe with lower power law indices have absolute magnitudes fainter than $-$19, putting them on the fainter end of the distribution of Type Ia luminosities. For comparison, the results for the \textit{Kepler} SNe \citep{Olling2015, Shappee2019} are shown in blue and cyan,corrected from the \textit{Kepler} to TESS bandpass based on the \citealt{Hsiao2007} template SED.

\floattable
\begin{deluxetable}{r|rr|rr|rr|rr|rr|rr|rr} \tablewidth{0pt}\tablecaption{TESS Light curves for Kasen (2010) models at 100 Mpc\tablenotemark{a}\label{tab:kasen_lc}}\tablehead{\colhead{Time}&\multicolumn{2}{c}{50 R$_\odot$  } &\multicolumn{2}{c}{25 R$_\odot$  } &\multicolumn{2}{c}{10 R$_\odot$  } &\multicolumn{2}{c}{5.0 R$_\odot$  } &\multicolumn{2}{c}{1.0 R$_\odot$  } &\multicolumn{2}{c}{0.5 R$_\odot$  } &\multicolumn{2}{c}{0.1 R$_\odot$  }  \\  \colhead{(days)}&\colhead{(e s$^{-1}$)} &\colhead{(mag)} &\colhead{(e s$^{-1}$)} &\colhead{(mag)} &\colhead{(e s$^{-1}$)} &\colhead{(mag)} &\colhead{(e s$^{-1}$)} &\colhead{(mag)} &\colhead{(e s$^{-1}$)} &\colhead{(mag)} &\colhead{(e s$^{-1}$)} &\colhead{(mag)} &\colhead{(e s$^{-1}$)} &\colhead{(mag)}   }\startdata
0.001023 & 0.006878 & 25.8464 & 0.005772 & 26.0367 & 0.004573 & 26.2895 & 0.003825 & 26.4835 & 0.002537 & 26.9293 & 0.002121 & 27.1239 & 0.001392 & 27.5807 \\
0.022525 & 0.164560 & 22.3992 & 0.136899 & 22.5990 & 0.106924 & 22.8673 & 0.088239 & 23.0758 & 0.056138 & 23.5669 & 0.045806 & 23.7877 & 0.027846 & 24.3281 \\
0.044027 & 0.322945 & 21.6672 & 0.267421 & 21.8720 & 0.207291 & 22.1485 & 0.169844 & 22.3649 & 0.105653 & 22.8803 & 0.085075 & 23.1155 & 0.049556 & 23.7023 \\
\ldots &\ldots &\ldots &\ldots &\ldots &\ldots &\ldots &\ldots &\ldots &\ldots &\ldots &\ldots &\ldots &\ldots &\ldots \\ \enddata\tablecomments{A machine readable version of this table is available in the online version of this article.\tablenotetext{a}{In the local universe, the effects of time dilation and K corrections are negligible, and these light curves can be rescaled based on luminosity distance alone.}}\end{deluxetable}

\begin{deluxetable}{rrrrrr} \tablewidth{0pt}\tablecaption{Results for fits to simulated light curves\label{tab:sim_params}}\tablehead{\colhead{Separation}&\colhead{Radius}&\colhead{$t_0$}&\colhead{$\beta$}&\colhead{Flux at $t=1$ day}&\colhead{Background}\\  \colhead{(cm)}&\colhead{($R_{\odot}$)}&\colhead{(days)}& &\colhead{(e s$^{-1}$)} &\colhead{(e s$^{-1}$)}  }\startdata Fits without noise \\ \hline
4.03$\times 10^{13}$ &  200 &  $-$0.008 &  0.91 &  47.66 &  0.41 \\  
2.02$\times 10^{13}$ &  100 &  $-$0.008 &  0.95 &  34.48 &  0.71 \\  
1.01$\times 10^{13}$ &  50 &  $-$0.008 &  1.02 &  23.17 &  1.06 \\  
0.50$\times 10^{13}$ &  25 &  $-$0.008 &  1.15 &  14.11 &  1.43 \\  
0.30$\times 10^{13}$ &  15 &  $-$0.008 &  1.28 &  9.18 &  1.65 \\  
0.20$\times 10^{13}$ &  10 &  $-$0.008 &  1.40 &  6.33 &  1.75 \\  
0.10$\times 10^{13}$ &  5 &  $-$0.008 &  1.61 &  3.48 &  1.69 \\  
0.02$\times 10^{13}$ &  1 &  $-$0.008 &  1.92 &  1.52 &  0.84 \\   \hline Fits with noise \\  \hline 4.03$\times 10^{13}$ &  200 &  $-$0.008 &  0.92 &  46.62 &  0.35 \\  
2.02$\times 10^{13}$ &  100 &  $-$0.008 &  0.94 &  35.02 &  0.37 \\  
1.01$\times 10^{13}$ &  50 &  $-$0.008 &  1.02 &  23.15 &  1.13 \\  
0.50$\times 10^{13}$ &  25 &  $-$0.008 &  1.15 &  14.10 &  1.41 \\  
0.30$\times 10^{13}$ &  15 &  $-$0.008 &  1.29 &  8.99 &  1.90 \\  
0.20$\times 10^{13}$ &  10 &  $-$0.008 &  1.40 &  6.32 &  1.51 \\  
0.10$\times 10^{13}$ &  5 &  $-$0.008 &  1.65 &  3.17 &  1.72 \\  
0.02$\times 10^{13}$ &  1 &  $-$0.008 &  1.93 &  1.44 &  1.38 \\  
\enddata\tablecomments{Recovered parameters from fits to simulated models of a $\beta=2$ power law rise (fireball) plus a \citet{Kasen2010} companion model.  Larger companions flatten the observed power law index.  The light curves, fits, and residuals are shown in Figure~\ref{fig:sim_fits_no_noise} and Figure~\ref{fig:sim_fits_with_noise}.  The parameters of the \citet{Kasen2010} models match the fiducial model described in \S\ref{sec:lc_shapes}.}\end{deluxetable}

\section{Comparison to Companion Models\label{sec:models}}
\begin{figure*}
\centering
\includegraphics[width=\textwidth]{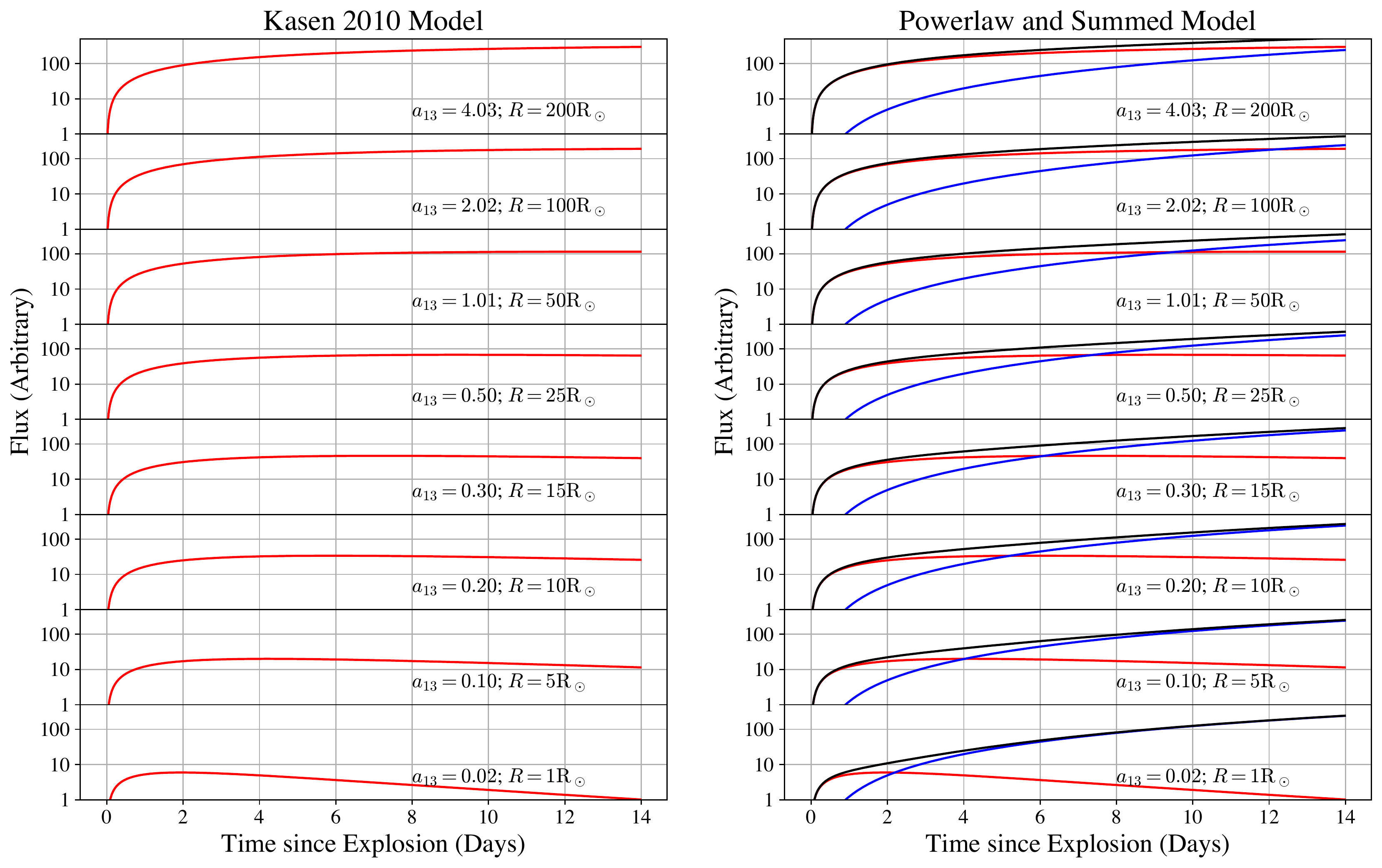}
\caption{Simulated light curves of the \citet{Kasen2010} companion models used to investigate the signatures of companion stars in our fits from \S\ref{sec:analysis}.  The left panel shows the \citet{Kasen2010} models alone (red lines).  The parameter $a_{13}$ is the separation between the exploded star and companion, in units of $10^{13}$ cm; $R$ is the equivalent Roche lobe radius of the companion for a 1.4/1.0 M$_{\odot}$ primary/secondary system.  The right panel shows the same models (red lines), a fireball $t^2$ component for the SNe ejecta (blue lines) and the total light curve (black lines).}
\label{fig:model_components}
\end{figure*}

\begin{figure*}
\centering
\includegraphics[width=\textwidth]{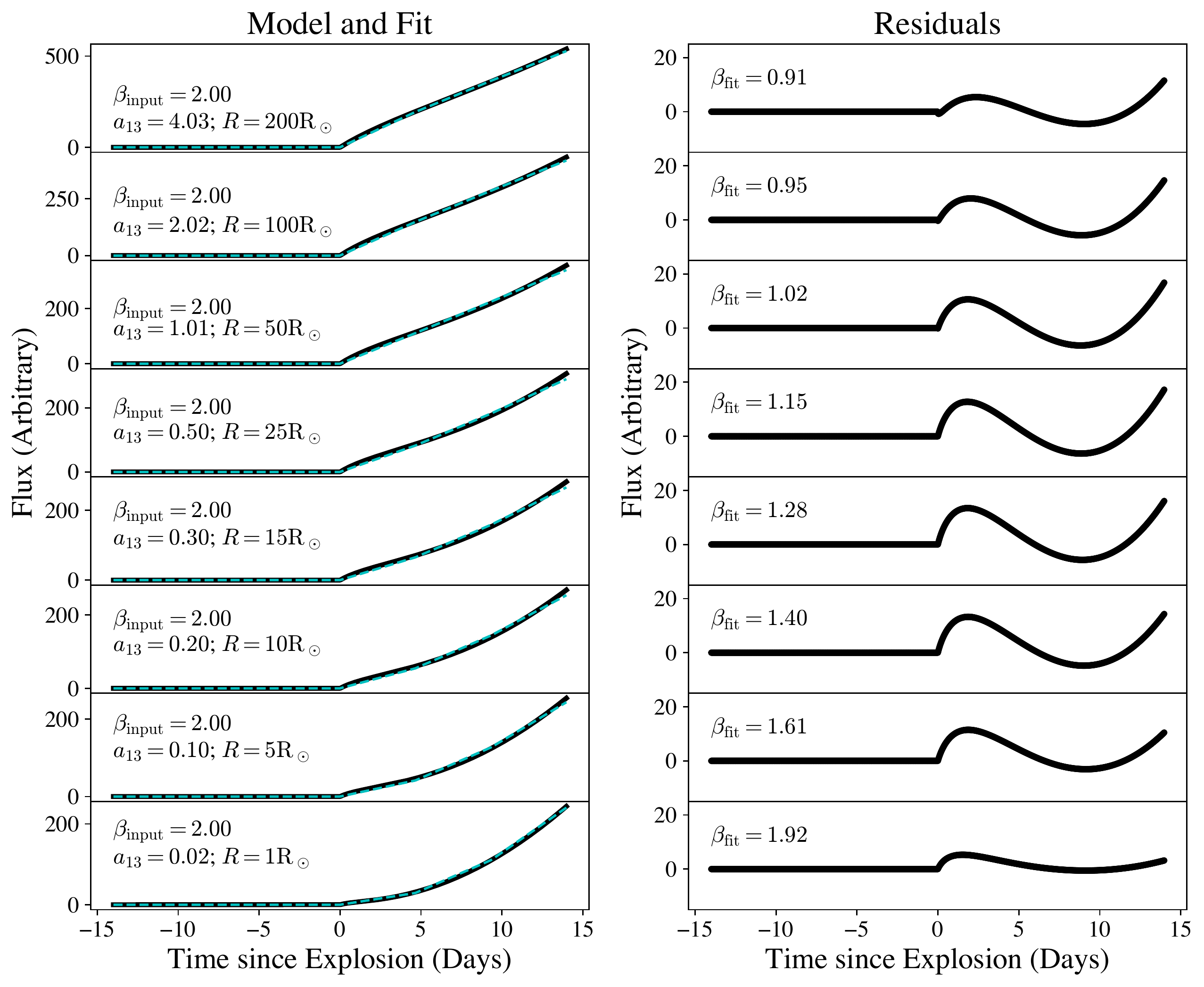}
\caption{Results for fits of Equation~\ref{equ:model} to the simulated light curves from Figure~\ref{fig:model_components}.  The simulated light curves are shown as black lines in the left panel, and the best fit power laws are shown as cyan lines.  The right panel shows the residuals from the fits.  Larger companion radii tend to bias the recovered power law index $\beta$ to low values and produce a characteristic pattern of residuals.  See Table~\ref{tab:sim_params} for details.}
\label{fig:sim_fits_no_noise}
\end{figure*}

\begin{figure*}
\centering
\includegraphics[width=\textwidth]{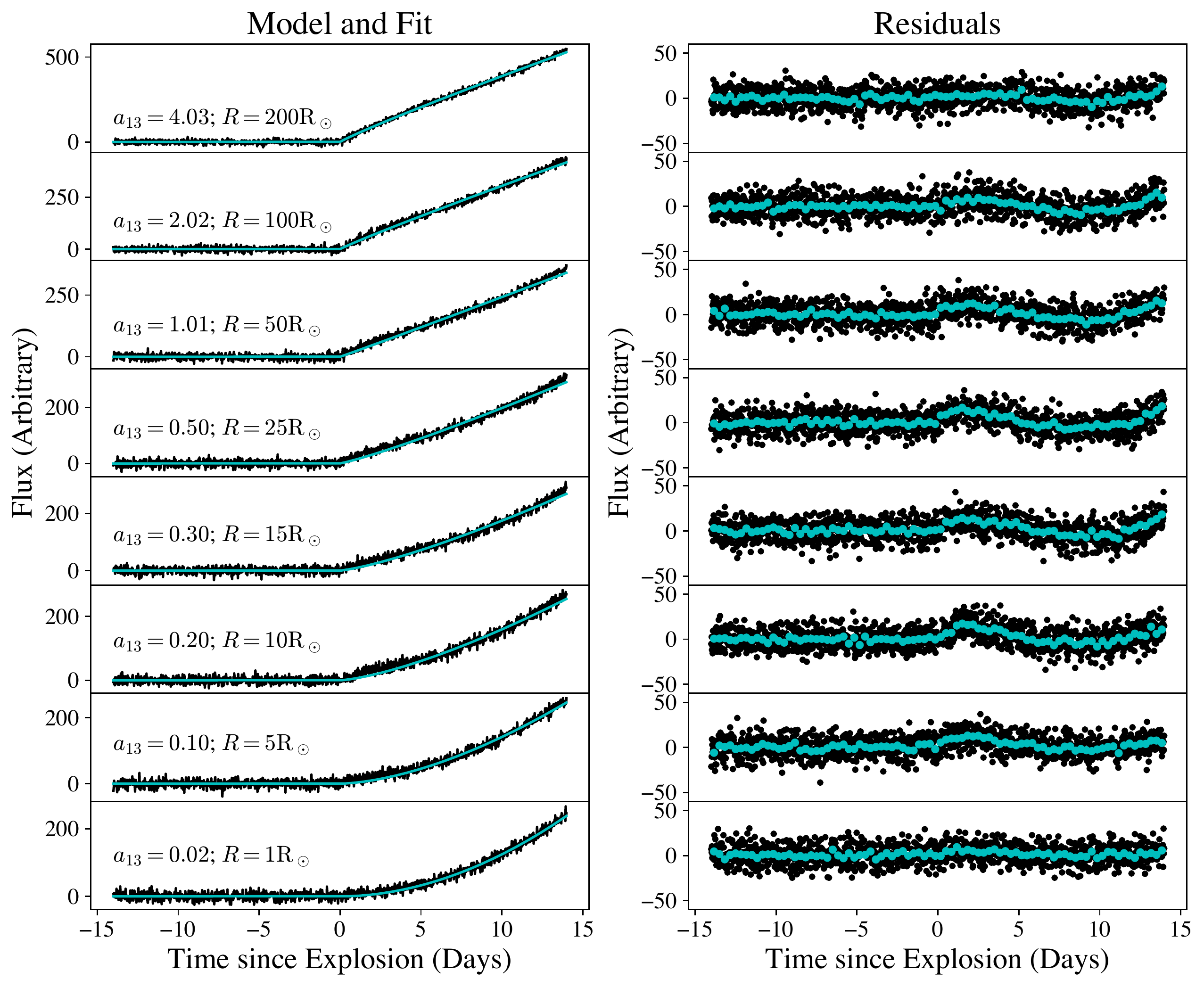}
\caption{Same as Figure~\ref{fig:sim_fits_no_noise}, but with random measurement noise added.  The cyan circles show the residuals binned to 8 hours.  The amplitude of the noise was set to 4\% of the flux of the fireball component after 14 days, which is comparable to the observed scatter of our best light curves.  Measurement noise washes out much of the pattern in the residuals from Figure~\ref{fig:sim_fits_no_noise}, but the pattern is still detectable on 8 hour time scales.}
\label{fig:sim_fits_with_noise}
\end{figure*}
Next, we compare the TESS light curves to analytic models from \citet{Kasen2010} that describe the interaction of the SN ejecta with a companion star.  In these models, the ejected material from the SN forms a bow shock as it encounters the companion star, producing a flash of X-ray emission followed by a quick rise and then slow decay of thermal emission at UV/optical wavelengths.  The observed light curve depends on the viewing angle of the observer because the viewing angle determines the solid angle of the shocked ejecta on the sky.  A viewing angle of $0^{\circ}$ places the companion in front of the SN explosion and maximizes the visibility of the shocked material, while a viewing angle of $180^{\circ}$ places the companion behind the exploded star so that no emission from the shocked material can be observed.
The other parameters of the \citet{Kasen2010} model are the separation of the white dwarf and secondary companion star (which sets the Roche lobe radius of the secondary), as well as the mass, velocity, and opacity of the SN ejecta.  In all calculations that follow, we adopt the fiducial model of \citet{Kasen2010}, with a viewing angle of $45^{\circ}$, an ejecta mass of 1.4 M$_\odot$, a velocity of 10$^4$ km s$^{-1}$, and an opacity of 0.2 cm$^2$ g$^{-1}$, while treating the companion separation/radius as a free parameter.   The \citet{Kasen2010} models assume that the companion fills its Roche-lobe, and so the companion separation can be used to estimate its radius.  The conversion between companion separation and radius depends weakly on the mass ratio of the white dwarf to companion star, and we assume masses of 1.4 M$_\odot$ and 1.0 M$_\odot$ respectively.  This choice is unimportant at the level of detail presented here.  

\citet{Maeda2018} derived similar light curve models as \citet{Kasen2010} for different assumptions about how energy from the SNe explosion is thermalized when colliding with the companion star.  In contrast, \citet{Kutsuna2015} used simulations to argue that the \citet{Kasen2010} models over-predict the companion signature, although their simulation did not include opacity from free-free emission and Compton scattering and so may have underestimated the luminosity of the shocked region.  More theoretical work might lead to a better understanding of the details of the companion signature models, and how to best distinguished companion signatures from other effects such as interactions with circumstellar material and inhomogeneities in the distribution of $^{56}$Ni in the SN ejecta \citep{Piro2016, Contreras2018, Maeda2018, Magee2020a, Magee2020b}.  For now, we use the \citet{Kasen2010} models, since they provide simple analytic prescriptions for the calculation of light curves and can easily be compared with observational studies that use the same models (e.g., \citealt{Cao2015, Hosseinzadeh2017, Shappee2019, Miller2020a}). 

We calculate light curves of these companion interaction models using synthetic photometry to combine the TESS wavelength-dependent response function, a blackbody spectrum with the temperature and photospheric radius defined by equations 24 and 25 of \citet{Kasen2010}, and the appropriate luminosity distance for each source (based on the redshifts in Table~\ref{tab:physical_data}).  Our calculations incorporate the effects of redshift on the rest-frame SED and time dilation on the light curves, although these effects are very small.  We do not include the effects of the travel time of the SNe ejecta to the companion, because the \citet{Kasen2010} models assume that the shocked ejecta expand homologously and are not valid at early times before a self-similar state is reached.  The ejecta travel-time scale is less than 15 minutes for the majority of the models we employ and so is shorter than the TESS FFI cadence.  However, the ejecta travel time scale can reach up to 3 to 12 hours for the largest separations that we consider ($>$$10^{13}$\, cm).  For convenience, we provide the TESS light curve models at a fiducial distance of 100~Mpc in Table~\ref{tab:kasen_lc}, and note that the flux can be rescaled with reasonable accuracy based on the luminosity distance alone.  

We investigate two issues relating to the signatures of companions in Type Ia SNe light curves.  First, we quantify the effect that the presence of a companion signature would have on the observed shape of the light curves.  Second, we place limits on the maximum companion signature detectable in the TESS light curves.

\floattable
\begin{deluxetable}{llrrrrrrr} \tablewidth{0pt}\tablecaption{Upper limits on Companion Parameters for \citet{Kasen2010} Models\label{tab:companion_params}}\tablehead{\multicolumn{2}{c}{Name}&\colhead{Separation}&\colhead{Radius} &\colhead{$t_0$}&\colhead{$\beta$} &\colhead{$A$} & \colhead{$dof$}&\colhead{$\chi^2/dof$} \\   & &\colhead{($10^{13}$ cm)}&\colhead{(R$_{\odot}$)}&\colhead{(BJD$-$2457000)} &   }\startdata\hline Best Fit\\ \hline
SN2018exc & ATLAS-18tne &  0.05 &   2.6 &    1333.09 & 2.74$\pm$0.09 & 7.32e-04$\pm$1.42e-04 &    755  &  0.83$\pm$0.05\\
SN2018fhw & ASASSN-18tb &  0.02 &   1.1 &    1344.72 & 1.37$\pm$0.03 & 3.53e-02$\pm$2.18e-03 &    947  &  0.99$\pm$0.05\\
SN2018fub & ASASSN-18ty &  0.25 &  13.7 &    1354.33 & 4.00$\pm$0.00 & 1.16e-05$\pm$6.56e-06 &    531  &  0.90$\pm$0.06\\
SN2018hib & Gaia-18czg &  0.02 &   1.1 &    1397.20 & 2.10$\pm$0.02 & 3.55e-03$\pm$1.77e-04 &    735  &  1.22$\pm$0.05\\
SN2018hkx & ASASSN-18yc &  0.06 &   3.4 &    1400.96 & 1.65$\pm$0.08 & 1.66e-02$\pm$1.98e-03 &    726  &  0.90$\pm$0.05\\
SN2018koy & ZTF-18adaifep &  0.33 &  18.1 &    1468.72 & 3.85$\pm$0.44 & 1.20e-04$\pm$1.05e-04 &    360  &  0.50$\pm$0.07\\
\hline  Max Companion (3$\sigma$ limit)\\ \hline
SN2018exc & ATLAS-18tne &  4.02 & 223.1 &    1327.09 & 4.00$\pm$0.33 & 4.35e-06$\pm$3.87e-06 &    755  &  1.17$\pm$0.05\\
SN2018fhw & ASASSN-18tb &  0.75 &  41.8 &    1345.82 & 4.00$\pm$1.18 & 6.60e-05$\pm$1.36e-04 &    947  &  1.18$\pm$0.05\\
SN2018fub & ASASSN-18ty &  1.32 &  73.1 &    1362.83 & 4.00$\pm$1.25 & 6.41e-04$\pm$9.45e-04 &    531  &  1.49$\pm$0.06\\
SN2018hib & Gaia-18czg &  0.43 &  23.9 &    1399.40 & 2.80$\pm$0.07 & 1.21e-03$\pm$1.60e-04 &    735  &  1.63$\pm$0.05\\
SN2018hkx & ASASSN-18yc &  1.00 &  55.3 &    1402.56 & 4.00$\pm$52.30 & 1.77e-05$\pm$1.06e-03 &    726  &  1.12$\pm$0.05\\
SN2018koy & ZTF-18adaifep &  0.43 &  23.9 &    1468.32 & 4.00$\pm$0.51 & 6.85e-05$\pm$7.00e-05 &    360  &  0.55$\pm$0.07\\
\hline  Max Companion (fixed $t_0$ from Table~\ref{tab:fit_params})\\ \hline
SN2018exc & ATLAS-18tne &  0.57 &  31.6 &    1329.09 & 4.00$\pm$0.30 & 1.01e-05$\pm$3.64e-06 &    755  &  1.17$\pm$0.05\\
SN2018fhw & ASASSN-18tb &  0.08 &   4.5 &    1343.12 & 2.54$\pm$0.08 & 2.06e-03$\pm$3.27e-04 &    947  &  1.27$\pm$0.05\\
SN2018fub & ASASSN-18ty &  0.19 &  10.4 &    1356.13 & 4.00$\pm$0.24 & 2.21e-05$\pm$1.20e-05 &    531  &  1.29$\pm$0.06\\
SN2018hib & Gaia-18czg &  0.05 &   2.6 &    1395.70 & 2.81$\pm$0.04 & 4.96e-04$\pm$4.14e-05 &    735  &  1.62$\pm$0.05\\
SN2018hkx & ASASSN-18yc &  0.33 &  18.1 &    1400.66 & 4.00$\pm$0.94 & 1.75e-04$\pm$1.42e-04 &    726  &  1.04$\pm$0.05\\
SN2018koy & ZTF-18adaifep &  0.33 &  18.1 &    1470.32 & 3.76$\pm$0.56 & 2.80e-04$\pm$2.69e-04 &    360  &  1.03$\pm$0.07\\
\enddata\tablecomments{Companion models are calculated for a fiducial model with a viewing angle of $45^{\circ}$, an ejecta mass of 1.4 M$_\odot$, a velocity of 10$^4$ km s$^{-1}$, and an opacity of 0.2 cm$^2$ g$^{-1}$.  See \S\ref{sec:models}  and \citet{Kasen2010} for details.  Uncertainties on $A$ and $\beta$ are the formal errors in the fits (see Equation~\ref{equ:model}).  Note that the reduced $\chi^2$ values in the middle and bottom panels differ because of the finite resolution of the grid search, or because of physical limits on the normalization for SN2018hkx and SN2018koy (see \S\ref{sec:companion_search} for details).}\end{deluxetable}

\subsection{Effect of companions on light curve shape}\label{sec:lc_shapes}

To begin, we simulated SN light curves  and investigated the effect that a range of \citet{Kasen2010} models would have on a single power law fit.  We simulated companion interaction models over a range of companion radii from  1 R$_{\odot}$ to 200 R$_{\odot}$.  Light curves were calculated by adding the \citet{Kasen2010} companion model light curves to a fireball $t^2$ rise.  The observed flux in the TESS band of the fireball was calculated using synthetic photometry on a $10^4$\, K blackbody spectrum, assuming a sphere with initial radius of $3\times 10^8$\, cm expanding at $10^9$\, cm s$^{-1}$.  To convert to flux, we arbitrarily placed the source at $z=0.01$ (43 Mpc).   We also experimented with the models of \citet{Piro2012}, which predict the temperature and photospheric radius of a hot shell of SNe ejecta powered by radioactive decay of $^{56}$Ni.  However, the normalizations of the \citet{Piro2012} models only slightly deviate from the $t^2$ fireball model used here, and the dependence of photospheric temperature on time in the \citet{Piro2012} models is very weak.  We therefore did not find any important differences when using the \citet{Piro2012} models compared to using a fireball.  The \citet{Kasen2010} companion light curve models are shown in the left-hand side of Figure~\ref{fig:model_components}, while the companion models, fireball rise, and their sum are shown in the right-hand panels.  Finally, we fit these model light curves using Equation~\ref{equ:model} described in \S\ref{sec:analysis}, and compared the resulting normalization and power law indices to the input values.  The results are given in Table~\ref{tab:sim_params} and the fits and residuals are shown in Figure~\ref{fig:sim_fits_no_noise}.

The parameter $A$ in equation~\ref{equ:model} can be interpreted as the flux of the model one day after the explosion ($t -t_0 = 1$ day).  Larger companions result in larger fluxes after one day owing to the significant contribution of flux from the companion component.  There is also a small bias in the inferred explosion time $t_0$ to earlier times ($-$0.008 days, about 11 minutes), although the bias is smaller than the 30 minute TESS FFI cadence.  The shapes of the companion light curves have a characteristic $t^{1/2}$ power law, which leads to a more linear light curve when combined with the intrinsic $\beta =2$ rise.  Table~\ref{tab:sim_params} shows that the fitted power law index ranges from 1.76 to 1.9 for small companions with radius $\lesssim 15$ R$_{\odot}$.  This range of companion radii will generally yield results consistent with a fireball model.  Larger companion radii (25 R$_{\odot}$ to 200 R$_{\odot}$) result in much smaller power law indices,  with $\beta=1.4$ to 1.1.  This result may be consistent with the SNe in this study that show power law indices close to unity (SN2018hkx, SN2018hpu, and SN2018koy).

However the residuals of these fits show a characteristic pattern where the linear rise under-predicts the companion flux at early times and over-predicts the fireball flux at late times (Figure~\ref{fig:sim_fits_no_noise}).  Thus, it is still possible to distinguish between an intrinsically linear rise in the SNe light curve and a large companion.

To  compare these simulations to observations, we next refit the models after adding noise.  We added Gaussian deviates to each point with a standard deviation equal to 4\% of the fireball rise after 14 days, which approximately reproduces our best light curves.  These results are shown in Figure~\ref{fig:sim_fits_with_noise}.  The measurement noise washes out most of the residual pattern on 30 minute time scales, but is often recoverable after binning to eight hours (cyan points in the right-hand panels of Figure~\ref{fig:sim_fits_with_noise}).

We are left with two main conclusions.  First, small companions lead to small biases in the power law index  and small systematic residuals.  For large companion stars, the power law rises are very flat, and the residuals show a characteristic pattern associated with the change in power law index from times when the companion emission dominates to times when the fireball dominates.  The detailed residuals can distinguish between an intrinsically linear light curve and a composite fireball-plus-companion light curve, given sufficiently high signal-to-noise ratios.

\begin{figure*}
    \centering
    \begin{tabular}{cc}
         \includegraphics[width=0.5\textwidth]{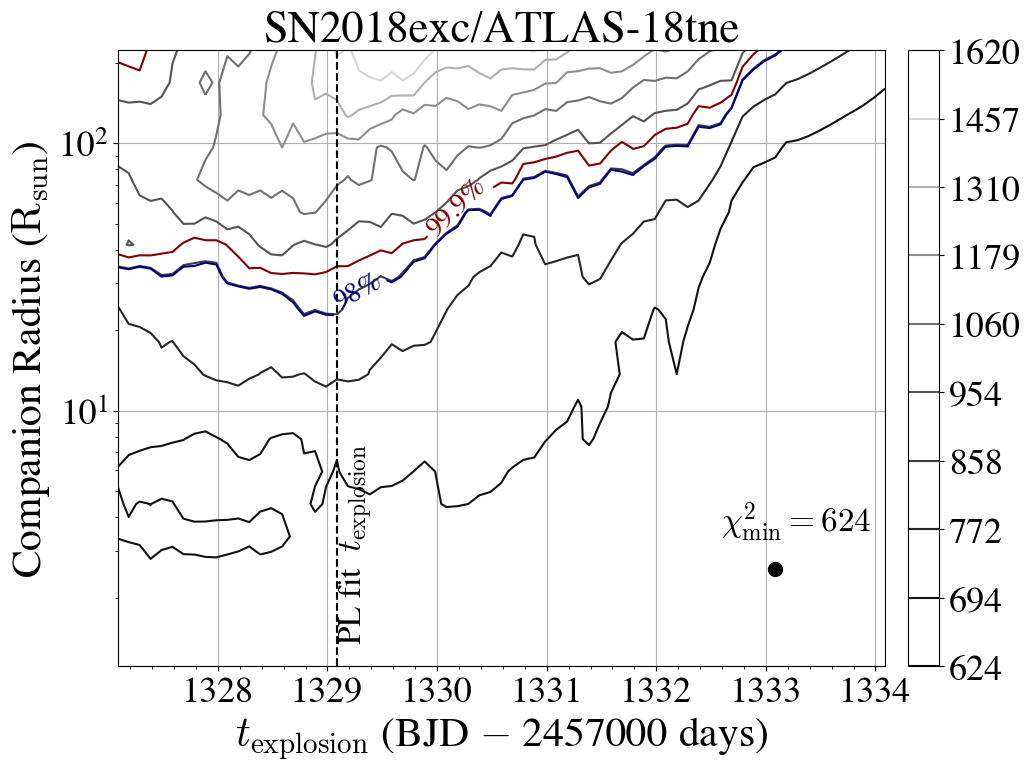}& \includegraphics[width=0.5\textwidth]{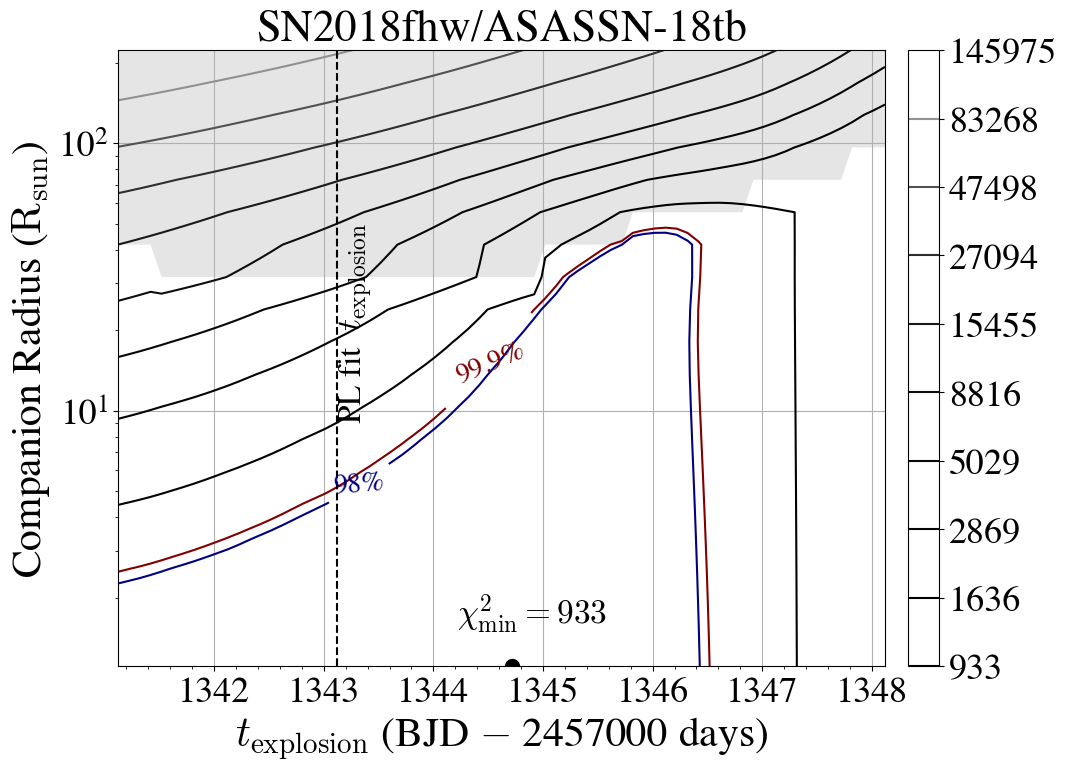}\\ 
         \includegraphics[width=0.5\textwidth]{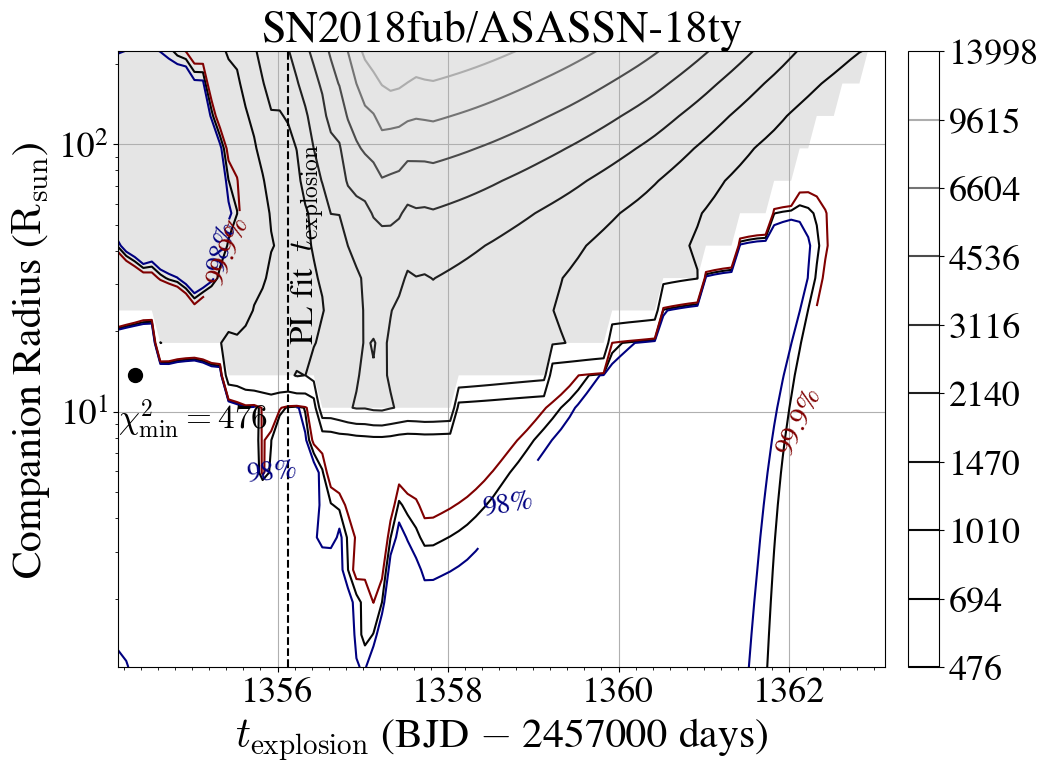}& \includegraphics[width=0.5\textwidth]{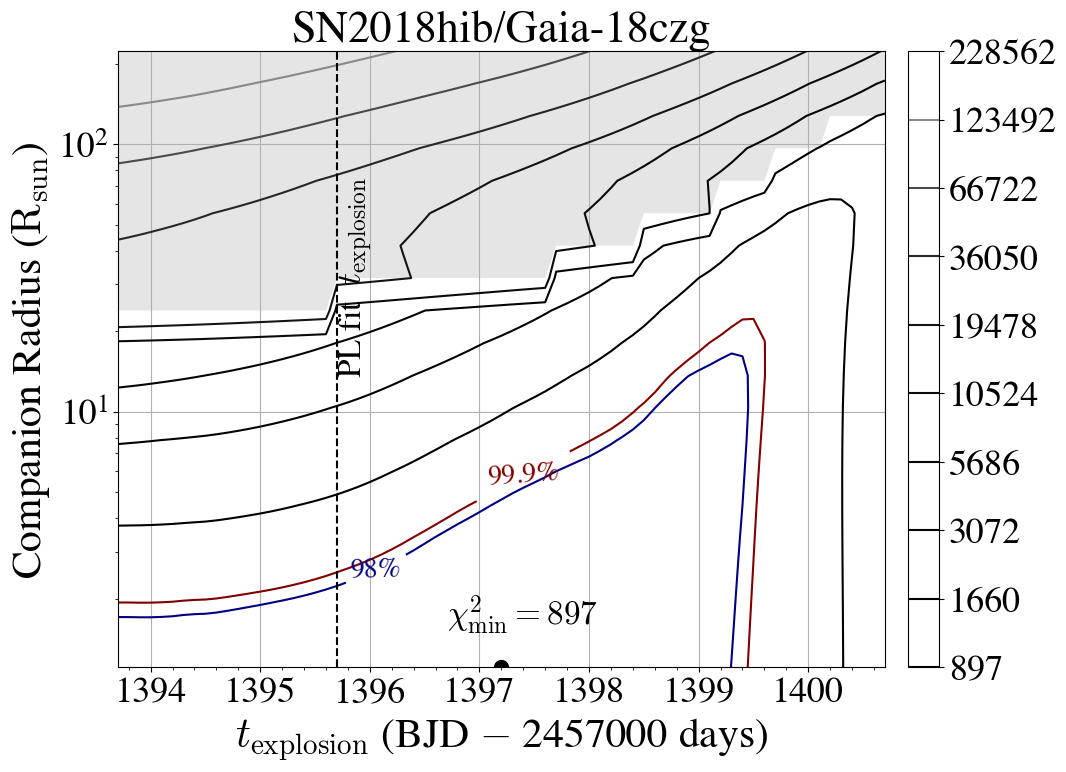}\\
         \includegraphics[width=0.5\textwidth]{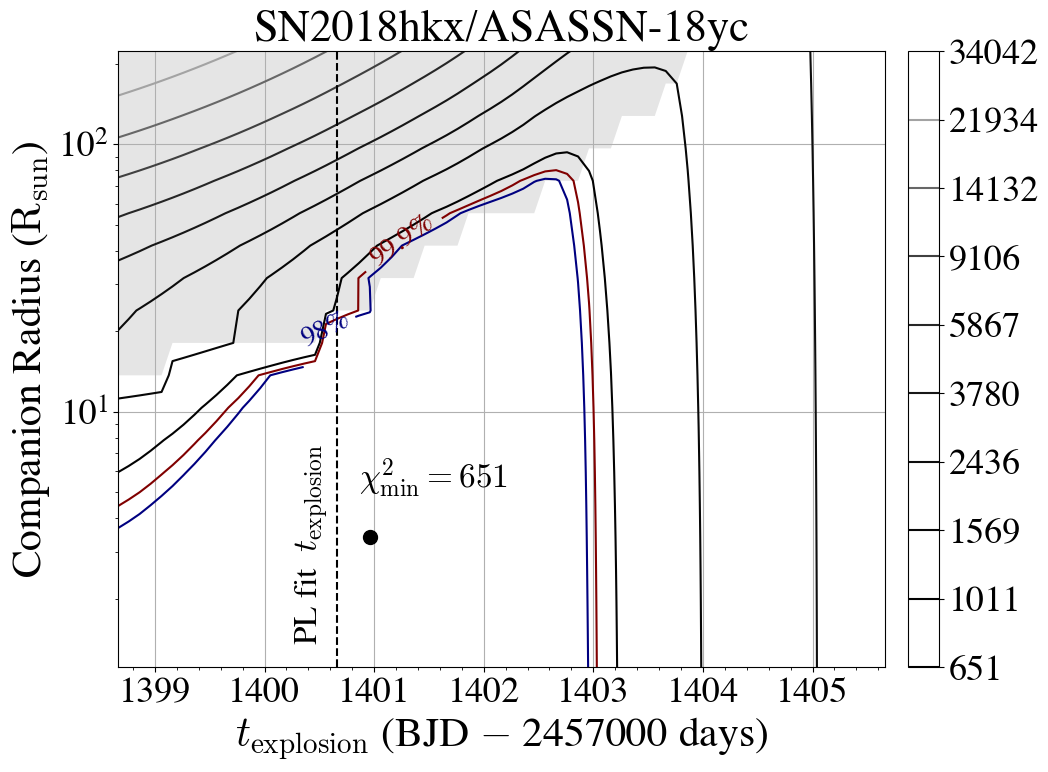}& \includegraphics[width=0.5\textwidth]{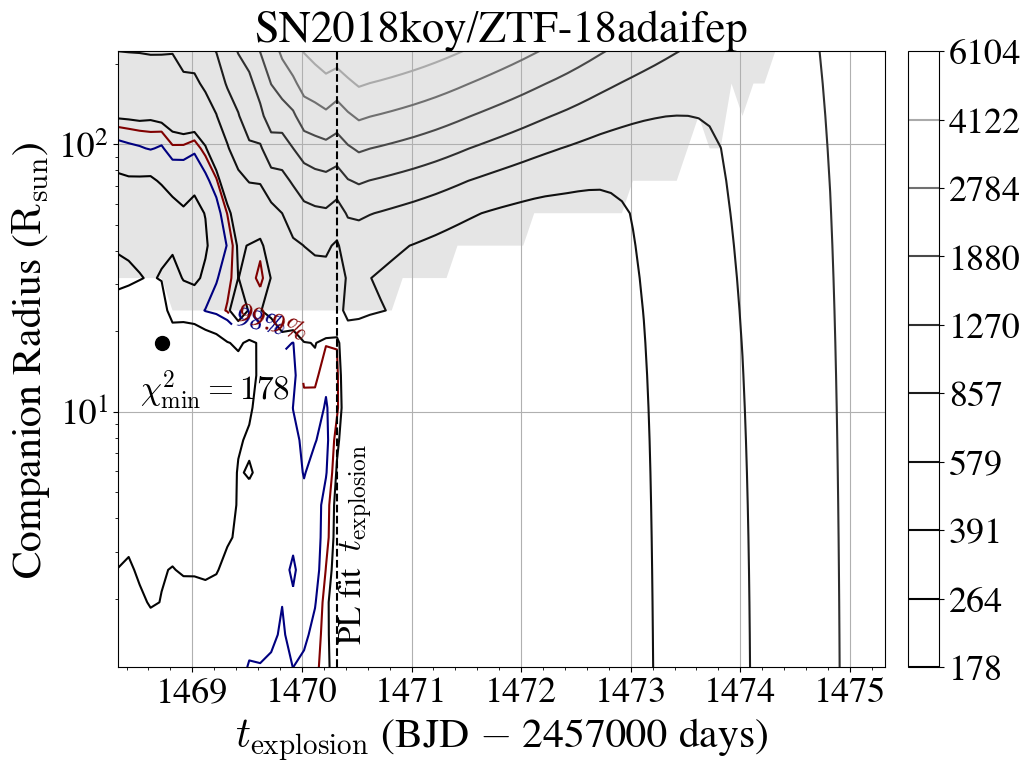} \\
    \end{tabular}
    \caption{$\chi^2$ contours for the grid-search of \citet{Kasen2010} companion models discussed in \S\ref{sec:companion_search}.  The minimum value of $\chi^2$ is marked for each SN.  The gray contours show values of $\chi^2$ equally spaced in logarithm, while the gray shaded regions correspond to non-physical solutions described in \S\ref{sec:companion_search}.  The blue and red contours mark the 98\% and 99.9\% confidence intervals of $\chi^2$ based on Monte Carlo simulations (see Appendix~C).  The vertical black dashed lines show the time of explosion derived from the simple power law fits in \S\ref{sec:analysis} (see Table~\ref{tab:fit_params}). Simulations suggest that a strong prior on $t_0$ to match the power law fits provides a more realistic constraint than the formally allowed maximum radius at the 99.9\% confidence interval.}
    \label{fig:companion_contours}
\end{figure*}

\begin{figure*}
    \centering
    \begin{tabular}{cc}
         \includegraphics[width=0.5\textwidth]{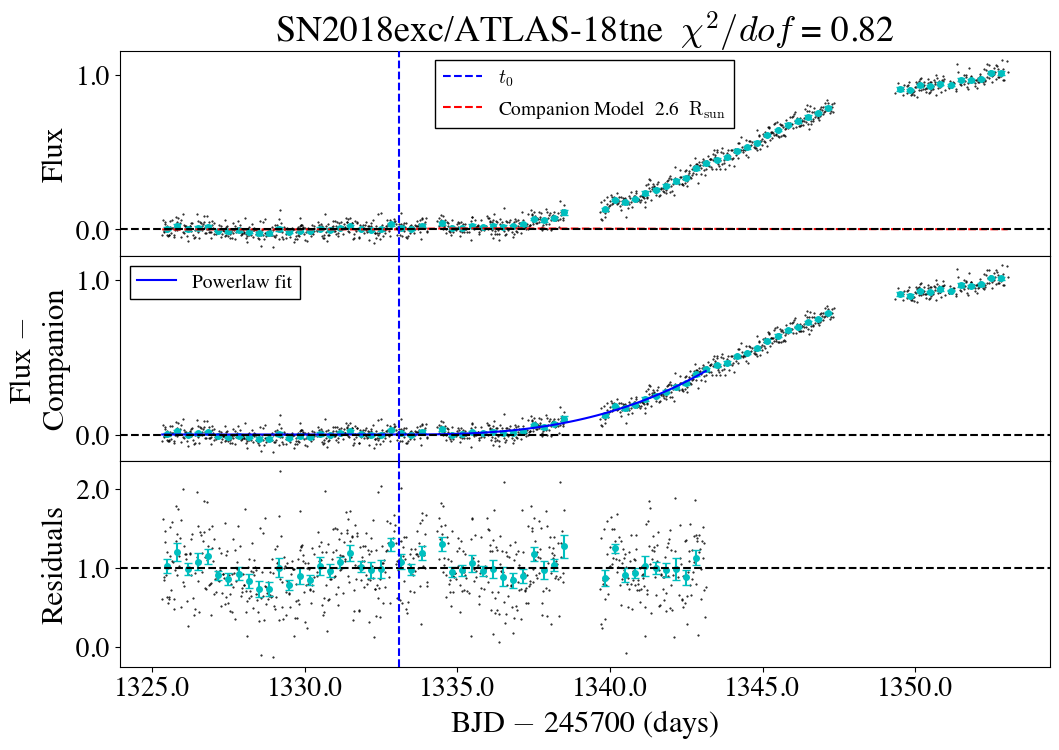}&
         \includegraphics[width=0.5\textwidth]{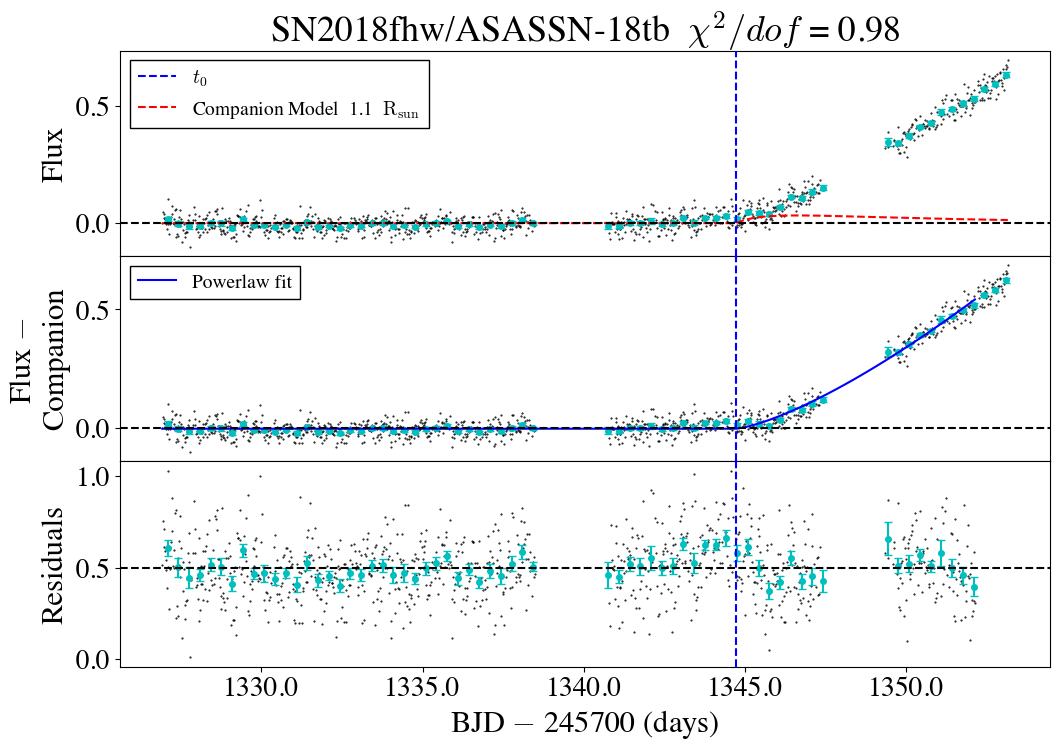}\\ 
         \includegraphics[width=0.5\textwidth]{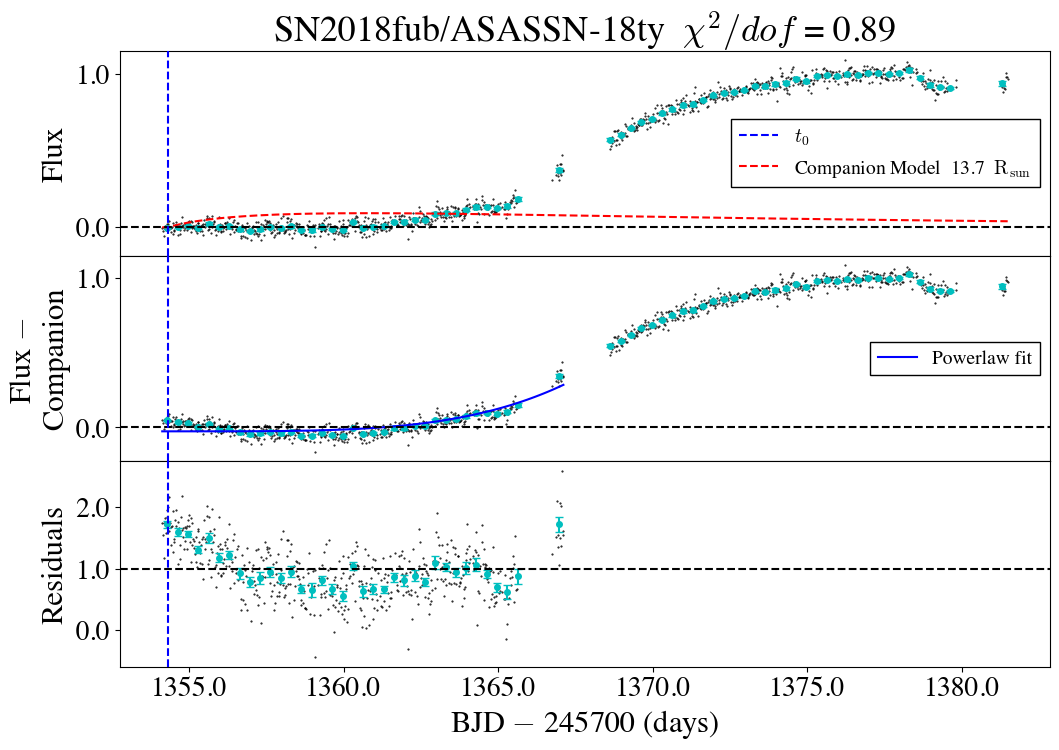}&
         \includegraphics[width=0.5\textwidth]{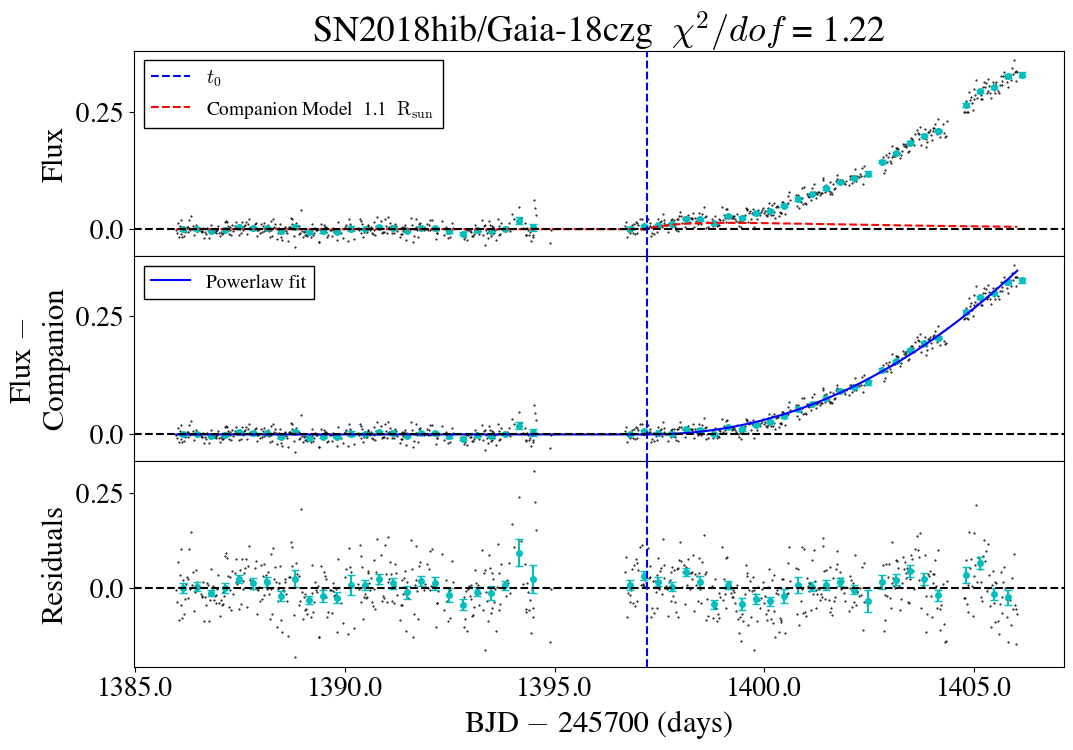}\\
         \includegraphics[width=0.5\textwidth]{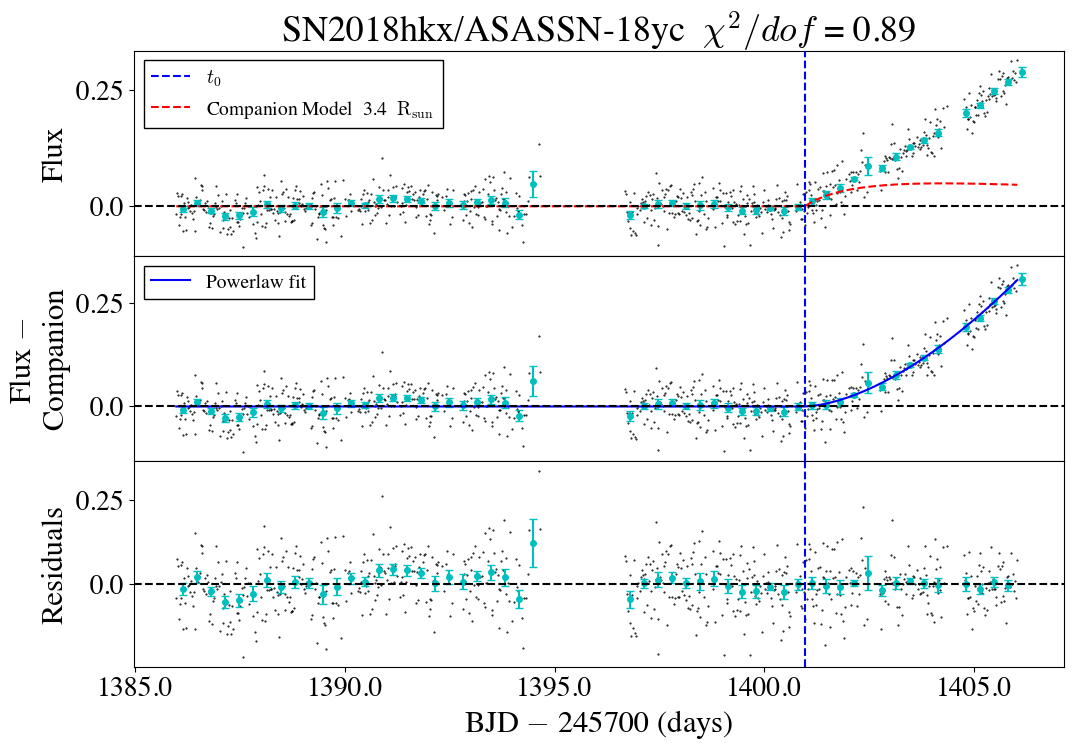}&
         \includegraphics[width=0.5\textwidth]{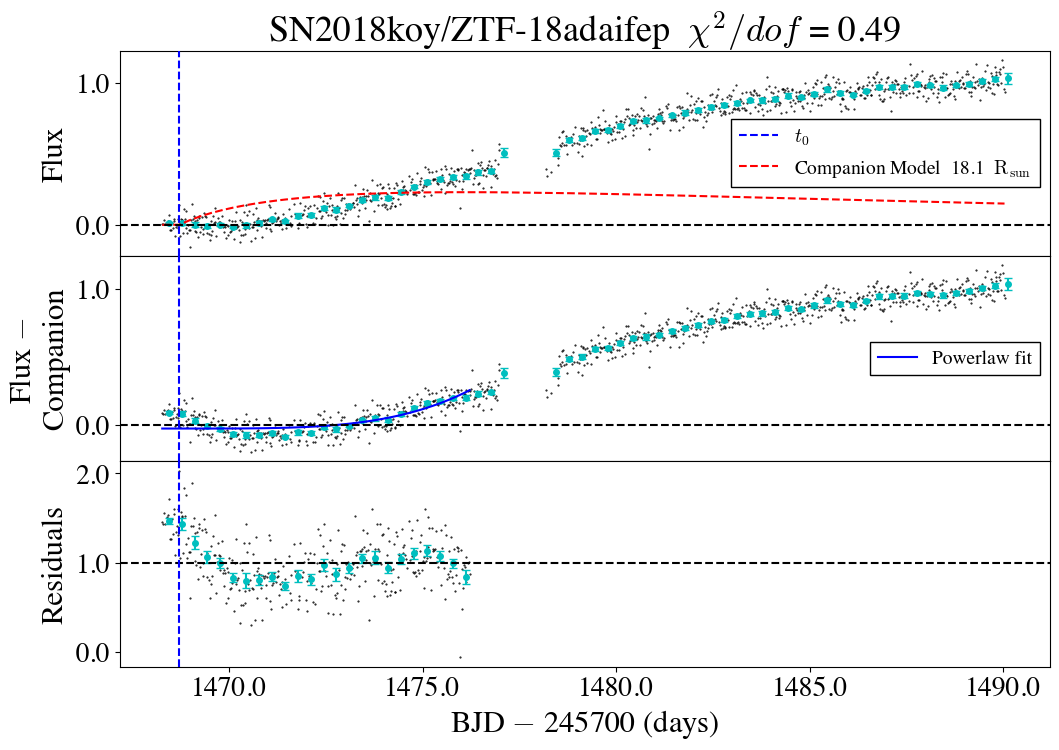} \\
    \end{tabular}
    \caption{Best-fit companion and power law models from the grid search in \S\ref{sec:companion_search}.  The top panels show the raw light curve (black dots) and data binned to 8 hours (cyan circles), with the best fit companion model from \citet{Kasen2010} in red.  The middle panels show the data with the \citet{Kasen2010} model subtracted out and the best-fit power law (with $t_0$ held fixed).  The bottom panel shows the residuals of the total fit.}
    \label{fig:companion_best_fit_chi2}
\end{figure*}

\begin{figure*}
    \centering
    \begin{tabular}{cc}
         \includegraphics[width=0.5\textwidth]{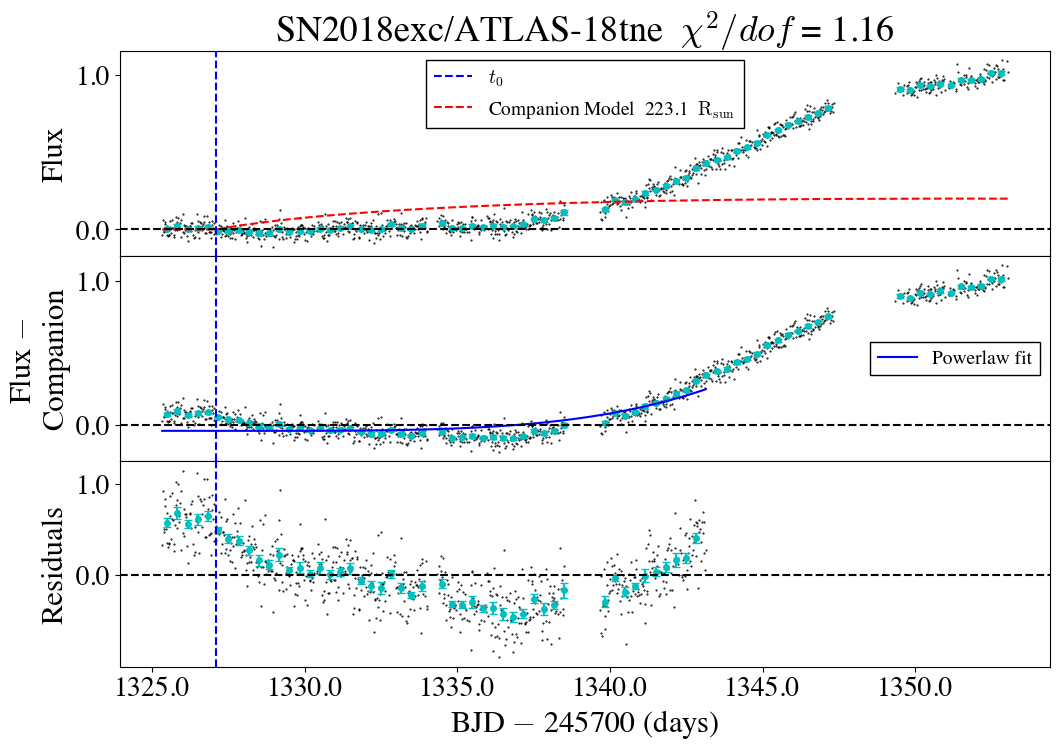}&
         \includegraphics[width=0.5\textwidth]{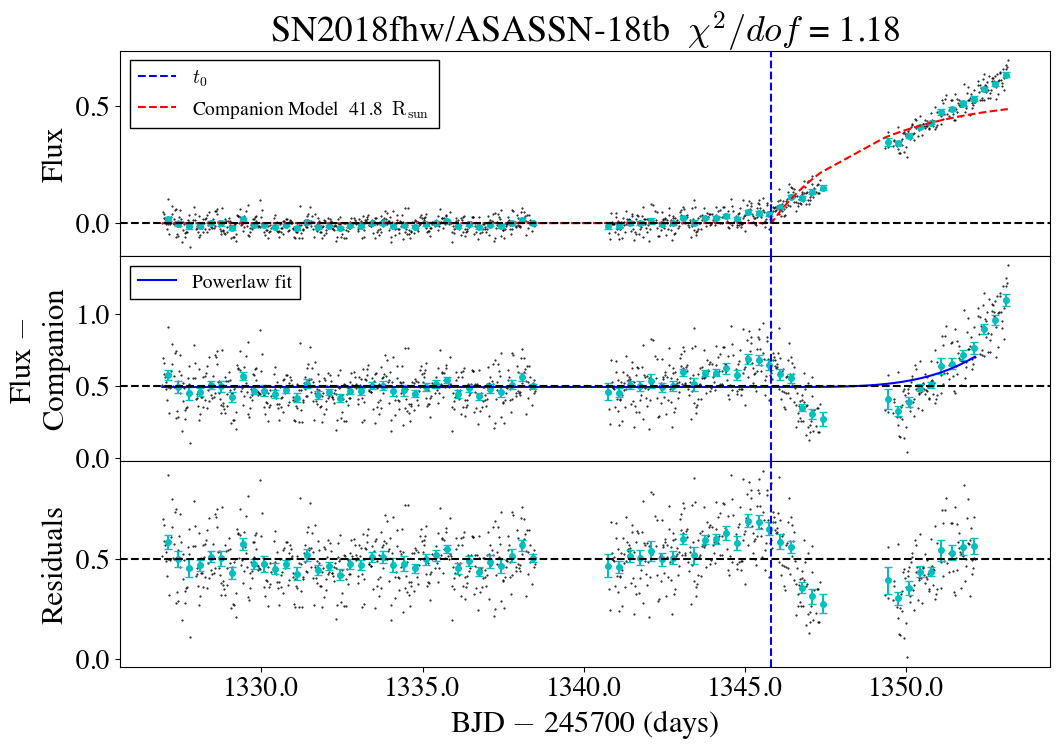}\\ 
         \includegraphics[width=0.5\textwidth]{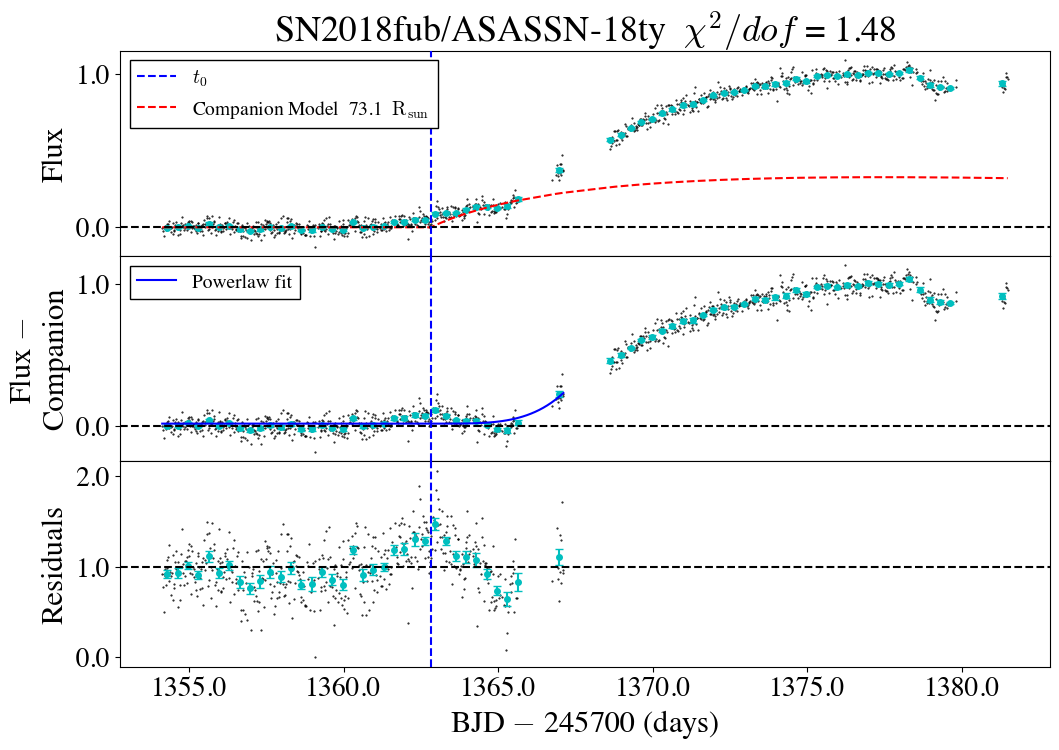}&
         \includegraphics[width=0.5\textwidth]{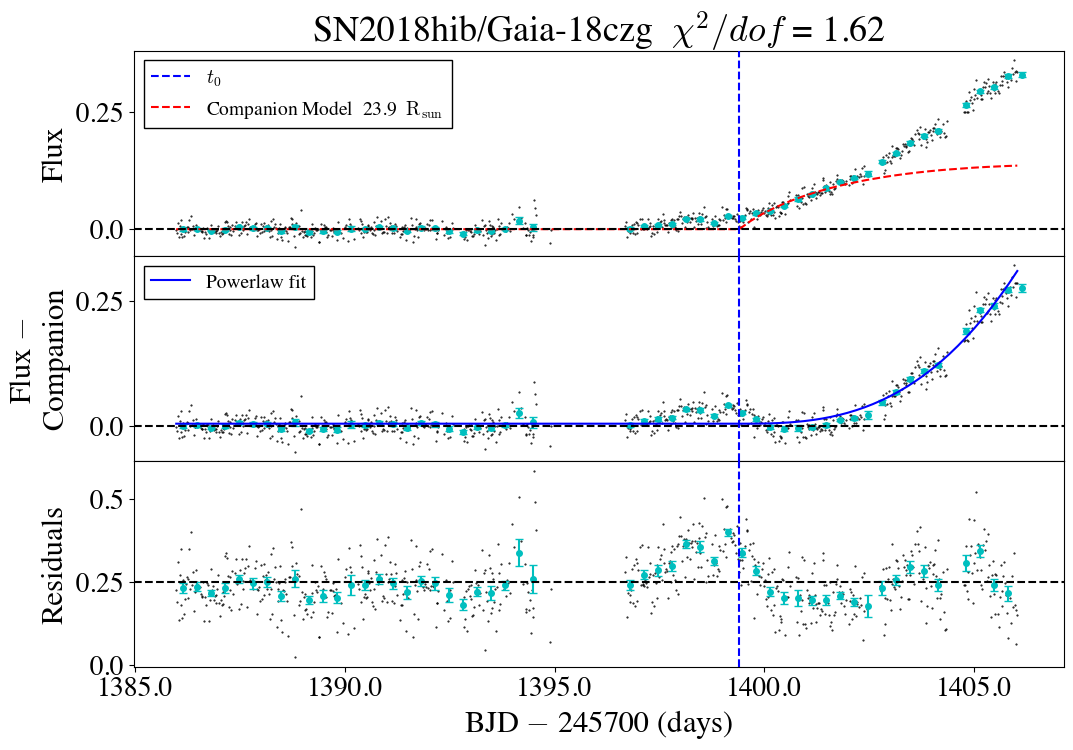}\\
         \includegraphics[width=0.5\textwidth]{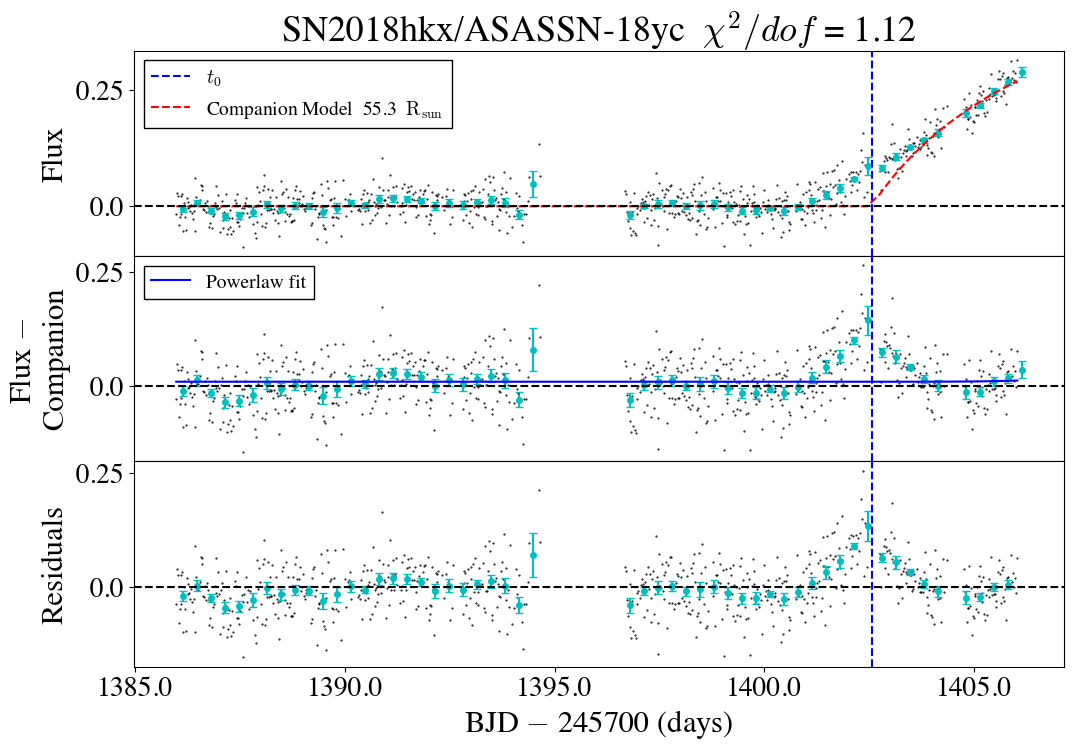}& 
         \includegraphics[width=0.5\textwidth]{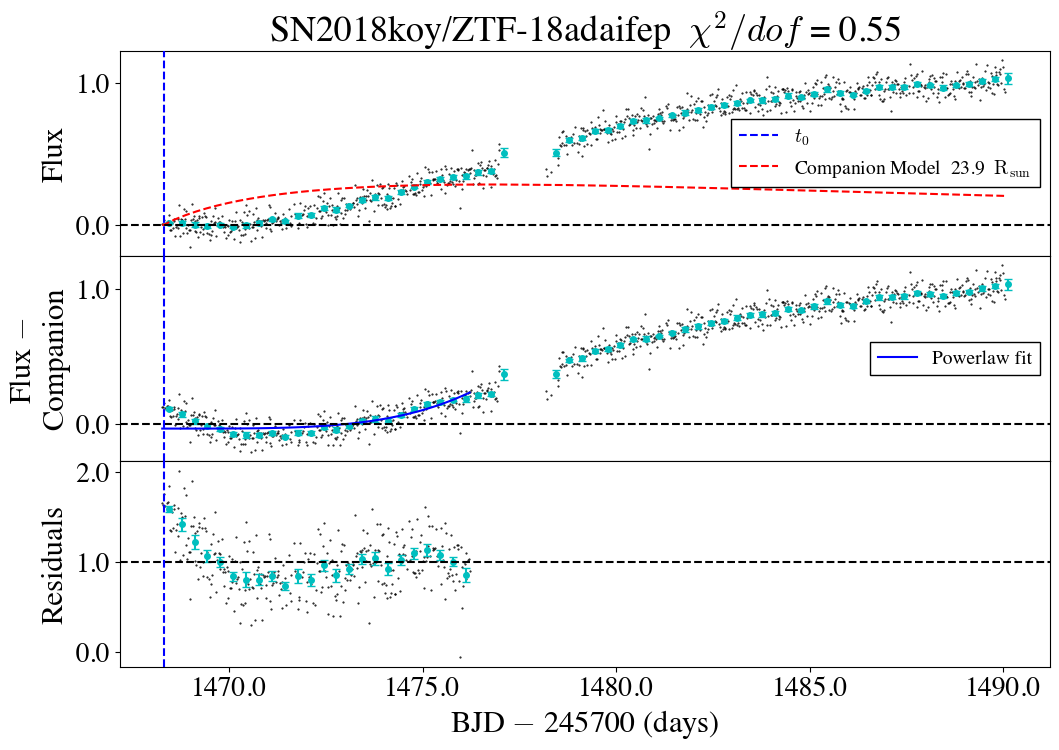} \\
    \end{tabular}
    \caption{Same as Figure~\ref{fig:companion_best_fit_chi2}, but for the maximum allowable \citet{Kasen2010} model at the 99.9\% confidence level (3$\sigma$ limit).}
    \label{fig:companion_max}
\end{figure*}

\begin{figure*}
    \centering
    \begin{tabular}{cc}
         \includegraphics[width=0.5\textwidth]{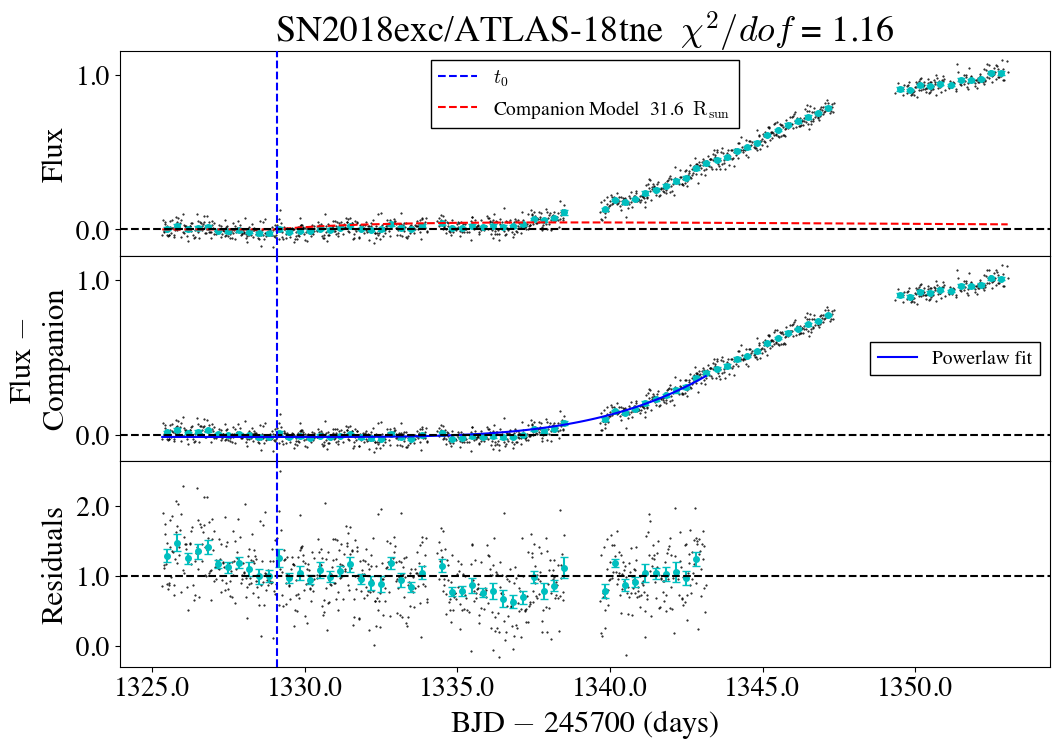}&
         \includegraphics[width=0.5\textwidth]{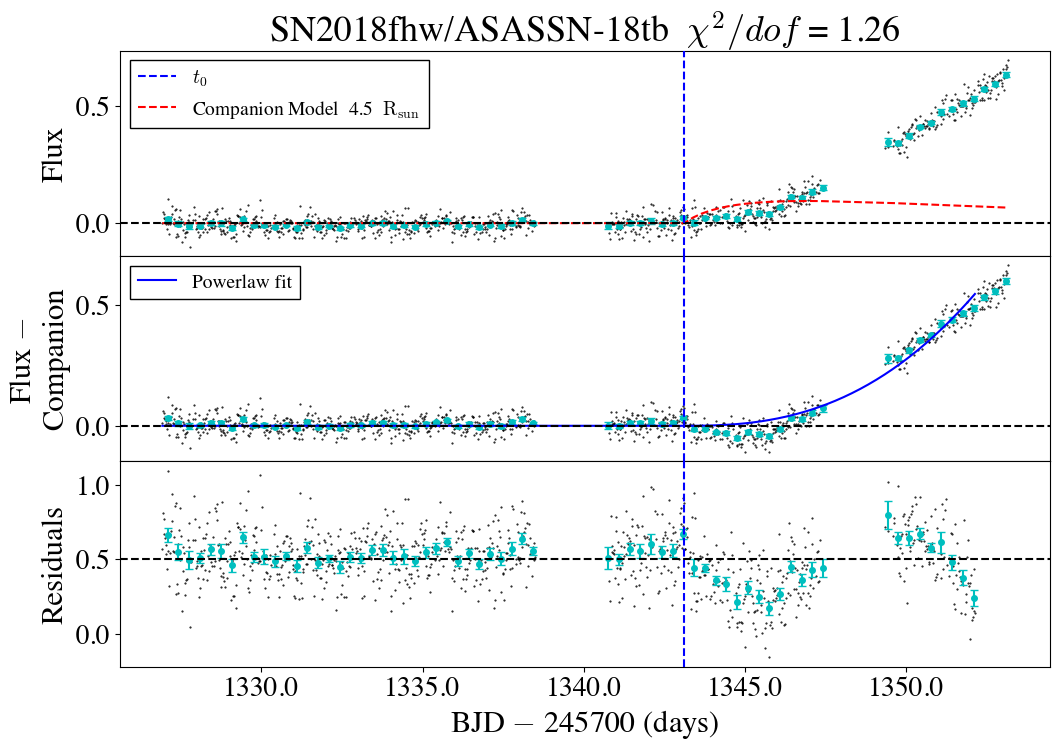}\\ 
         \includegraphics[width=0.5\textwidth]{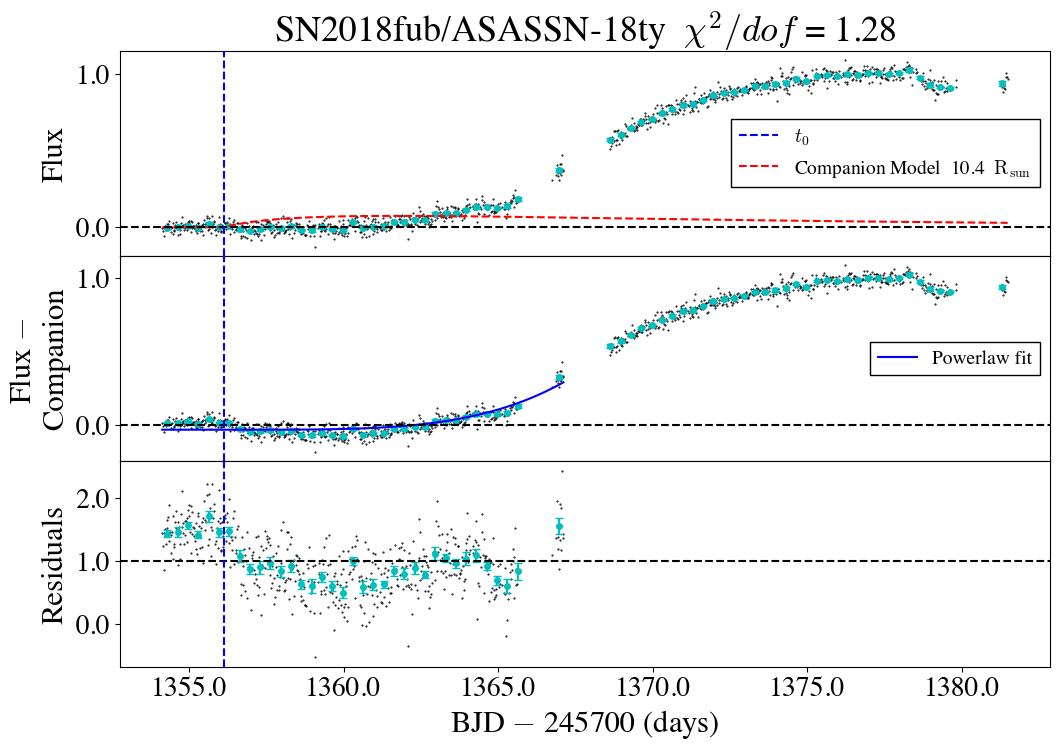}&
         \includegraphics[width=0.5\textwidth]{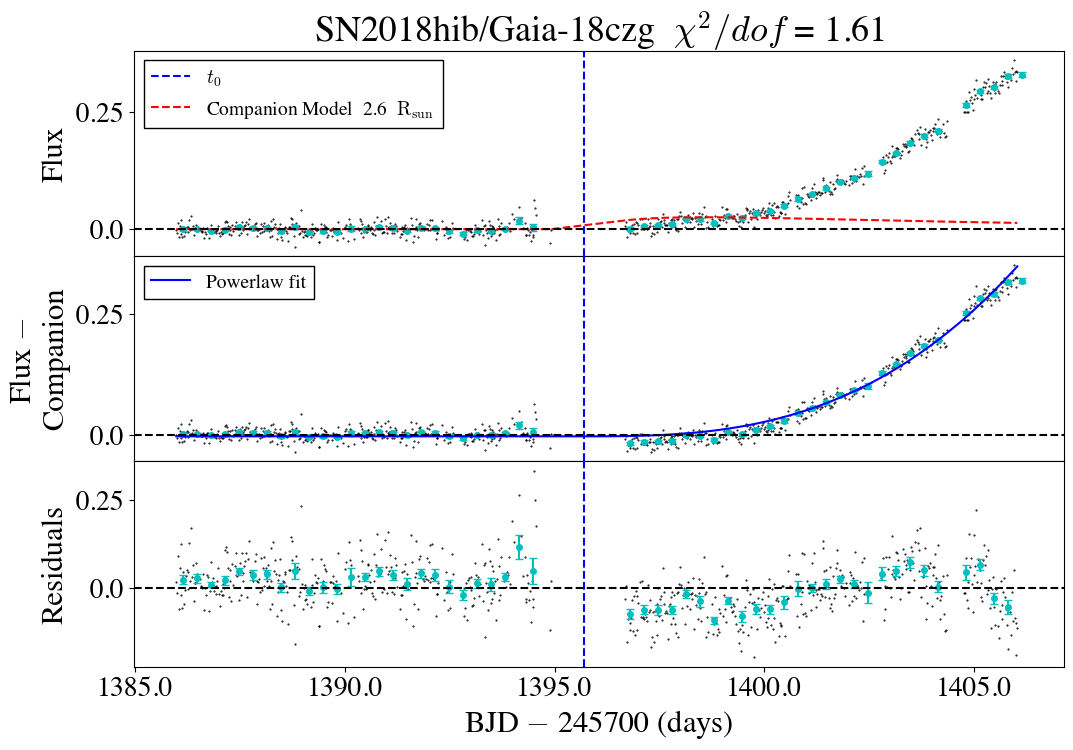}\\
         \includegraphics[width=0.5\textwidth]{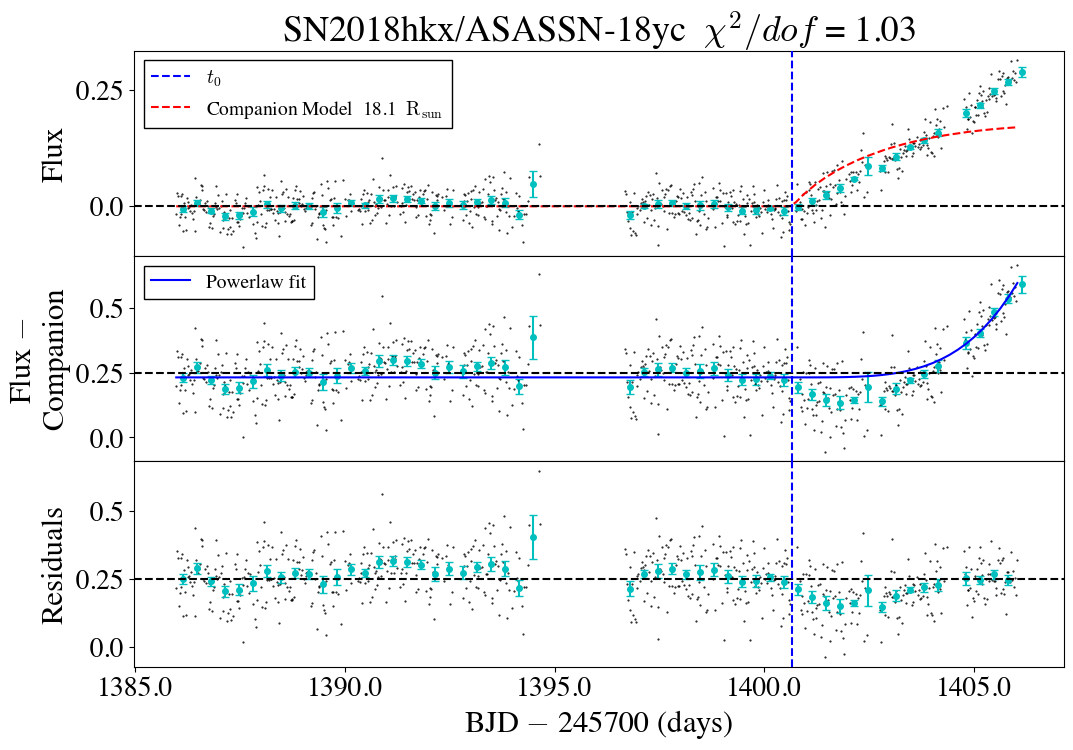}& 
         \includegraphics[width=0.5\textwidth]{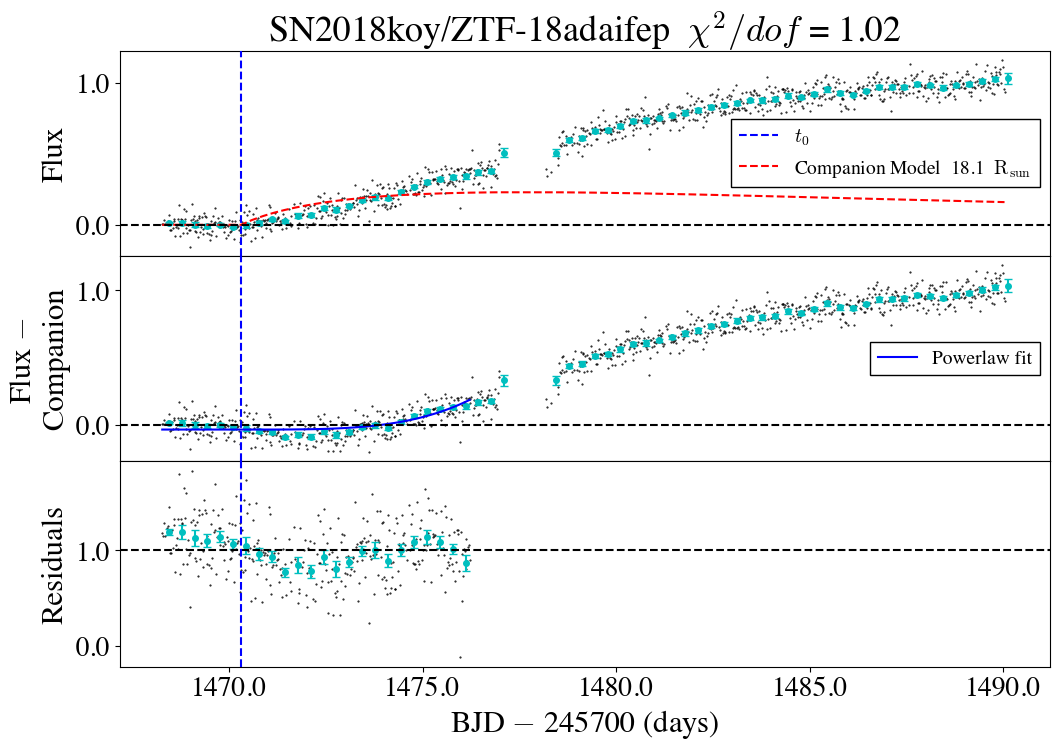} \\
    \end{tabular}
    \caption{Same as Figure~\ref{fig:companion_best_fit_chi2}, but for the maximum allowable \citet{Kasen2010} model at the 99.9\% confidence level (3$\sigma$ limit) with $t_0$ held fixed to the value in Table~\ref{tab:fit_params}.  Simulations show that $t_0$ is not significantly biased when fitting a power law model, while the models in Figure~\ref{fig:companion_max} push $t_0$ to significantly later times. We therefore adopt these models as our final 3$\sigma$ limits.}
    \label{fig:companion_max_texp}
\end{figure*}

\subsection{Search for companions in the TESS SNe light curves}\label{sec:companion_search}

Finally, we  searched  each SN light curve for companion signatures that match the \cite{Kasen2010} models.  We searched a grid of companion models with 20 separations/radii over 70 initial explosion times $t_0$ from $-$2 to $+$5 days relative to the value of $t_0$ found in \S\ref{sec:analysis}.  For each combination of companion radius and $t_0$, we first subtract the \citet{Kasen2010} model from the observed light curve, and then fit a power law model, optimizing $A$, $\beta$, and $B$ in Equation~\ref{equ:model} (with $t_0$ held fixed)  to minimize $\chi^2$.    The companion separations/radii are equally spaced in logarithm and range from  $2.2\times 10^{11}$ cm to $4\times 10^{13}$ cm ($1.1$ R$_\odot$ to $223$ R$_\odot$), while the values of $t_0$ are evenly spaced every 2.4 hours.  We verified that this fitting procedure and grid spacing will accurately recover the parameters of any companion signatures by performing Monte Carlo simulations, the details of which are discussed in Appendix~C.  We also used these simulations to identify the $\chi^2$ values that correspond to the 98\% and 99.9\% confidence limits on the presence of any companion signatures.

We restrict our analysis to the six SNe where continuous observations are available through a significant portion of the rising light curve (SN2018exc, SN2018fhw, SN2018fub, SN2018hib, SN2018hkx, and SN2018koy). This removes SN2018hzh, SN2018hpu, and SN2018jwi from further consideration.  Figure~\ref{fig:companion_contours} shows the  $\chi^2$ contours for the six SN on the grid of companion radii and $t_0$, and Figure~\ref{fig:companion_best_fit_chi2} shows the best-fit companion and power law models.

The gray regions in Figure~\ref{fig:companion_contours} show unphysical solutions where subtracting a large companion model causes the light curve flux to become negative.  However, a formally reasonable fit can be achieved by forcing the model normalization $A$ to a very small value.  In these cases, we have flagged solutions where $A <  10^{-7}$ (in flux units relative to peak) as unphysical.  A similar example of this family of solutions is shown in the bottom left panel of Figure~\ref{fig:companion_max} for SN2018hkx; in this case, the normalization is $A\approx 1.8\times10^{-5}$.  We tried refitting the data with a prior on the normalization $A$ of $1/A$ (such that small $A$ increases $\chi^2$), but simulations show that this biases the recovered companion separation/radius to the smallest value on the grid in order to keep $A$ as large as possible, even when large companion signatures ($>$5 \rsun) are injected into the light curves.  Thus, in Figure~\ref{fig:companion_contours} we simply mark unphysical solutions with small values of $A$.  In cases where the 3$\sigma$ contours extend into non-physical regions of parameter space, we adopt the the solution at fixed $t_0$ with maximum companion separation/radius and $A>10^{-7}$ instead of the 3$\sigma$ contour.  This issue only affects the companion limits for SN2018hkx and SN2018koy.

For each SN, there are three solutions of particular interest:  the best-fit, the 3$\sigma$ limit for the maximum companion separation/radius, and the 3$\sigma$ limit at the time of explosion inferred from the initial fit in \S\ref{sec:analysis}.  The parameters for these solutions are shown in Table~\ref{tab:companion_params}, where the uncertainties on $A$ and $\beta$ are from the formal errors in the fits.  Figures~\ref{fig:companion_best_fit_chi2}, \ref{fig:companion_max}, and \ref{fig:companion_max_texp} show each of these solutions for each SN.  For simplicity, we will only discuss the results in terms of the companion model radii, although this is equivalent to using the companion separation because the \citet{Kasen2010} models assume Roche-lobe overflow.  The best-fit companion radii of SN2018exc, SN2018fhw, SN2018hib, and SN2018hkx are all less than 4 \rsun.  The best fit radii for SN2018fhw and SN2018hib are the minimum value for the search of 1.1 \rsun.  SN2018fub and SN2018koy have larger best fit values of 13.7 \rsun and 18.1 \rsun, respectively.  However, these models over-predict the flux at early times and are unlikely to be physical even if they formally have the best $\chi^2$ (see Figure~\ref{fig:companion_best_fit_chi2}).

In all cases, the 3$\sigma$ limits allow for larger companion radii. Larger companion models also push the explosion time $t_0$ to significantly later times than found in \S\ref{sec:analysis}, and Figure~\ref{fig:companion_max} shows that later values of $t_0$ result in a characteristic flux excess around the time when the light curves start rising.  These fits also drive the power law index $\beta$ to the maximum allowed value of 4.0.  This is because the large companion model pushes the rising SN flux to later times, and the power law must have a high value of $\beta$ at fixed $t_0$ to capture the rise.  The combination of the residual flux pattern, later value of $t_0$, and high value of $\beta$ are unlikely to be physical, although we cannot statistically rule out these solutions.  For these cases, we use the 3$\sigma$ companion radius limit at the value of $t_0$ that matches the initial fits from \S\ref{sec:analysis}.  This is equivalent to imposing a strong prior on $t_0$.  Appendix~C suggests that such a prior is reasonable, and provides a more detailed discussion of uncertainties in $t_0$.

We are able to rule out giant companions with radii $\gtrsim$\,32 R$_\odot$ in all six SNe, assuming the viewing angles are less than $45^\circ$.   Ignoring SN2018exc (which is at higher redshift than the other SNe), the limit improves to $\gtrsim$\,20 R$_\odot$. The strongest limits are for SN2018fhw and SN2018hib, at 4.5 \rsun and 2.6 \rsun, respectively.  We note that these fits still display coherent residual structures near the times of explosion which the simpler power law fits do not.  In all cases, the power law index is steeper than the fits in \S\ref{sec:analysis}.  Four out of five of these fits have $\beta \gtrsim 3.5$, which reflects the same issue described above ($\beta$ must increase at fixed $t_0$ as larger companion signatures are subtracted from the light curves).  Having good priors for $\beta$ would greatly improve the limits on companions in these light curves.

Errors in the adopted distances  affect these results by changing the observed flux of the companion signatures relative to the noise properties of the light curves.   Three of these sources, SN2018fhw, SN2018hib, and SN2018hkx, have $\Delta m_{15}$ measurements, and the differences in distance estimates using $\Delta m_{15}$ instead of the redshift are $-$0.23, $-$0.19, and  0.35 magnitudes, respectively. These would only change the \citet{Kasen2010} companion model normalizations by 10--16\% and have little effect on our results.  

Our calculation for SN2018fhw also agrees  with that of \citet{Vallely2019}, although our upper limit on the companion radius is higher (4.5\,\rsun\ rather than $\sim$1\,\rsun) because the redshift-based distance is 14\% larger than their adopted distance (74 instead of 65 Mpc, about a 0.3 mag difference). Our results also strongly depend on the assumed ejecta mass/velocity (1.4 M$_\odot$/$10^4$ km s$^{-1}$), and, too a lesser extent, the assumed ejecta opacity (0.2 cm$^2$ g$^{-1}$).  Finally, SN2018fhw is remarkable for the presence of broad H$\alpha$ emission in the nebular spectrum, discovered 139 days after maximum light \citep{Kollmeier2019}.  The presence of broad H$\alpha$ emission in the nebular phase of the SN is widely expected to be a signature of single degenerate progenitor systems.  However, \citet{Kollmeier2019} find that the mass of hydrogen inferred from the integrated H$\alpha$ luminosity is a factor of 100 smaller than the expected mass of hydrogen stripped from the companion star ($\gtrsim 0.1$\,M$_{\odot})$.  \citet{Vallely2019} analyze additional late time spectra, and find that the time evolution of the H$\alpha$ line profile and luminosity do not match the SN ejecta luminosity or velocity structure, as traced by Fe \textsc{iii} $\lambda$4660\,\AA\ emission.  Combined with the early time TESS light curve, \citet{Vallely2019} interpret the H$\alpha$ emission in SN2018fhw as a result of interactions with circumstellar material, and our results do nothing to change this interpretation.   H$\alpha$ emission associated with circumstellar emission is a hallmark of the so-called Type Ia-CSM SNe, although SN2018fhw is importantly different than typical objects of this class.  A more detailed discussion can be found in \citet{Kollmeier2019} and \citet{Vallely2019}.

\subsection{Implications for Type Ia SN progenitors\label{sec:progenitor_discussion}}
Failure to detect companion signatures can always be attributed to unfavorable viewing angles where the shocked ejecta are hidden by the optically thick SN ejecta.  However, if no companion signature is found in a large sample of SN light curves, confidence intervals for limits on companion radii can be inferred by calculating the probability that all the SNe are viewed at unfavorable angles.  For a uniform distribution of the cosine of the viewing angle, the probability that all five of the SNe with companions $\lesssim 20$\,\rsun are viewed at angles greater than 45$^\circ$ is 44\%.  The probability that both SN2018fhw and SN2018hib are viewed at angles greater than 45$^\circ$ is 72\%. 

Thus, this sample is  too small to put any meaningful constraints on the progenitor systems of Type Ia SNe.   To push these limits to the 10\%  or 1\% level requires 14 or 28 SNe, respectively.  
As of 2020 July, there have been 134 Type Ia in the TESS fields discovered at optical magnitudes brighter than 20th.  Not all of these objects will have early time observations in TESS, but the total number is very close to the expectation based on scaling the parent sample in this work (34 SNe) from six to 26 sectors. We therefore expect to have about $22\pm 5$\ SN light curves with limits $\lesssim 20$~\rsun and $9\pm 3$\ with limits $\lesssim 5$~\rsun.  
Analysis of SNe from the two years of the TESS primary mission will be presented in future work.

From stacking analyses of several hundred SNe from SDSS II and the Supernova Legacy Survey, \citet{Hayden2010} and  \citet{Bianco2011} did not detect any early time companion interactions.  \citet{Hayden2010} was able to rule out companion interactions up to 9\% of the peak SN flux, while \citet{Bianco2011} infer that less than 20\% of Type Ia SNe have companions with radii $> 110$ R$_\odot$.  These results disfavor red giants as typical companions in Type Ia SN progenitor systems.  Our results are in agreement, although there is a reasonable probability that any such companions were all viewed at unfavorable angles with only six supernovae.  However, all six of our objects have tighter flux limits than these studies, showing that TESS will improve the limits on typical Type Ia progenitor systems as the sample of early time TESS SN light curves grows.

There are six SNe from the literature that show evidence for a blue excess at early times that might be associated with a non-degenerate companion star: SN2012cg, iPTF14atg, SN2017cbv, HSC17bmhk, ASASSN-18bt, and SN2019yvq \citep{Marion2016, Cao2015, Hosseinzadeh2017, Jiang2020, Dimitriadis2019, Miller2020a}.  If these SNe are associated with single degenerate progenitors, a large TESS sample would automatically provide estimates of the relative rates of single and double degenerate Type Ia SNe.  However, the interpretation of these signals is debated---for SN2012cg, iPTF14atg, and ASASSN-18bt,  potential issues with the companion interaction model are discussed by \citet{Shappee2018}, \citet{Kromer2016}, and \citet{Shappee2019}, respectively.  For SN2017cbv and SN2019yvq, \citet{Hosseinzadeh2017} and \citet{Miller2020a} show that \citet{Kasen2010} models cannot fit both the UV and optical light curves of these sources.  For HSC17bmhk, the early time light curve is very sparse, and the blue excess can be reproduced by decay of $^{56}$Ni on the surface of the SN ejecta \citep{Jiang2020}.

\section{Future Prospects and Conclusion \label{sec:conclusion}}
\begin{figure}
    \centering
    \includegraphics[width=0.48\textwidth]{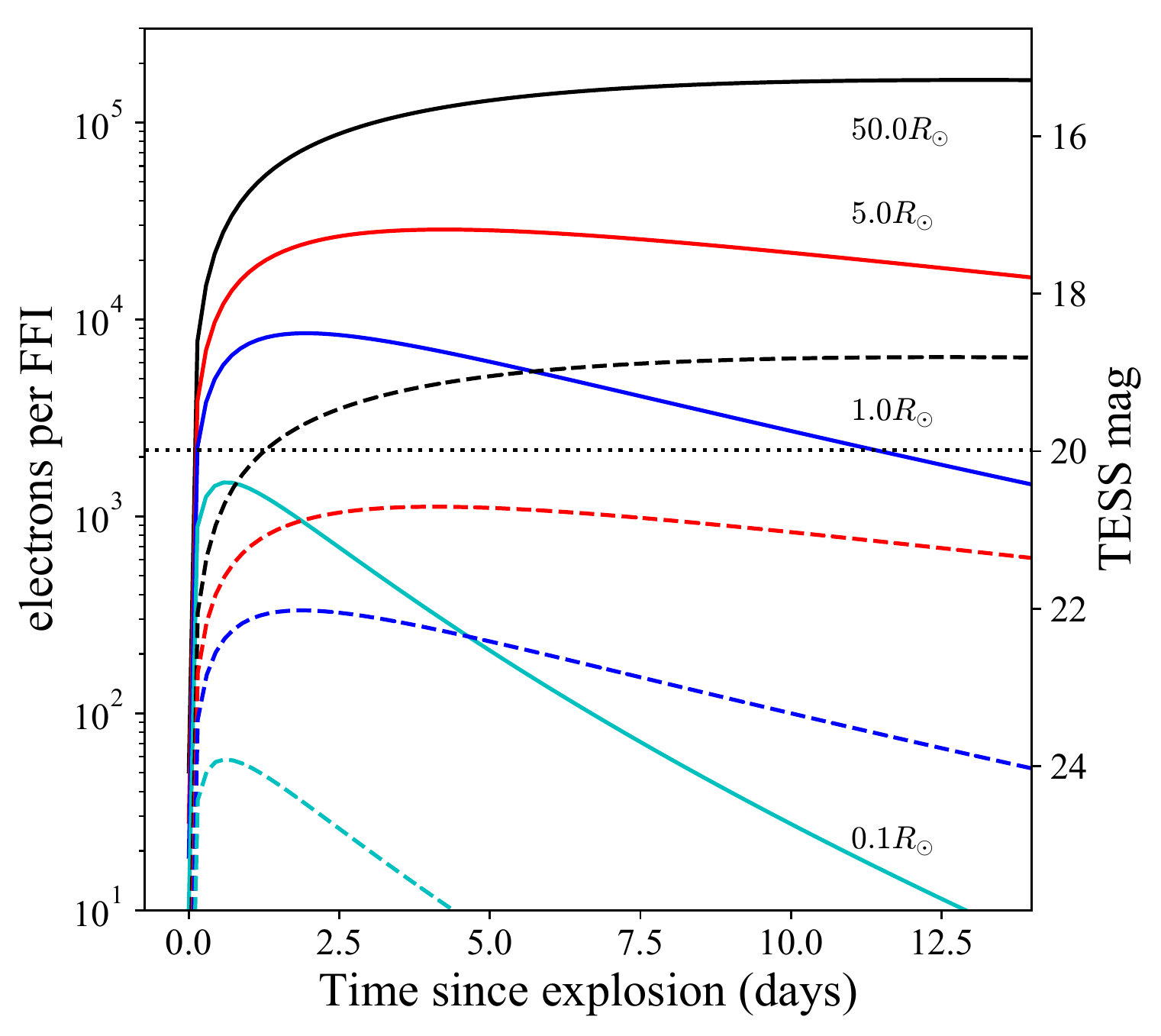}
    \caption{Light curves of \citet{Kasen2010} models, converted to electrons per TESS FFI and TESS magnitudes.  The models show the expected response of the SNe ejecta shocked by colliding with a companion star---flux from the ejecta itself (powered by the decay of $^{56}$Ni) is not included.  The light curves are calculated for a fiducial model with a viewing angle of $45^{\circ}$, an ejecta mass of 1.4 M$_\odot$, a velocity of 10$^4$ km s$^{-1}$, and an opacity of 0.2 cm$^2$ g$^{-1}$ (see \S\ref{sec:models}). We show light curves for R $=$ 0.1, 1.0, 5.0, and 50 R$_\odot$ companions (from faintest to brightest) at either 50 Mpc ($z \approx 0.01$, solid curves) or 200 Mpc ($z \approx 0.05$, dashed curves).  The horizontal line shows the typical 3$\sigma$ detection limit for an 8 hour average of the FFIs.  Within about $\sim$50 Mpc ($z \approx 0.01$), we expect to be able to detect emission from an 0.1--1.0 R$_\odot$ companion (see \S\ref{sec:conclusion}).}
    \label{fig:tess_obs}
\end{figure}

This work shows the impact that TESS can have on transient and SN science.  With its unique combination of continuous monitoring and survey area, TESS will be able to characterize the early time light curves of a large number of bright SNe in great detail, and provide a sample of sources with which to constrain progenitor models.  

With regard to the progenitor systems of Type Ia SNe, we have calculated the observable properties of companion signatures in order to identify the most promising SNe for future TESS analysis.  Figure~\ref{fig:tess_obs} shows the expected TESS light curves for the \citet{Kasen2010} models (with no SN component).  As shown in \S\ref{sec:lc_shapes}, emission from the collision with a companion star would dominate the TESS light curve for the first $\sim$2 days after explosion, at least for the fiducial parameters adopted in \S\ref{sec:models}.  We also provide tabulations of the models for different companion radii at a distance of 100 Mpc in Table~\ref{tab:kasen_lc}.  For any SNe within $\sim$50~Mpc ($z \approx 0.01$) observed by TESS, we expect to be able to detect emission signatures from companions with radii between a 0.1--1 R$_\odot$.  Such SNe would be quite bright, at least $T_{\rm peak} \approx 14.5$ mag.  The constraints fall off with increasing distance, until little can be said about any companions at distances greater than 200~Mpc ($z\approx 0.05$).  These statements of course depend on the adopted detection limit---since TESS improved its pointing stability in Sector~4, we have noticed a tendency for the detection limits to push fainter than the average value $\langle T_{\rm limit} \rangle \approx 20$ mag of the SNe light curves presented here.  It remains to be seen if this represents a global improvement in the 8 hour 3$\sigma$ detection threshold, or if the limit is driven primarily by effects unique to each SN.  However, we expect that TESS will provide constraints on companion stars for any SNe brighter than 16--17 mag at peak.  As discussed in \S\ref{sec:models}, the sample of Type Ia SNe observed by TESS in two years is likely to constrain companion stars with radii $>20$ \rsun\ at greater than 99\% confidence.  After six years, TESS may observe up to $\sim$24 Type Ia SNe with constraints similar to SN2018fhw or SN2018hib, and might eventually put limits on the occurrence rate of main sequence companions.

In summary, we have presented early time light curves of Type Ia SNe from the first six sectors of the TESS primary mission, nine of which were amenable to analysis of the early rise.  We fit rising power law models and searched for signatures of any companions stars in these sources.
Our main results are:
\begin{enumerate}
    \item We find a diversity of early time light curve shapes, although most of our SNe are consistent with fireball models with power law index $\beta \approx 2.0$.  Three out of nine SNe have a much smaller rising power law index with $\beta \sim 1.0$.  
    \item We do not find obvious evidence for additional structure in the early time light curves or companion signatures.
    \item We place upper limits on the radii of any companion stars of $\lesssim$\,20 \rsun\ for five SNe and $\lesssim$\,4 \rsun\ for two SNe.  These constraints assume a favorable viewing angle ($<$45$^\circ$), as well as specific SN properties (ejecta mass, velocity, and opacity).  The viewing angle assumptions imply that the odds of non-detections in this sample are 44\% for companions $\gtrsim$20 \rsun\ and 72\% for companions $\gtrsim$4 \rsun, if such systems are actually commonplace.  Thus, we cannot make inferences about the fraction of Type Ia SN with non-degenerate companions.  
 However, after its two year mission, TESS has a 99\% chance of observing enough SNe to either detect the signature of a companion star with $R > 20$\,\rsun in a Type Ia SN, or set strong limits on the occurrence rate of such systems.
    \item TESS is able to observe a companion signature for small companion stars ($R < 4$\,\rsun) in a Type Ia SN within 50 Mpc, and has a reasonable chance of doing so after about six years of operations.
\end{enumerate}

\facility{TESS}

\software{ 
  Matplotlib \citep{matplotlib},
  Numpy \citep{numpy},
  Scipy \citep{scipy}, Astropy \citep{astropy},
  limfit\footnote{\url{https://lmfit.github.io/lmfit-py/index.html}}, pysynphot \citep{pysynphot}, SNooPy \citep{Burns2011},
  ISIS \citep{Alard1998}
 }
\bigskip

\noindent{ACKNOWLEDGEMENTS}
\acknowledgments

We thank Saul Rappaport for discussions about binary systems and mass transfer, and Marco Montalto for discussion of TESS scattered light signals at the TESS Science Conference I (August 2019).

This paper includes data collected by the TESS mission, which are publicly available from the Mikulski Archive for Space Telescopes (MAST) and described in \citet{Jenkins2016}. Funding for the TESS mission is provided by NASA's Science Mission directorate. This research has made use of NASA's Astrophysics Data System, as well as the NASA/IPAC Extragalactic Database (NED) which is operated by the Jet Propulsion Laboratory, California Institute of Technology, under contract with the National Aeronautics and Space Administration.    

PJV is supported by the National Science Foundation Graduate Research Fellowship Program Under Grant No. DGE-1343012. KZS and CSK are supported by NSF grants AST-1515876, AST-1515927, and AST-1814440. CSK was also supported by a fellowship from the Radcliffe Institute for Advanced Studies at Harvard University. 
BJS, CSK, and KZS are supported by NSF grant AST-1907570. BJS is also supported by NASA grant 80NSSC19K1717 and NSF grants AST-1920392 and AST-1911074.
MAT acknowledges support from the DOE CSGF through grant DE-SC0019323.  The work of AP was supported in part by the GINOP-2.3.2-15-2016-00033 project which is funded by the Hungarian National Research, Development and Innovation Fund together with the European Union.  TD acknowledges support from MIT's Kavli Institute as a Kavli postdoctoral fellow.

\section*{Appendix A}
\begin{figure*}
    \centering
    \includegraphics[width=0.8\textwidth]{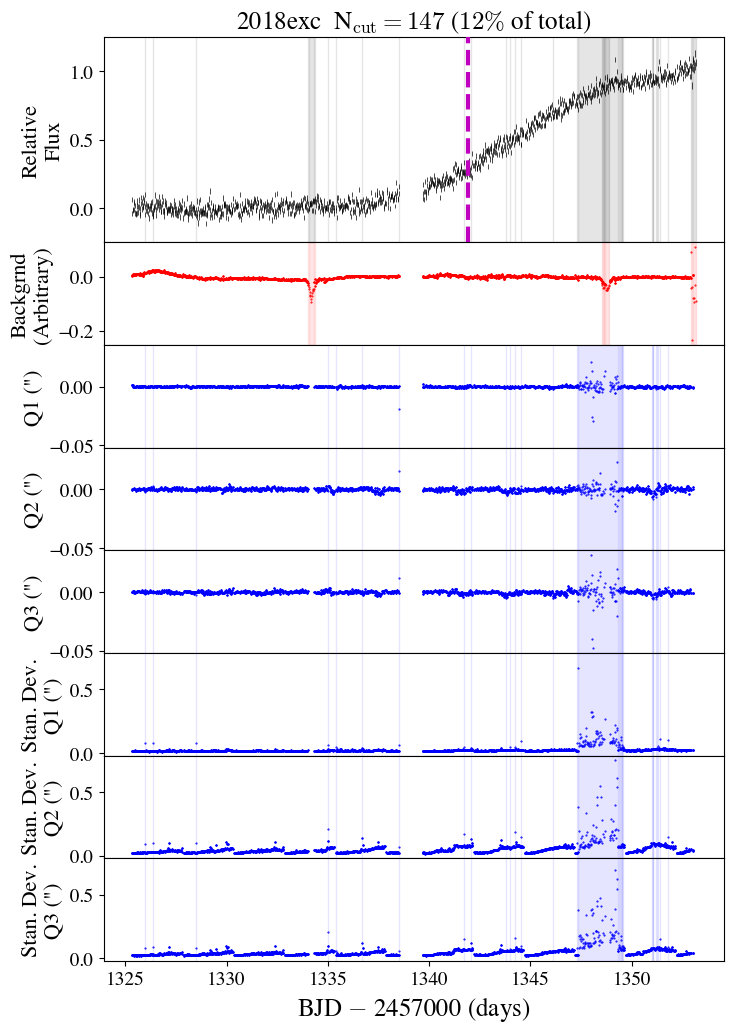}
    \caption{The light curve of SN2018exc and the associated timeseries data used to identify unreliable photometry.  The flux units are normalized relative to peak.  The top panel shows the light curve (black lines) and 1$\sigma$ uncertainties.  The vertical purple line is the time of discovery.  The shaded gray regions show the epochs that were removed based on the procedures defined in \S\ref{sec:systematics} (the same epochs are shown in the panels below).  The next panel shows the timeseries of the local background estimated in a 4/8 pixel inner/outer annular region in the difference images.  Three rounds of 5$\sigma$ clipping are used to identify periods with rapidly changing or elevated backgrounds (see \S\ref{sec:systematics}), which are highlighted by the shaded red regions.  The panels labeled "Q1," "Q2," and "Q3" show the guiding offset quaternions in units of arcseconds; the Q1/Q2 components represent offsets along the TESS detectors' rows/columns, while the Q3 direction represents spacecraft roll.    The guiding offset data has been rebinned from 0.5 seconds to the 30 minute time periods of each FFI, and therefore show the average pointing of the spacecraft during that exposure.  The last three panels show the standard deviation of the guiding offsets in each FFI bin.  Three rounds of 5$\sigma$ clipping are used to identify periods with increased jitter or large pointing offsets (see \S\ref{sec:systematics}), and are highlighted by the shaded blue regions.}
    \label{fig:2018exc}
\end{figure*}

In Appendix A, we show the original light curves of each SN and the auxiliary  data used to identify and remove systematic errors.  Figures~12 to 35 show the data for each SN---see \S\ref{sec:data_analysis} and \S\ref{sec:systematics}, and Figure~\ref{fig:2018exc} for more details.

\begin{figure*}
    \centering
    \includegraphics[width=0.8\textwidth]{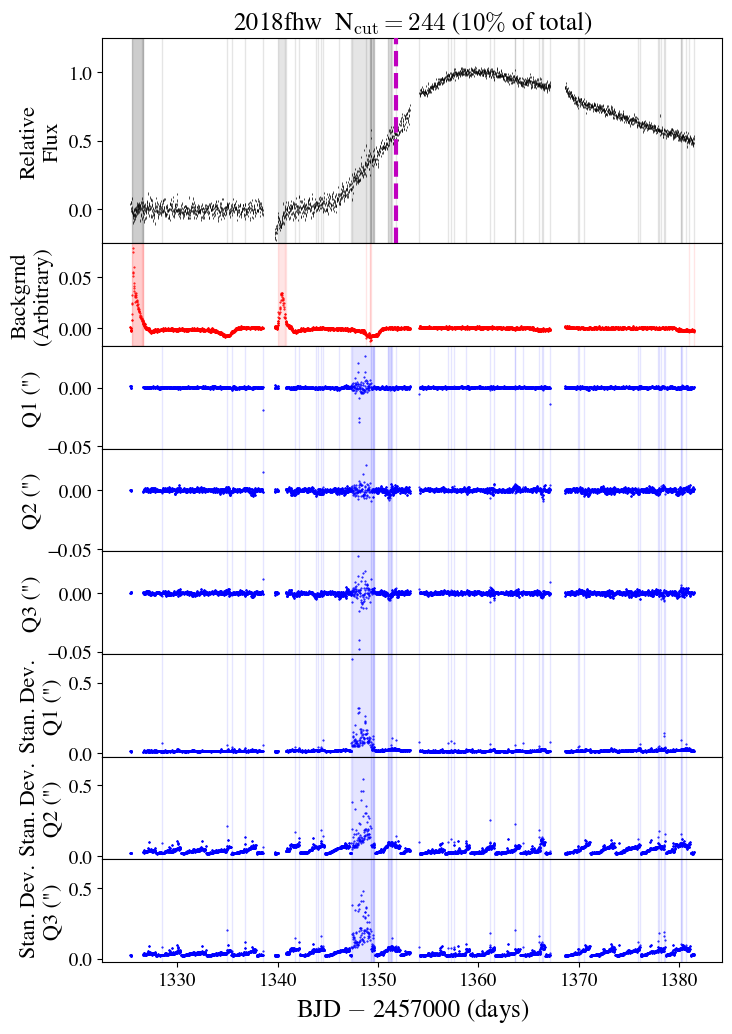}
    \caption{Same as Figure~\ref{fig:2018exc}, but for SN2018fhw.}
    \label{fig:2018fhw}
\end{figure*}

\begin{figure*}
    \centering
    \includegraphics[width=0.8\textwidth]{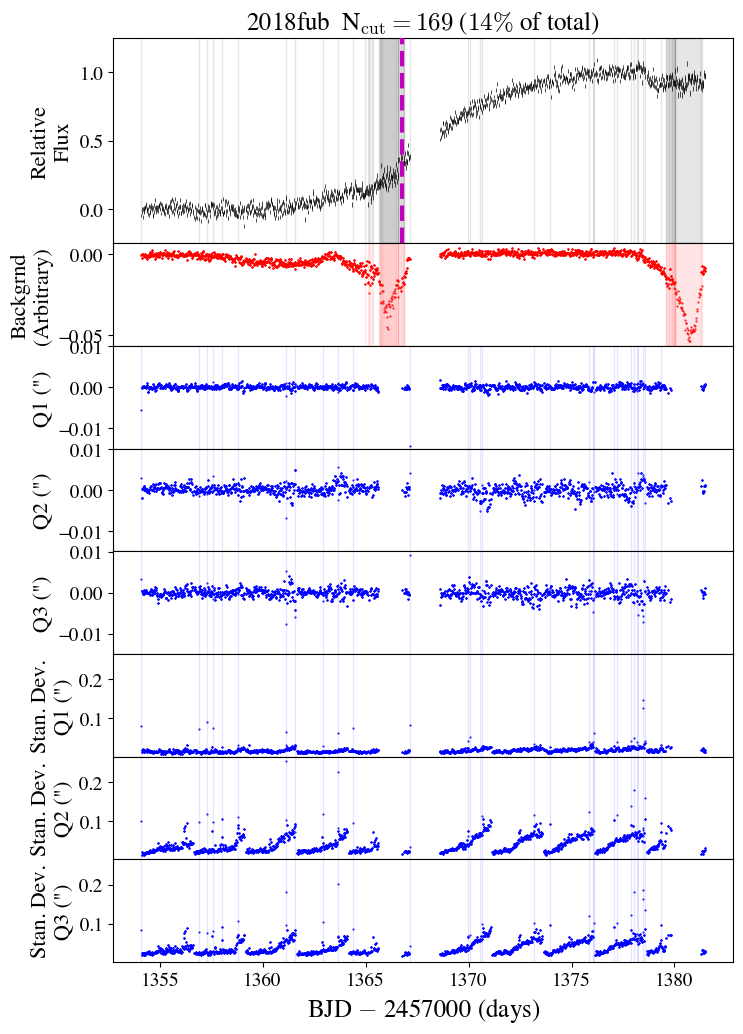}
    \caption{Same as Figure~\ref{fig:2018exc}, but for SN2018fub.}
    \label{fig:2018fub}
\end{figure*}

\begin{figure*}
    \centering
    \includegraphics[width=0.8\textwidth]{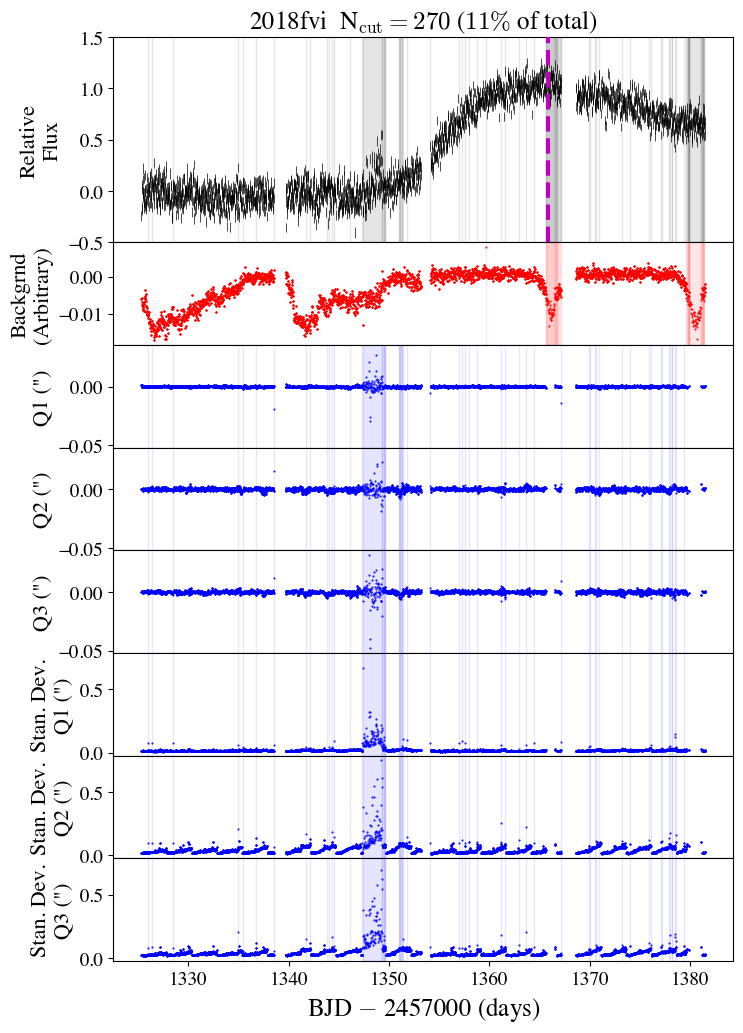}
    \caption{Same as Figure~\ref{fig:2018exc}, but for SN2018fvi.}
    \label{fig:2018fvi}
\end{figure*}

\begin{figure*}
    \centering
    \includegraphics[width=0.8\textwidth]{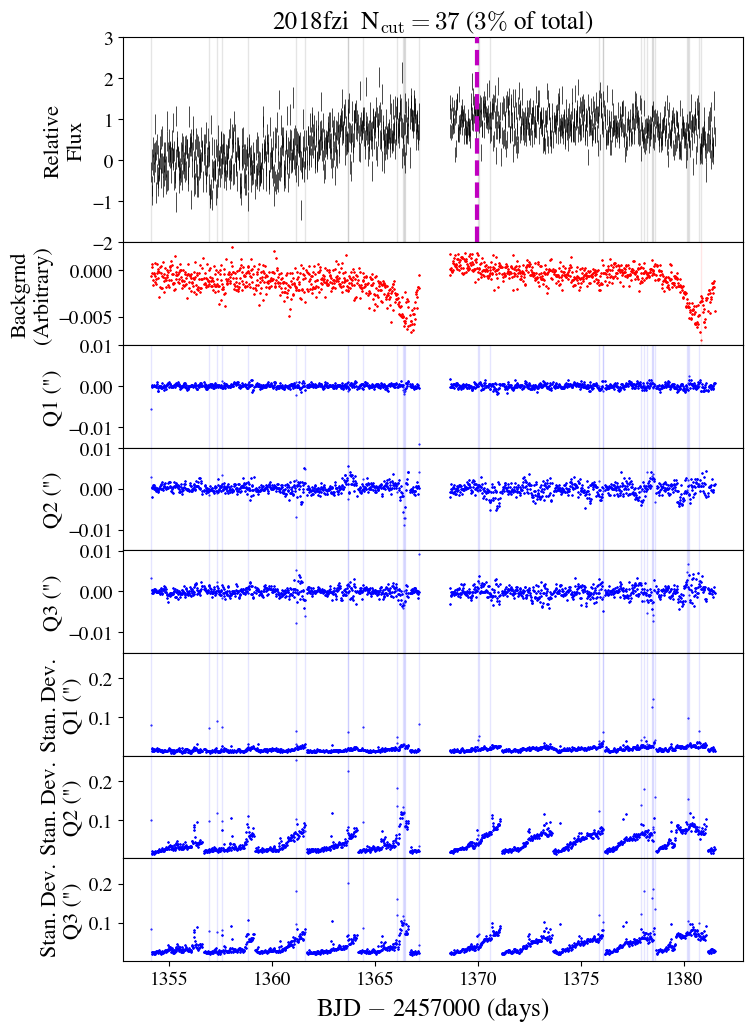}
    \caption{Same as Figure~\ref{fig:2018exc}, but for SN2018fzi.}
    \label{fig:2018fvi}
\end{figure*}

\begin{figure*}
    \centering
    \includegraphics[width=0.8\textwidth]{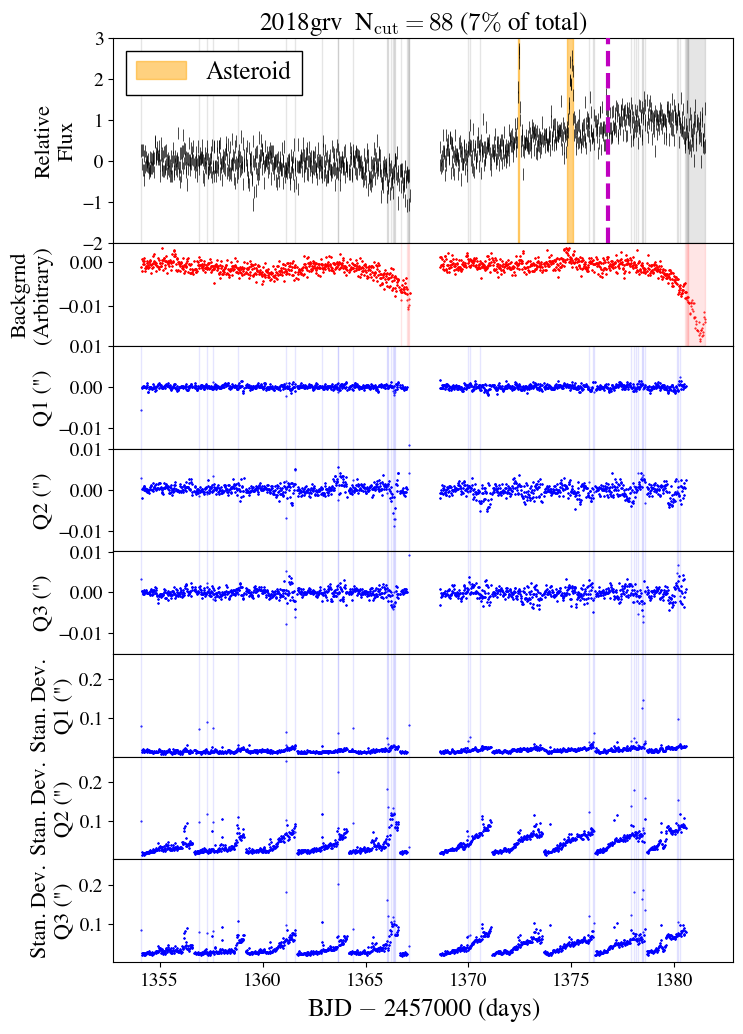}
    \caption{Same as Figure~\ref{fig:2018exc}, but for SN2018grv.}
    \label{fig:2018grv}
\end{figure*}

\begin{figure*}
    \centering
    \includegraphics[width=0.8\textwidth]{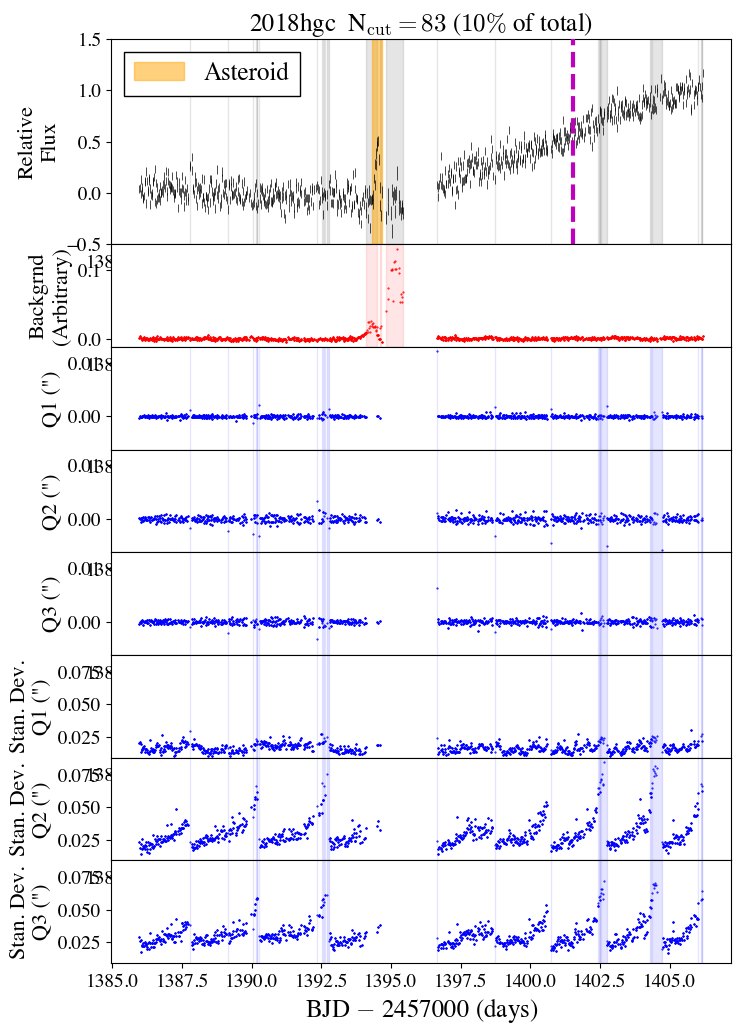}
    \caption{Same as Figure~\ref{fig:2018exc}, but for SN2018hgc.}
    \label{fig:2018hgc}
\end{figure*}

\begin{figure*}
    \centering
    \includegraphics[width=0.8\textwidth]{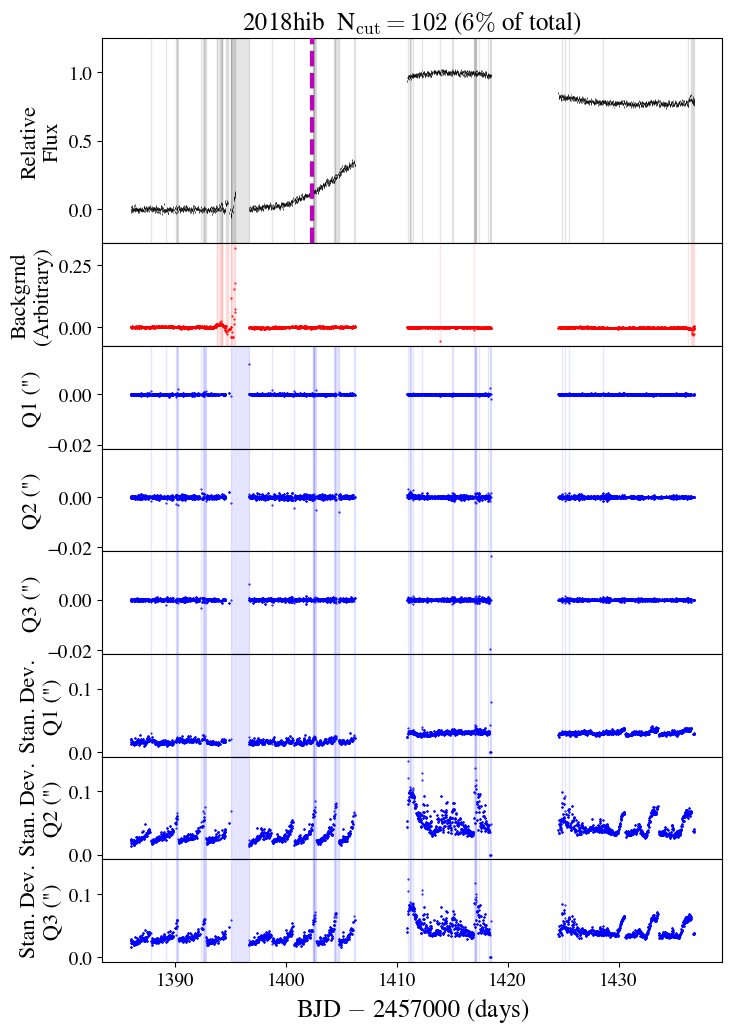}
    \caption{Same as Figure~\ref{fig:2018exc}, but for SN2018hib.}
    \label{fig:2018hib}
\end{figure*}

\begin{figure*}
    \centering
    \includegraphics[width=0.8\textwidth]{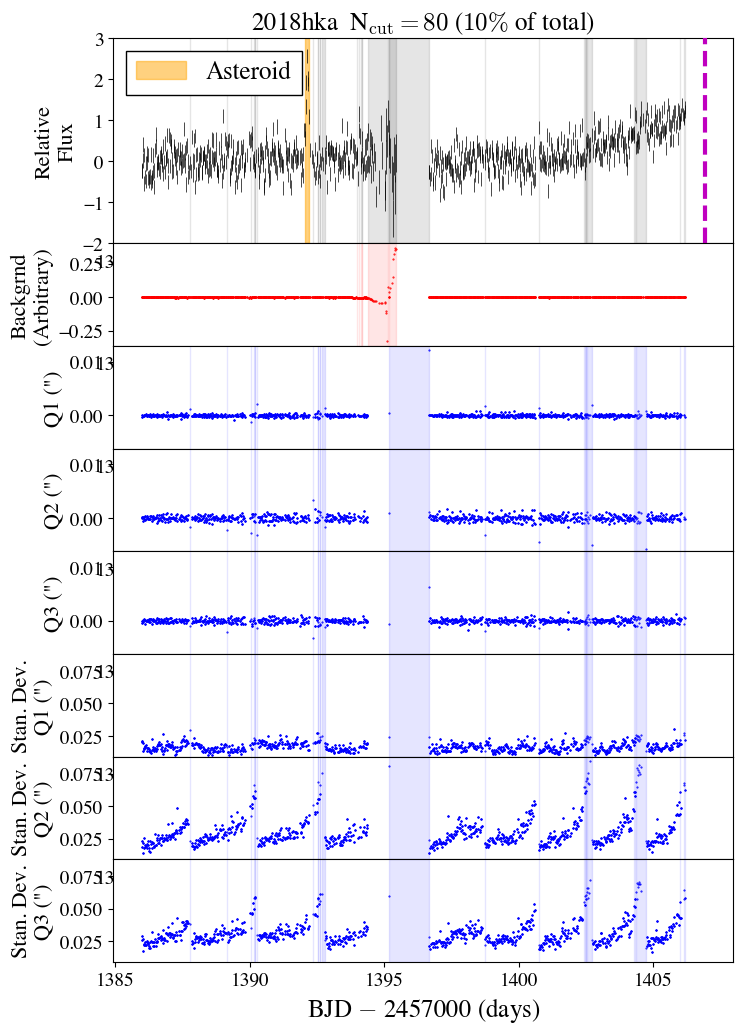}
    \caption{Same as Figure~\ref{fig:2018exc}, but for SN2018hkz.}
    \label{fig:2018hka}
\end{figure*}

\begin{figure*}
    \centering
    \includegraphics[width=0.8\textwidth]{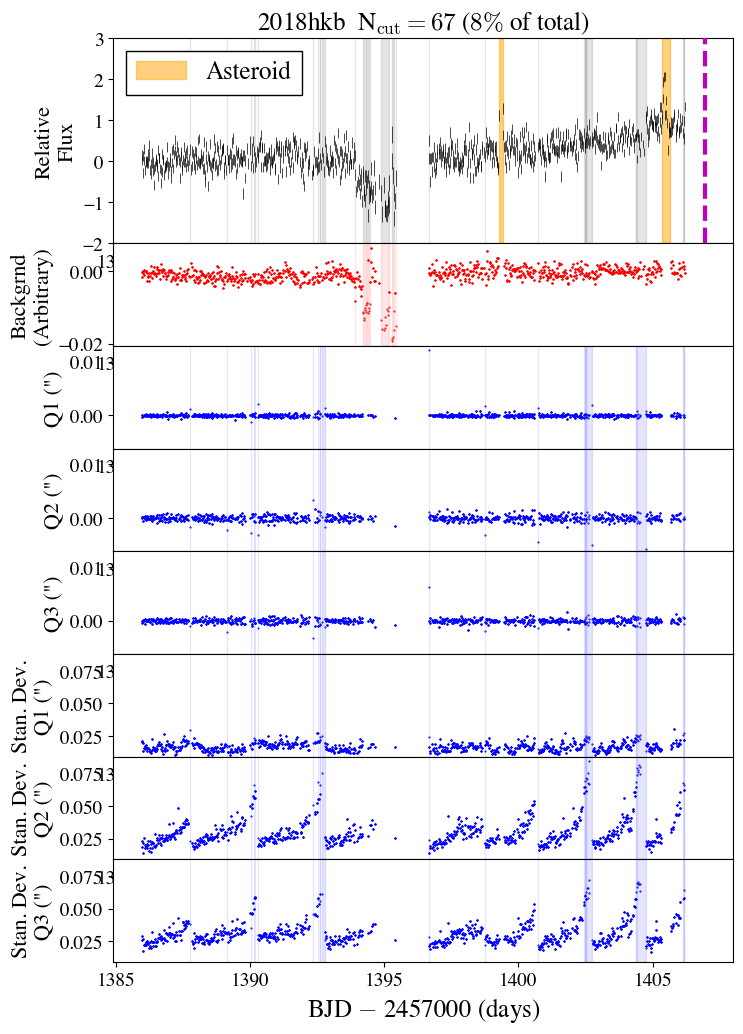}
    \caption{Same as Figure~\ref{fig:2018exc}, but for SN2018hkb.}
    \label{fig:2018hka}
\end{figure*}

\begin{figure*}
    \centering
    \includegraphics[width=0.8\textwidth]{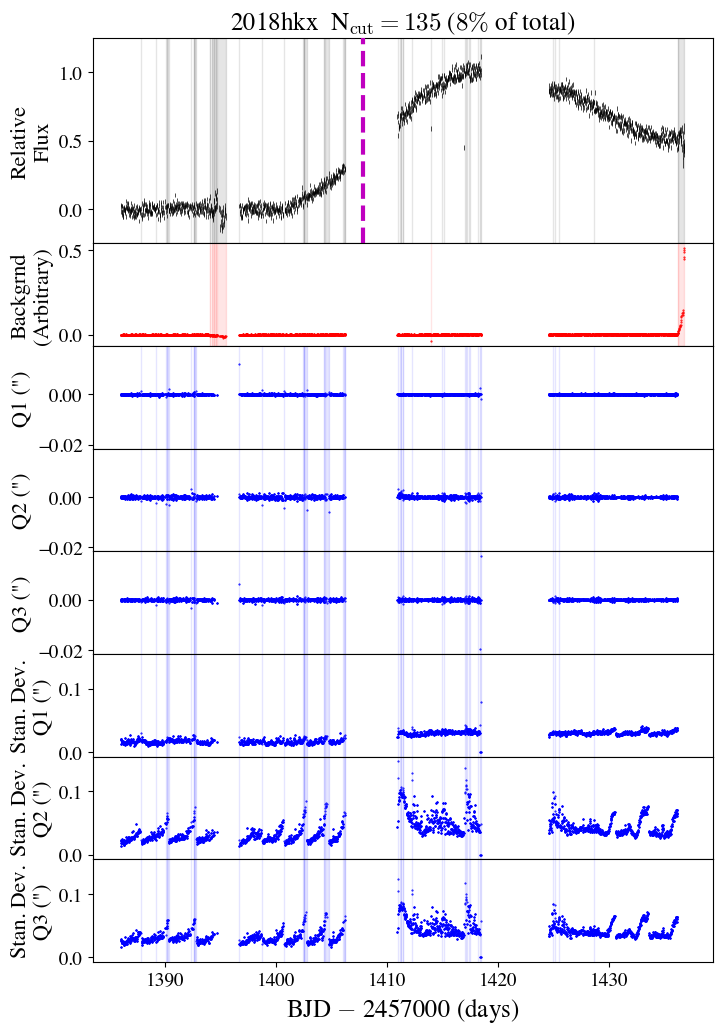}
    \caption{Same as Figure~\ref{fig:2018exc}, but for SN2018hkx.}
    \label{fig:2018hkx}
\end{figure*}

\begin{figure*}
    \centering
    \includegraphics[width=0.8\textwidth]{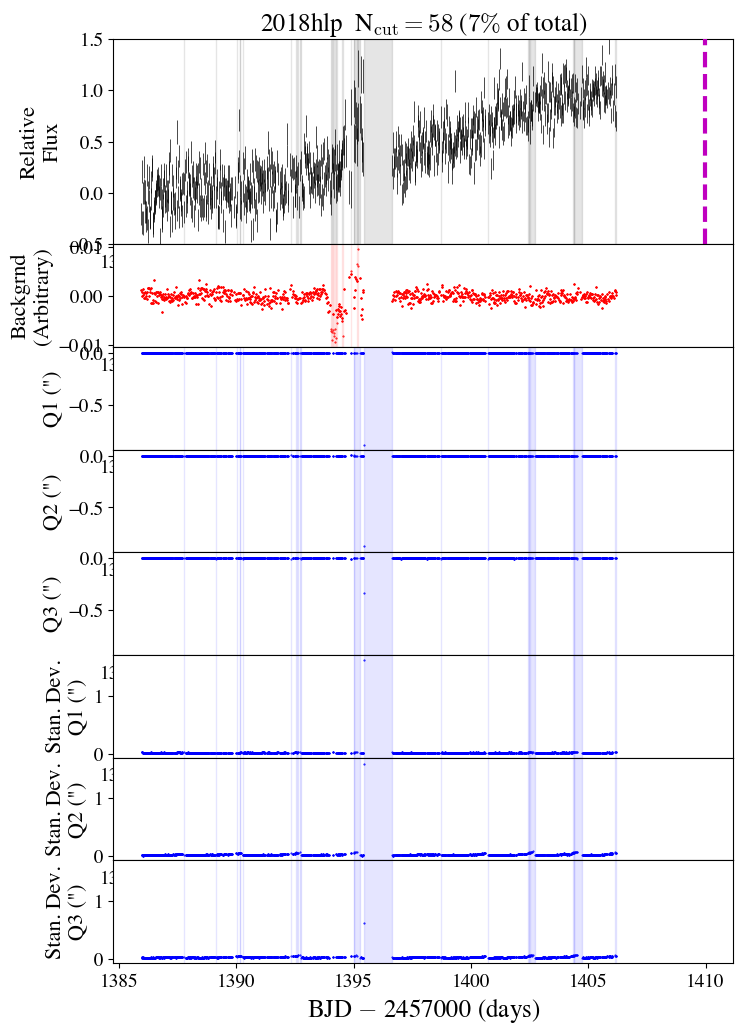}
    \caption{Same as Figure~\ref{fig:2018exc}, but for SN2018hlp.}
    \label{fig:2018hzh}
\end{figure*}

\begin{figure*}
    \centering
    \includegraphics[width=0.8\textwidth]{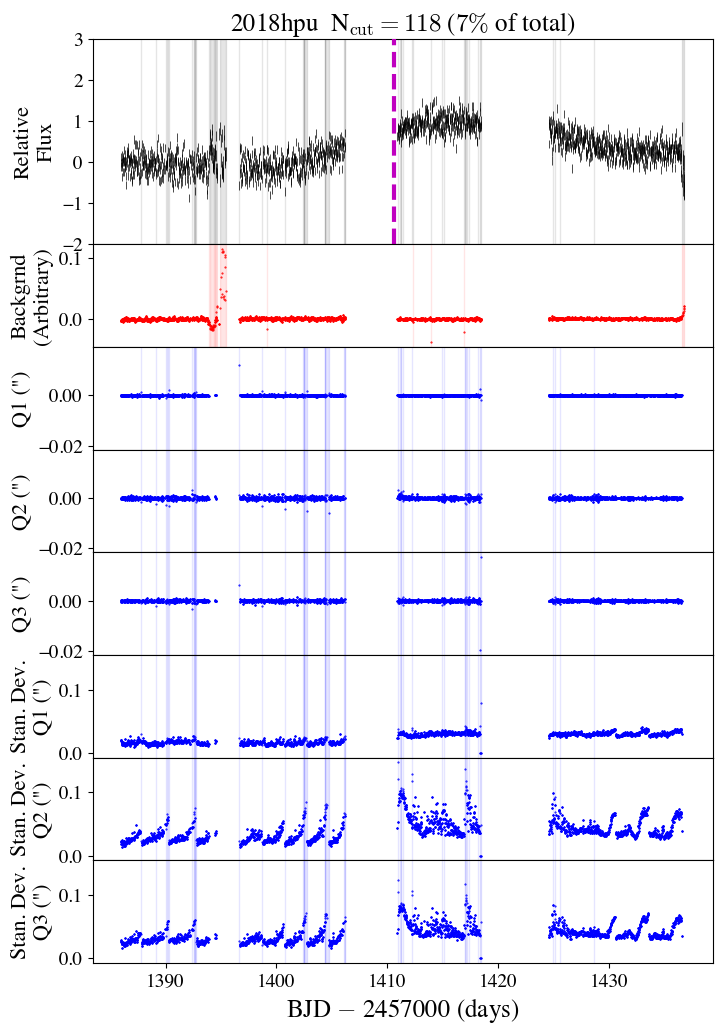}
    \caption{Same as Figure~\ref{fig:2018exc}, but for SN2018hpu.}
    \label{fig:2018hpu}
\end{figure*}

\begin{figure*}
    \centering
    \includegraphics[width=0.8\textwidth]{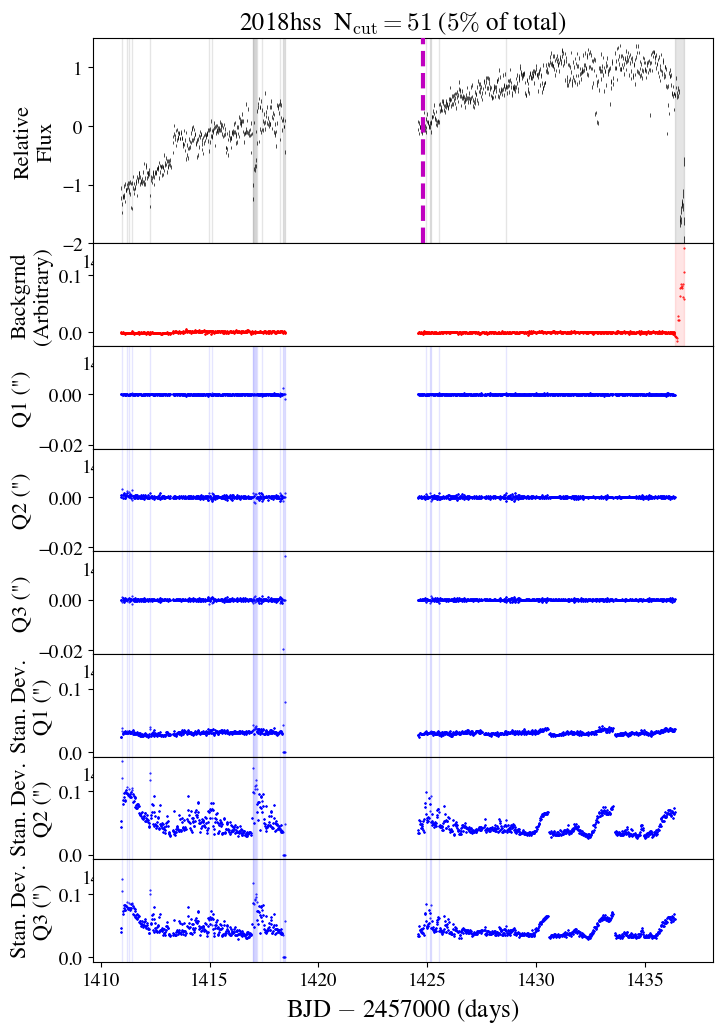}
    \caption{Same as Figure~\ref{fig:2018exc}, but for SN2018hss.}
    \label{fig:2018hss}
\end{figure*}

\begin{figure*}
    \centering
    \includegraphics[width=0.8\textwidth]{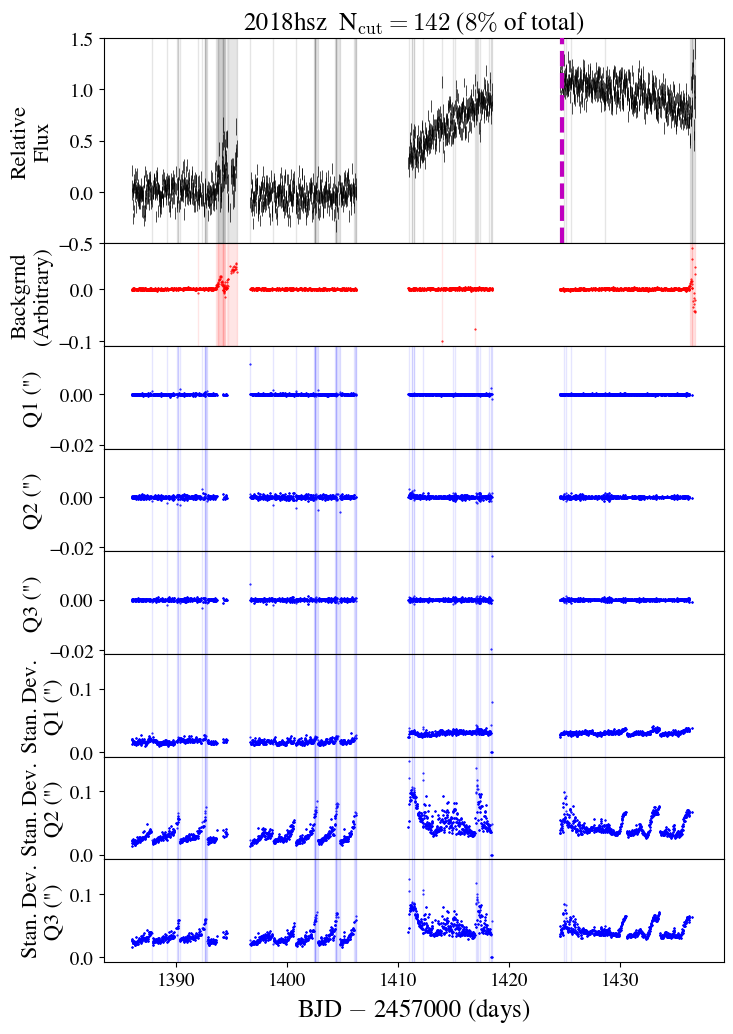}
    \caption{Same as Figure~\ref{fig:2018exc}, but for SN2018hsz.}
    \label{fig:2018hzh}
\end{figure*}

\begin{figure*}
    \centering
    \includegraphics[width=0.8\textwidth]{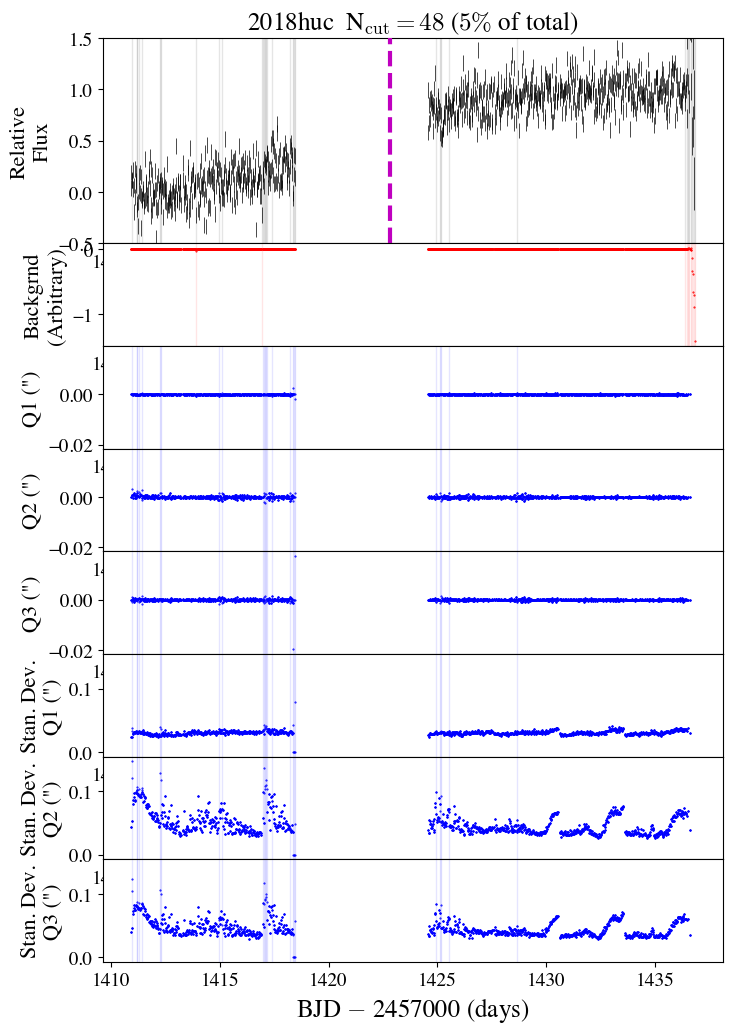}
    \caption{Same as Figure~\ref{fig:2018exc}, but for SN2018huc.}
    \label{fig:2018hzh}
\end{figure*}

\begin{figure*}
    \centering
    \includegraphics[width=0.8\textwidth]{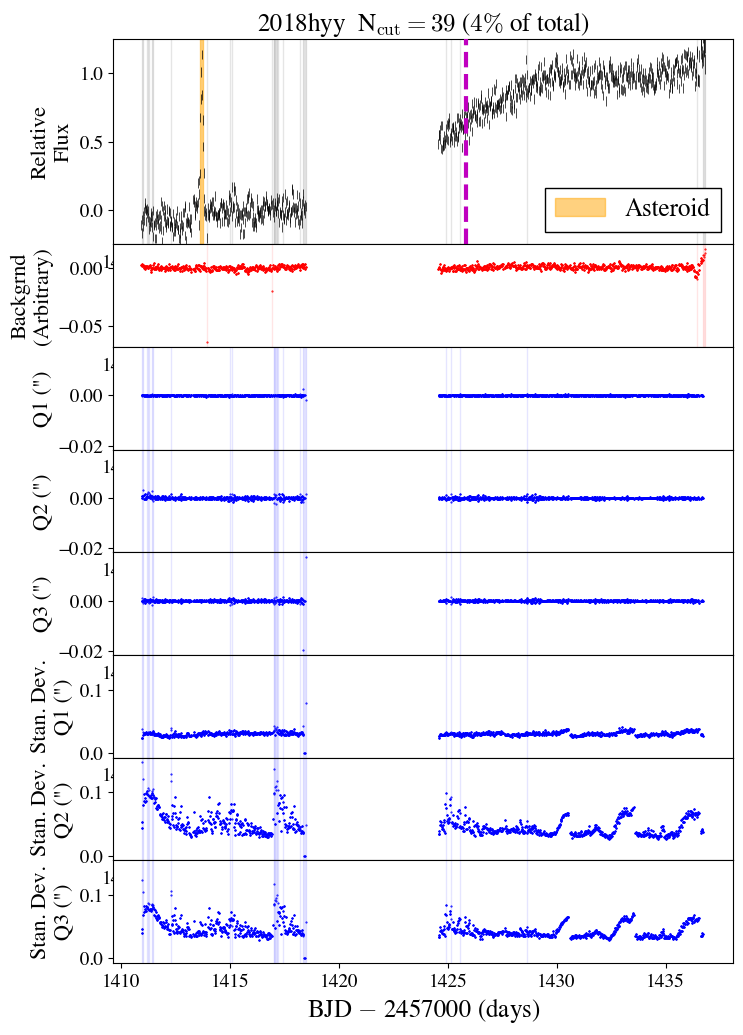}
    \caption{Same as Figure~\ref{fig:2018exc}, but for SN2018hyy.}
    \label{fig:2018hyy}
\end{figure*}

\begin{figure*}
    \centering
    \includegraphics[width=0.8\textwidth]{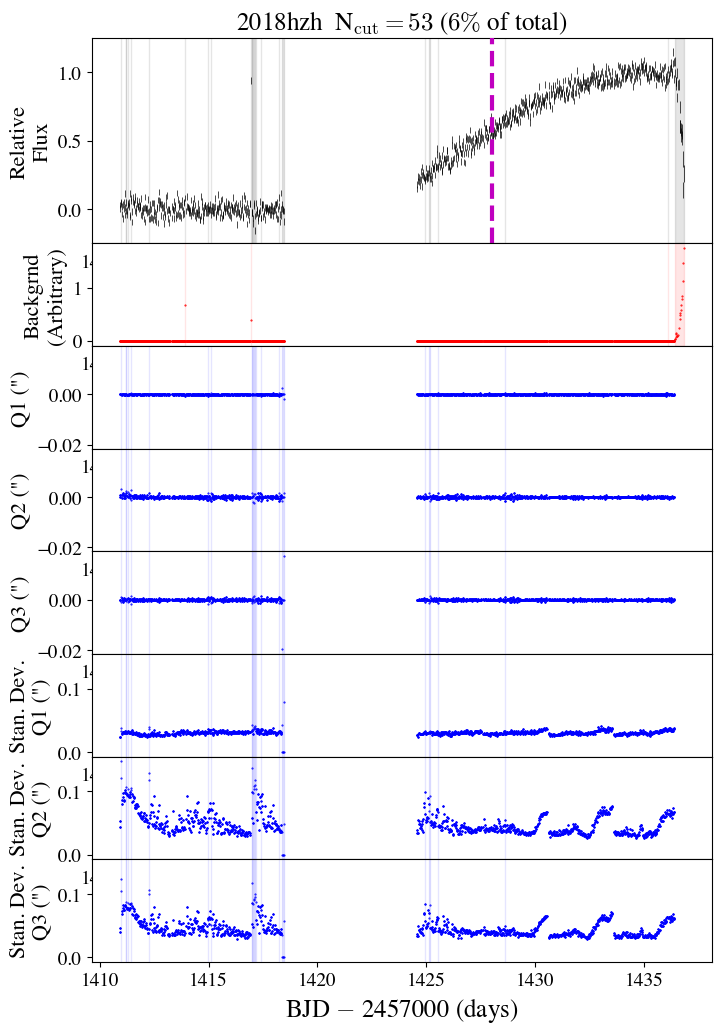}
    \caption{Same as Figure~\ref{fig:2018exc}, but for SN2018hzh.}
    \label{fig:2018hzh}
\end{figure*}

\begin{figure*}
    \centering
    \includegraphics[width=0.8\textwidth]{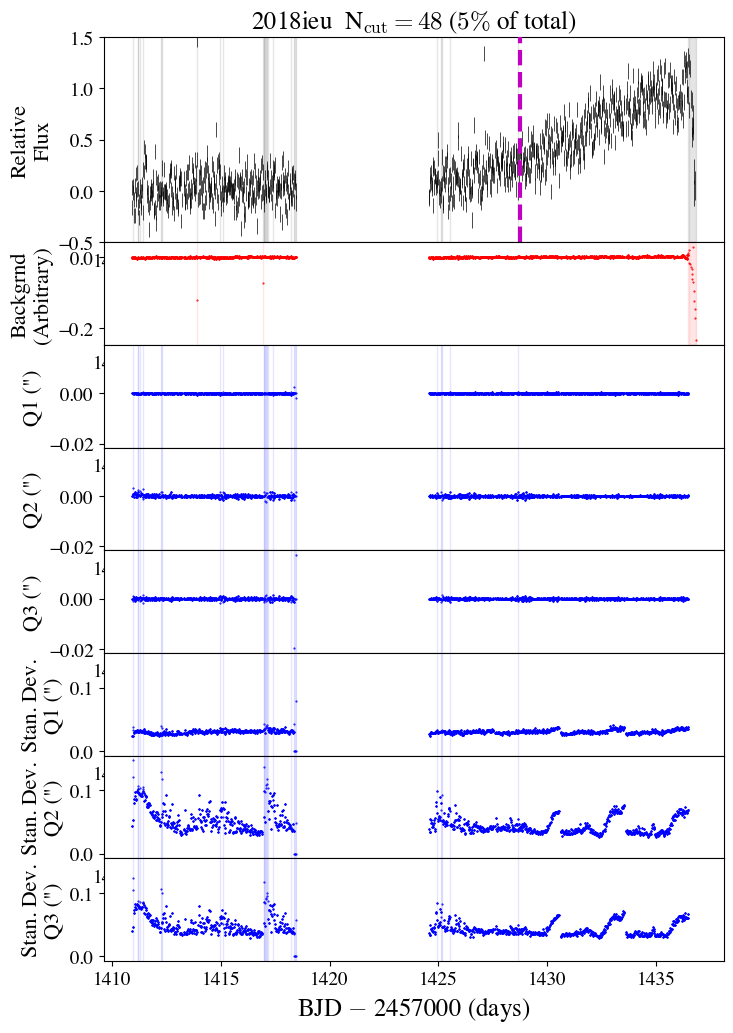}
    \caption{Same as Figure~\ref{fig:2018exc}, but for SN2018ieu.}
    \label{fig:2018ieu}
\end{figure*}

\begin{figure*}
    \centering
    \includegraphics[width=0.8\textwidth]{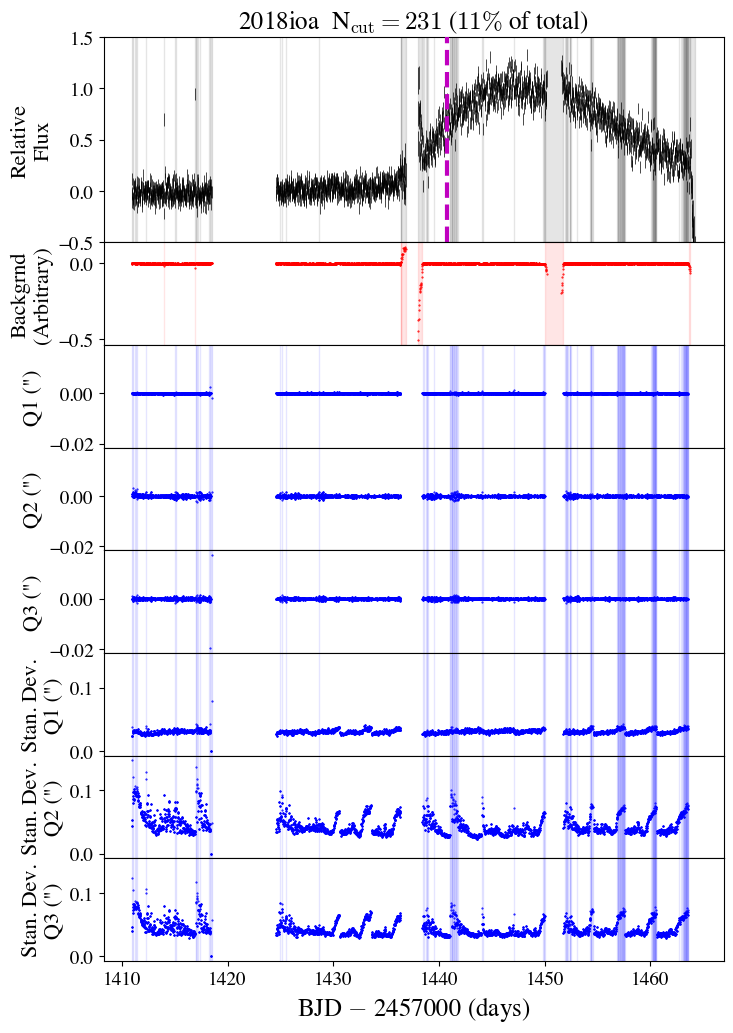}
    \caption{Same as Figure~\ref{fig:2018exc}, but for SN2018ioa.}
    \label{fig:2018ioa}
\end{figure*}

\begin{figure*}
    \centering
    \includegraphics[width=0.8\textwidth]{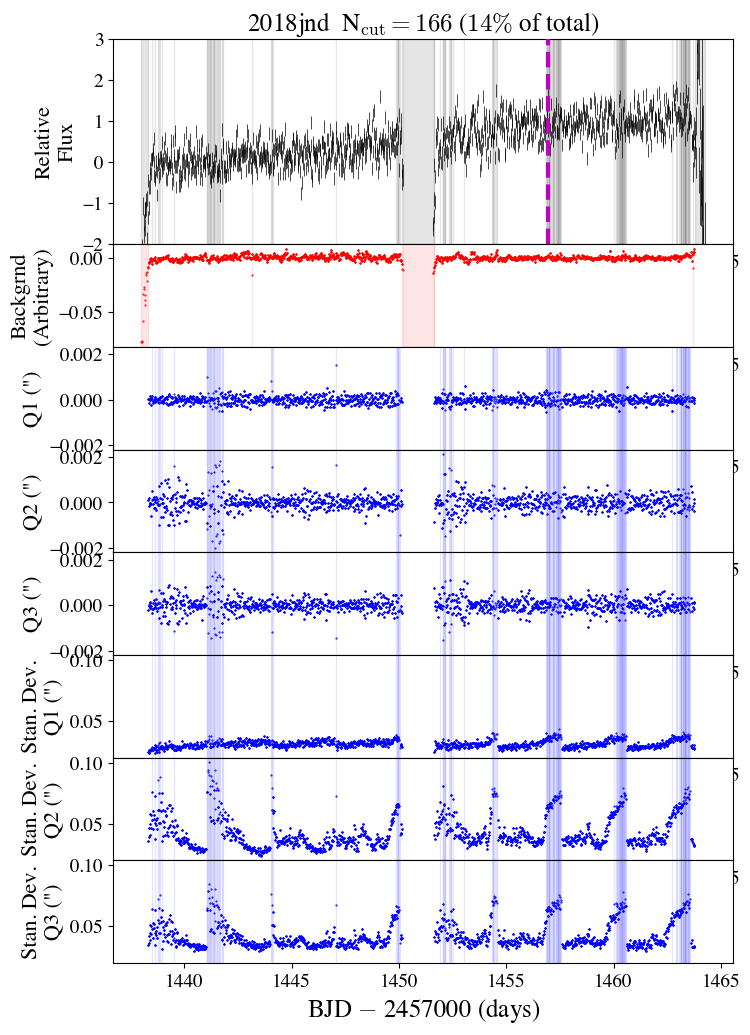}
    \caption{Same as Figure~\ref{fig:2018exc}, but for SN2018jnd.}
    \label{fig:2018jnd}
\end{figure*}

\begin{figure*}
    \centering
    \includegraphics[width=0.8\textwidth]{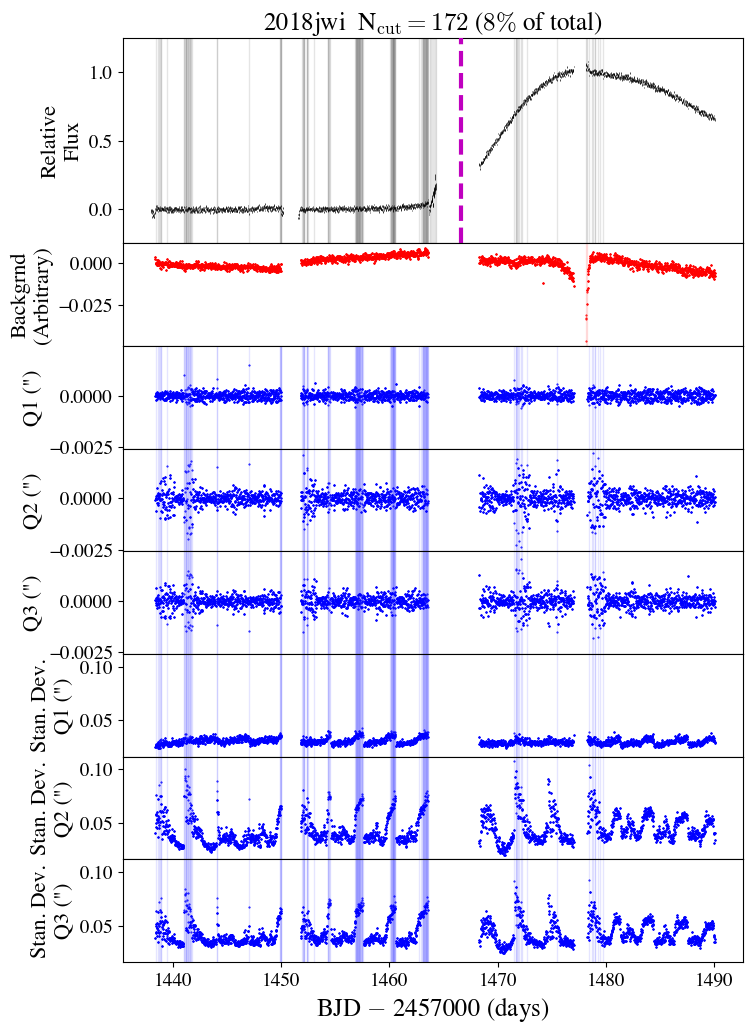}
    \caption{Same as Figure~\ref{fig:2018exc}, but for SN2018jwi.}
    \label{fig:2018jnd}
\end{figure*}

\begin{figure*}
    \centering
    \includegraphics[width=0.8\textwidth]{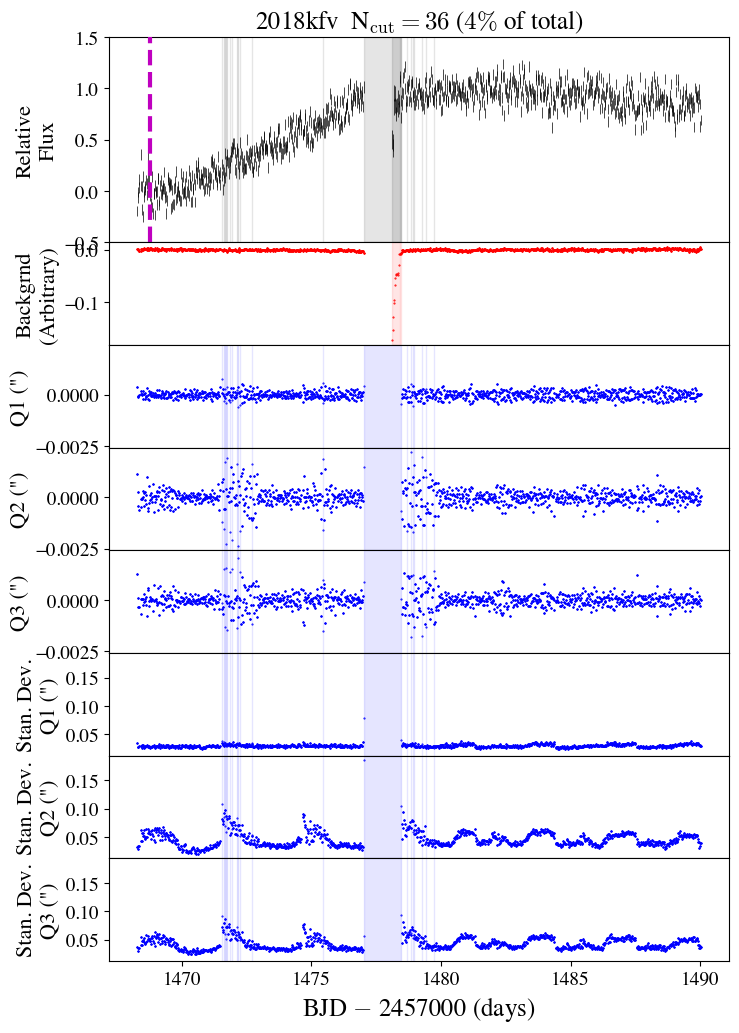}
    \caption{Same as Figure~\ref{fig:2018exc}, but for SN2018kfv.}
    \label{fig:2018hzh}
\end{figure*}

\begin{figure*}
    \centering
    \includegraphics[width=0.8\textwidth]{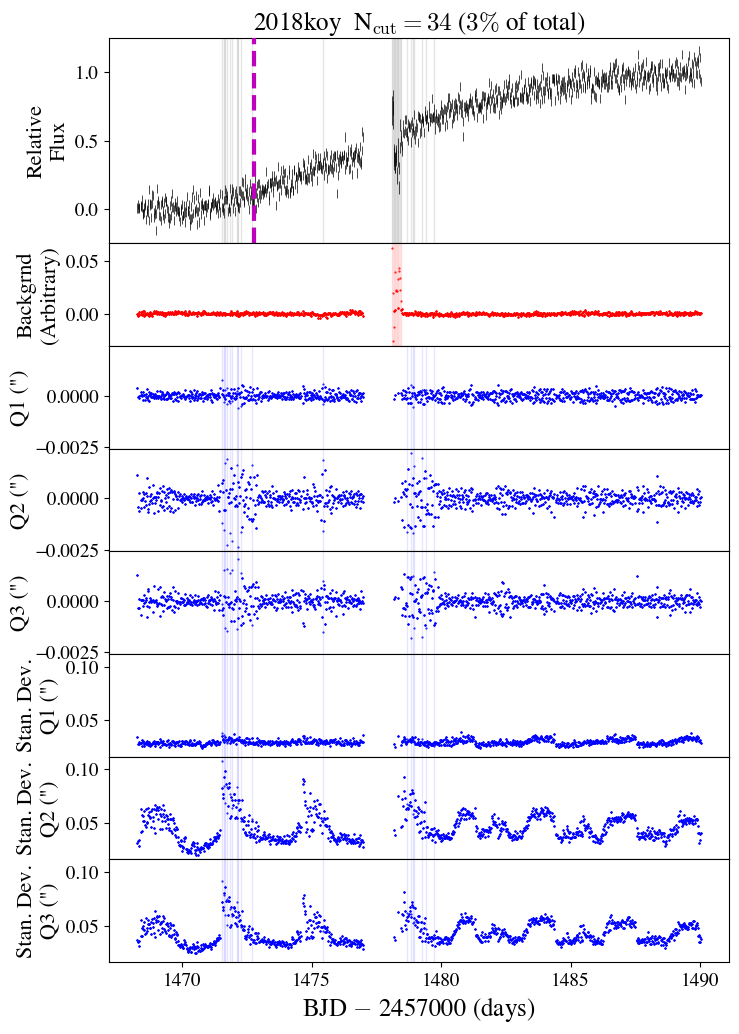}
    \caption{Same as Figure~\ref{fig:2018exc}, but for SN2018koy.}
    \label{fig:2018koy}
\end{figure*}

\section*{Appendix B}
In Appendix B, Figures~\ref{fig:detrend_2018fhw}, \ref{fig:detrend2}, and \ref{fig:detrend3}, we show SN light curves affected by nearby bright stars, as well as the light curves of the stars and fits used to correct the data.  
\begin{figure*}
    \centering
    \includegraphics[width=0.8\textwidth]{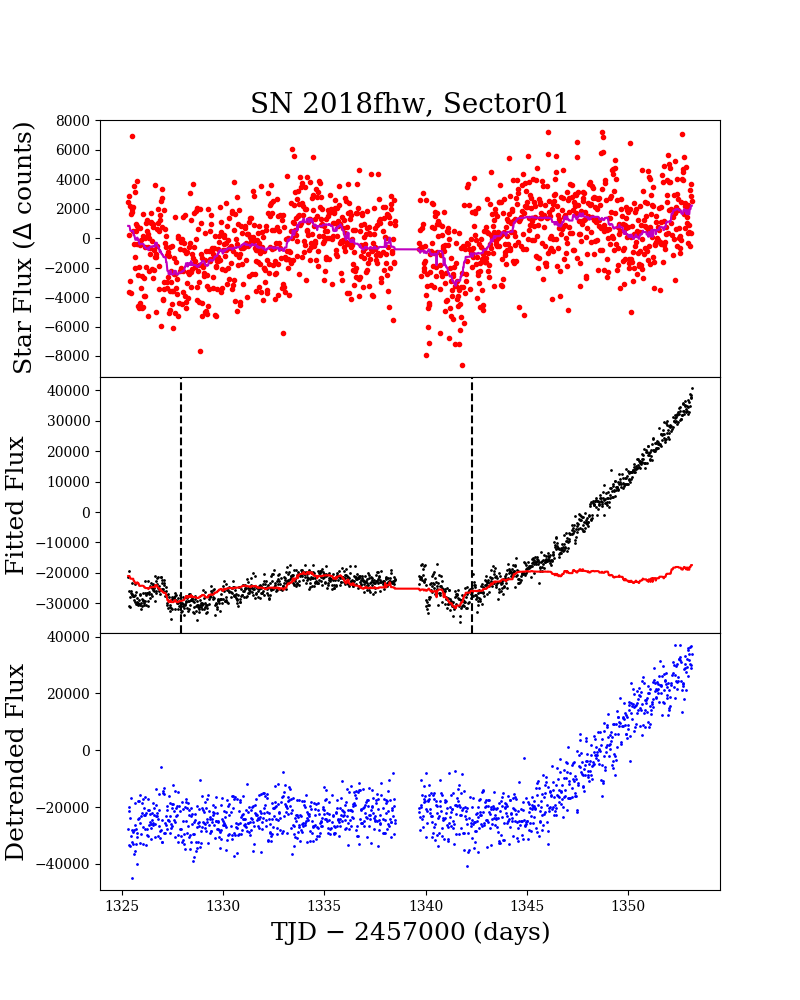}
    \caption{Detrending procedure for SN2018fhw, Sector~1.  This SN lands near a variable star that contaminates the photometric aperture.  In the top panel, the red points show the star light curve, and the magenta line shows a smoothed model using a median filter.  In the middle panel, the black points show the raw SN light curve, and the red line is the smoothed star model.  The vertical dashed black lines show the period of time over which the star model was fit (with offset and scaling parameters).  The blue points in the bottom panel show the corrected SN light curve.  See \S\ref{sec:systematics} for more details.\label{fig:detrend_2018fhw}}
\end{figure*}

\begin{figure*}

    \centering
    \begin{tabular}{cc}
    \includegraphics[width=0.45\textwidth]{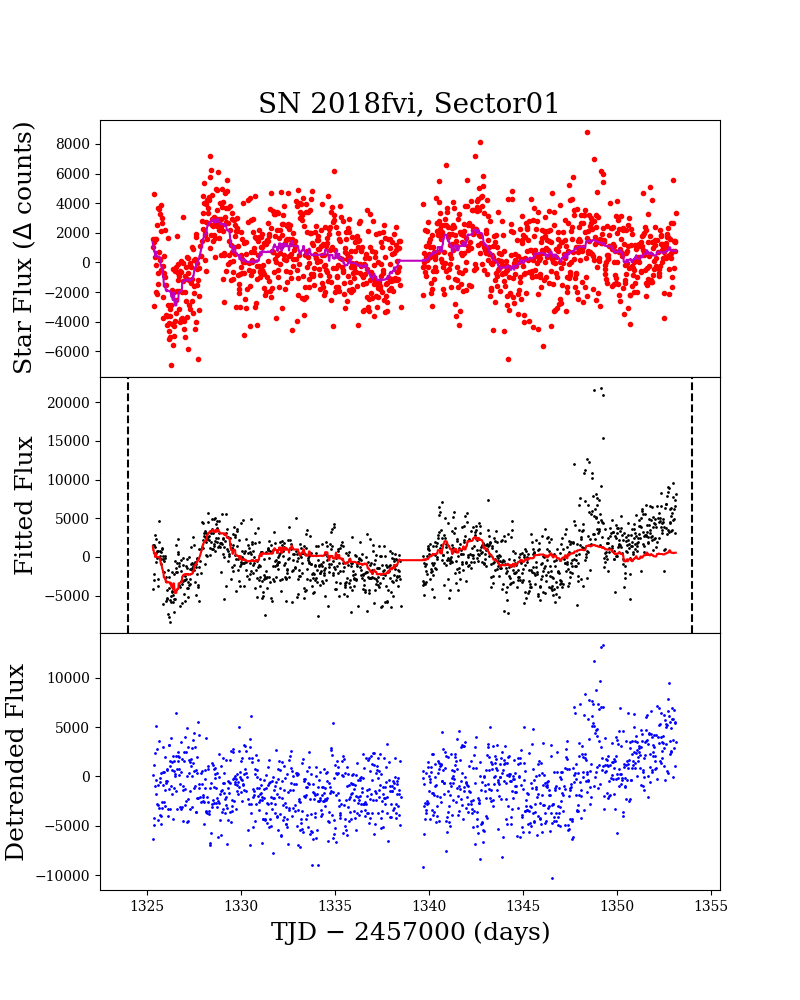}&
    \includegraphics[width=0.45\textwidth]{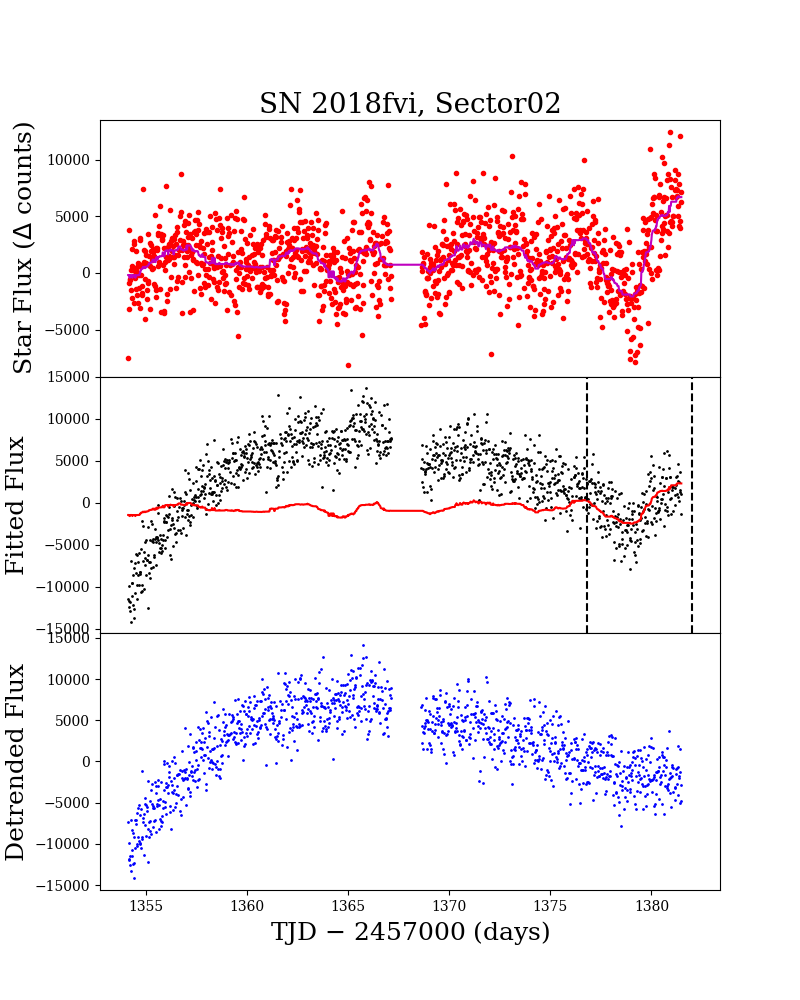}    \\\
    \includegraphics[width=0.45\textwidth]{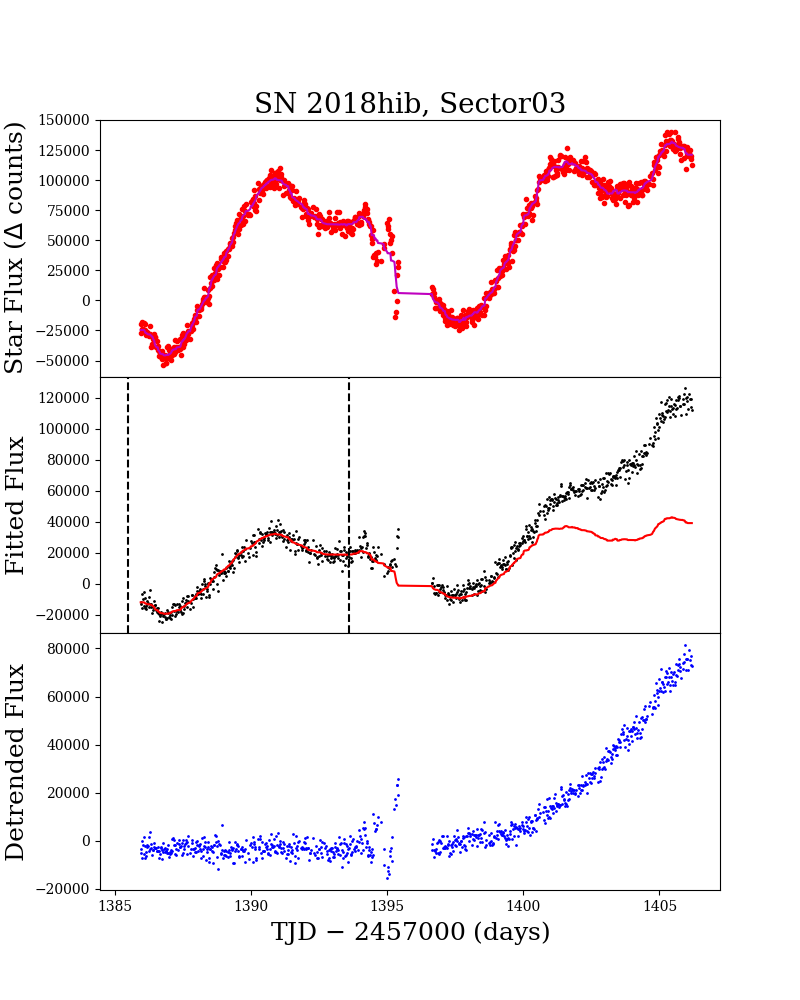}&
    \includegraphics[width=0.45\textwidth]{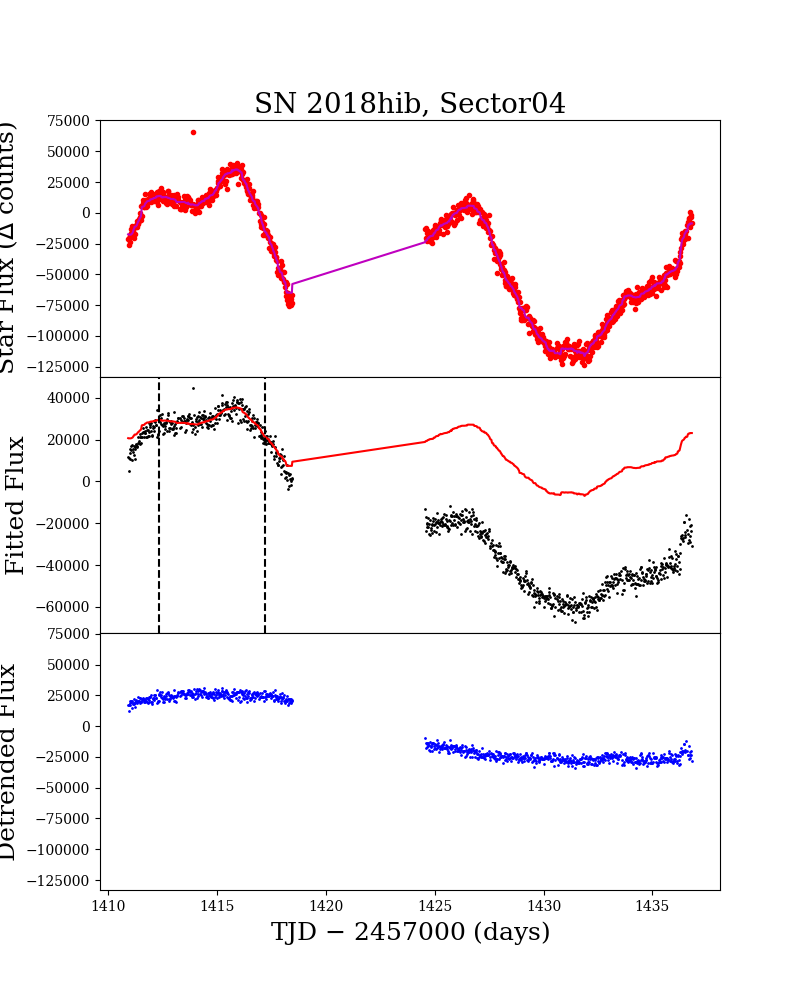}    \\\
        \end{tabular}
    \caption{Same as Figure~\ref{fig:detrend_2018fhw}, but for 2018fvi and 2018hib.\label{fig:detrend2}}
\end{figure*}
\clearpage

\begin{figure*}
    \centering
    \begin{tabular}{cc}
    \includegraphics[width=0.45\textwidth]{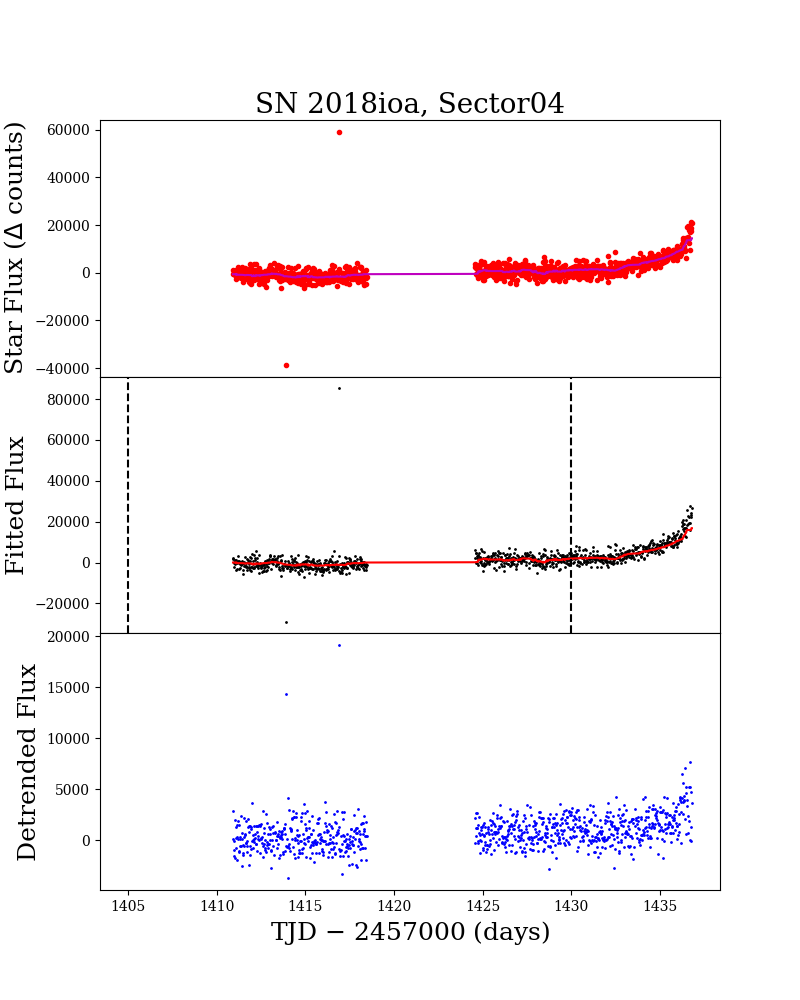}&
        \includegraphics[width=0.45\textwidth]{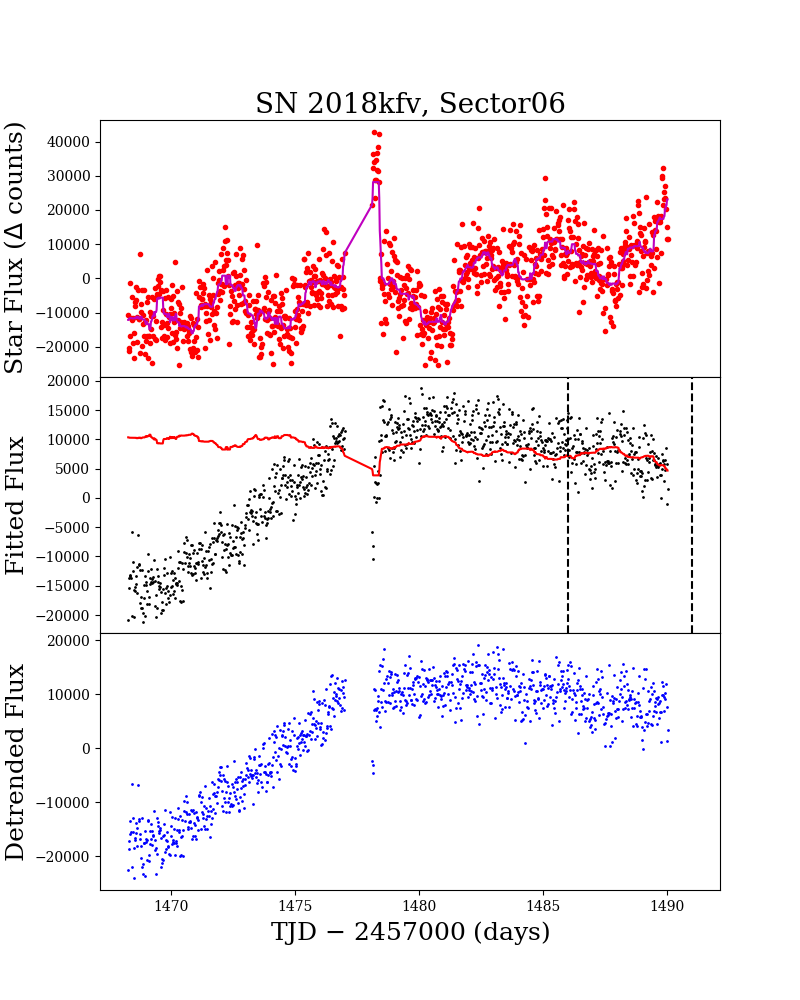}\\
 \end{tabular}
    \caption{Same as Figure~\ref{fig:detrend_2018fhw}, but for SN2018ioa and SN2018kfv.\label{fig:detrend3}}
\end{figure*}

\section*{Appendix C}

\begin{deluxetable}{ccc}
\tablewidth{0pt}
\tablecaption{$\Delta\chi^2$ Confidence Intervals  \label{tab:chi2_confidence}}
\tablehead{
\colhead{Name} & \colhead{98\% $\Delta\chi^2$} & \colhead{99.9\% $\Delta\chi^2$} }
\startdata
SN2018exc &230& 277\\
SN2018fhw &285& 325\\
SN2018fub  &200& 236\\
SN2018hib  &239& 285\\
SN2018hkx &232& 274\\
SN2018koy &157& 190\\
\enddata
  \end{deluxetable}
  
In this section, we describe the Monte Carlo simulations used to test the grid search for companion signatures described in \S\ref{sec:companion_search}, and we define the 98\% and 99.9\% confidence intervals for $\chi^2$ used to place 2 and 3$\sigma$ limits on the separations/radii of possible companion stars in the TESS light curves.
 
We use the same fireball model for the SNe explosion as in \S\ref{sec:lc_shapes}, which consists of a sphere with a constant temperature of $10^4$\, K, an initial radius of $3\times 10^8$\, cm, and an expansion velocity of $10^9$\, cm s$^{-1}$.  The \citet{Kasen2010} models have the same fiducial parameters used else where in this work,  with a viewing angle of $45^{\circ}$, an ejecta mass of 1.4 M$_\odot$, a velocity of 10$^4$ km s$^{-1}$, and an opacity of 0.2 cm$^2$ g$^{-1}$.  We simulate each of the separations/radii of companion models as in \S\ref{sec:lc_shapes} (see Figure~\ref{fig:model_components}), with radii of 200 \rsun, 100 \rsun, 50 \rsun, 25 \rsun, 15 \rsun, 10 \rsun, 5 \rsun, and 1 \rsun (or separations of 4.03, 2.02, 1.01, 0.50, 0.30, 0.20, 0.10, and 0.02 $\times 10^{13}$ cm respectively).  We also simulate the observed properties (such as redshift and data gaps) of each SNe presented in \S\ref{sec:companion_search}, namely, SN2018exc, SN2018fhw, SN2018fub, SN2018hib, SN2018hkx, and SN2018koy.  In total, there are eight companion models for each of six SN, resulting in a total of 48 simulations.

For each simulation, we convert the fireball and \citet{Kasen2010} models to flux units by placing the object at the luminosity distance appropriate for each SNe (see Table~\ref{tab:physical_data}), calculating the TESS instrument response using synthetic photometry, and summing the two components to produce the simulated light curve.  The light curves are then sampled at the TESS 30 minute FFI cadence, and sections of the light curve are removed to match the gaps in the observed light curves presented in \S\ref{sec:data_analysis}.  Finally, random Gaussian deviates are added to the synthetic light curves with amplitudes matching the observed pre-explosion light curve dispersion.  In units relative to peak, these scatter amplitudes are 0.04, 0.04, 0.03, 0.01, 0.03, and 0.05
for SN2018exc, SN2018fhw, SN2018fub, SN2018hib, SN2018hkx, and SN2018koy respectively.  In this way, the synthetic data closely match the observed properties of the light curves, including their noise properties and sampling pattern.
 
After constructing the synthetic light curves, we perform three fits.   First, we fit a power law model of the form given in Equation~\ref{equ:model}.  Second, we add a \citet{Kasen2010} companion component to the power law model, and perform a grid search over the companion separation/radius and $t_0$ to find the best-fit $\chi^2$.   For each combination of companion separation/radius and $t_0$, we first subtract the \citet{Kasen2010} model from the observed light curve, and then fit a power law model optimizing $A$, $\beta$, and $B$ in Equation~\ref{equ:model} (with $t_0$ held fixed) so as to minimize $\chi^2$.  We experimented with the spacing of the grid in companion separation/radii and $t_0$, and found that the following parameters work for our fits:
 \begin{itemize}
 \item 10 logarithmically spaced companion models between 1.1 \rsun and 223 \rsun (separations between 0.02 and 4.00 $\times10^{13}$ cm, respectively), which translates into a grid of  1.1, 2.0, 3.6, 6.4, 11.7, 21.0, 38.0, 68.3, 123.2, and 223 \rsun.
 \item  $t_0$ spaced every 0.1 days (2.4 hours) from $-$1 to $+$1 days relative to the value of $t_0$ determined in the initial power law fit.  
 \end{itemize}
 In \S\ref{sec:companion_search}, we used a finer grid of 20 logarithmically spaced companion separations, and expanded the range of $t_0$ from $-$2 to $+$5 days around the initial estimate of $t_0$.  In the third and final fit, we perform the same grid search but on a synthetic light curve with no companion model.  We perform 10\,000 iterations for each of the 48 simulations with different realizations of the measurement noise, and record the best fit $\chi^2$ from our grid search and the corresponding model parameters.
   
\begin{figure*}
    \centering
    \begin{tabular}{cc}
         \includegraphics[width=0.5\textwidth]{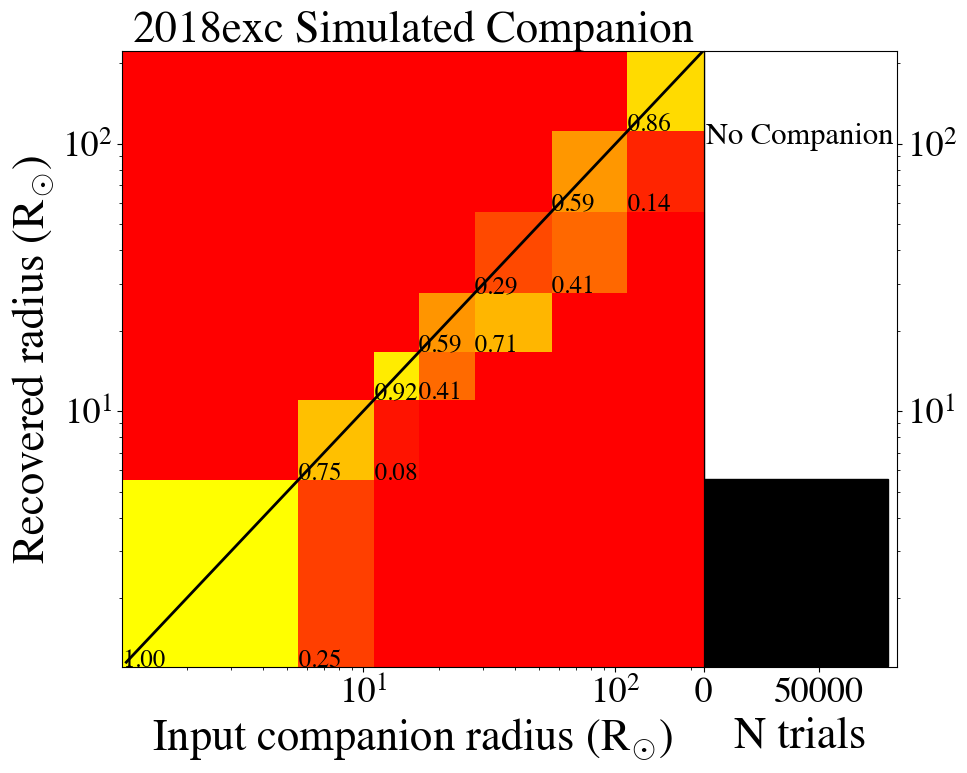}&
         \includegraphics[width=0.5\textwidth]{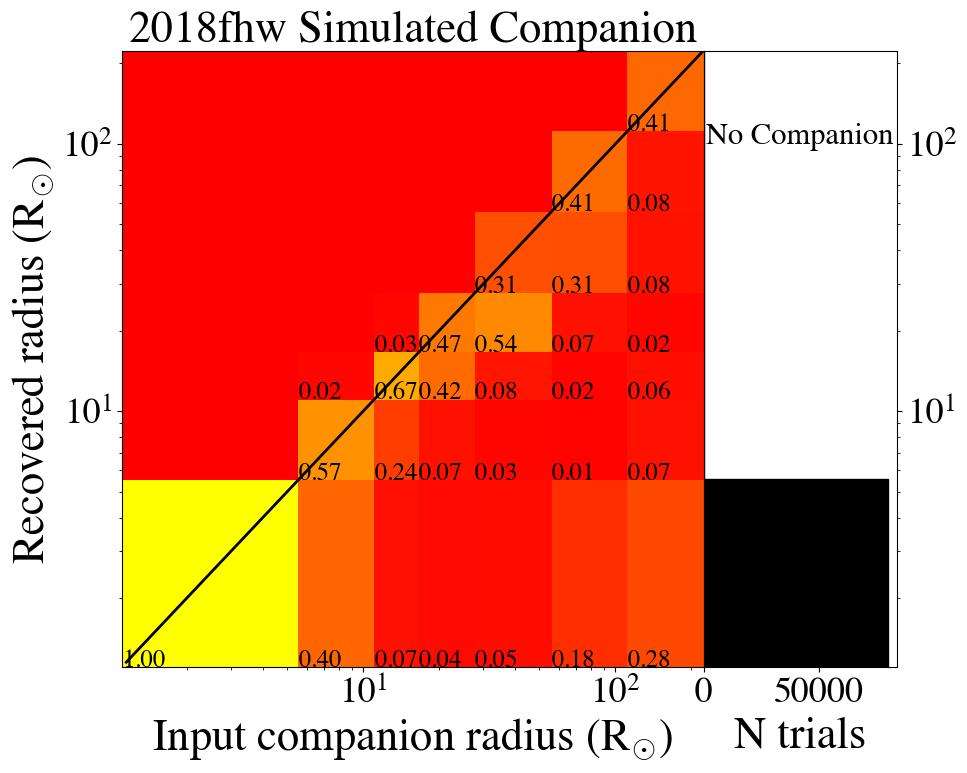}\\ 
         \includegraphics[width=0.5\textwidth]{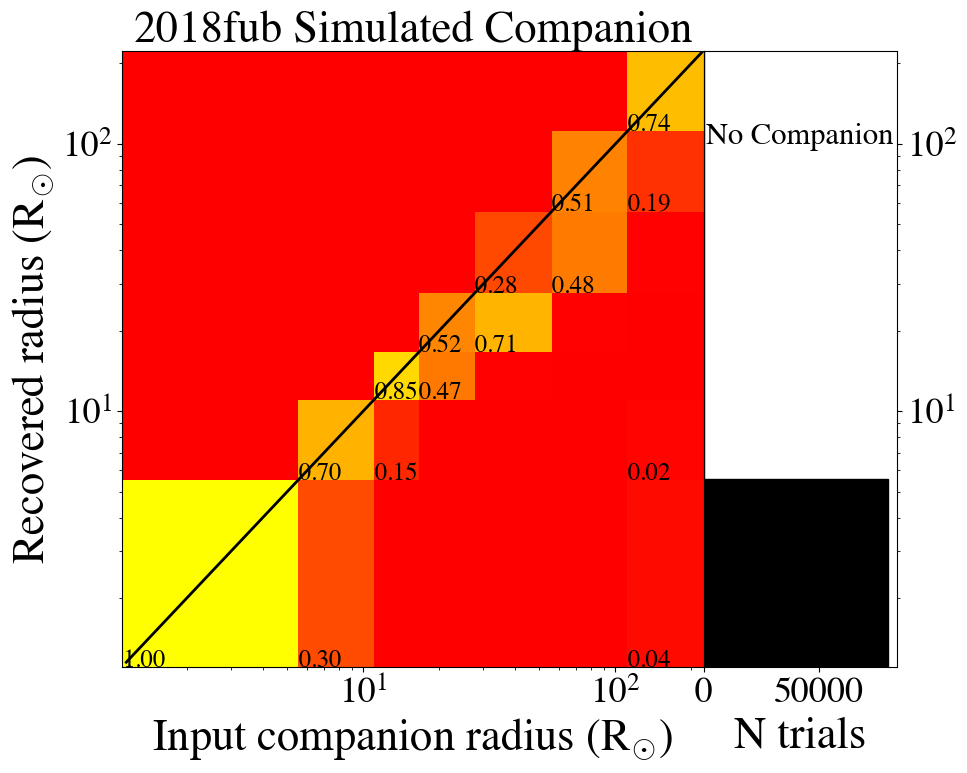}&
         \includegraphics[width=0.5\textwidth]{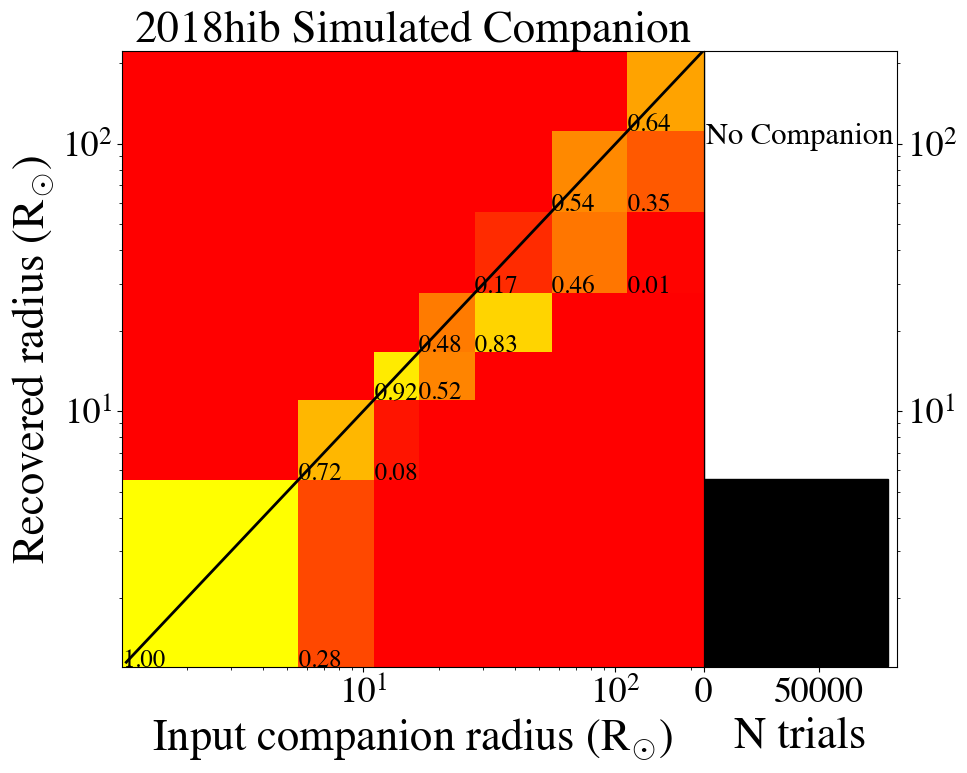}\\
         \includegraphics[width=0.5\textwidth]{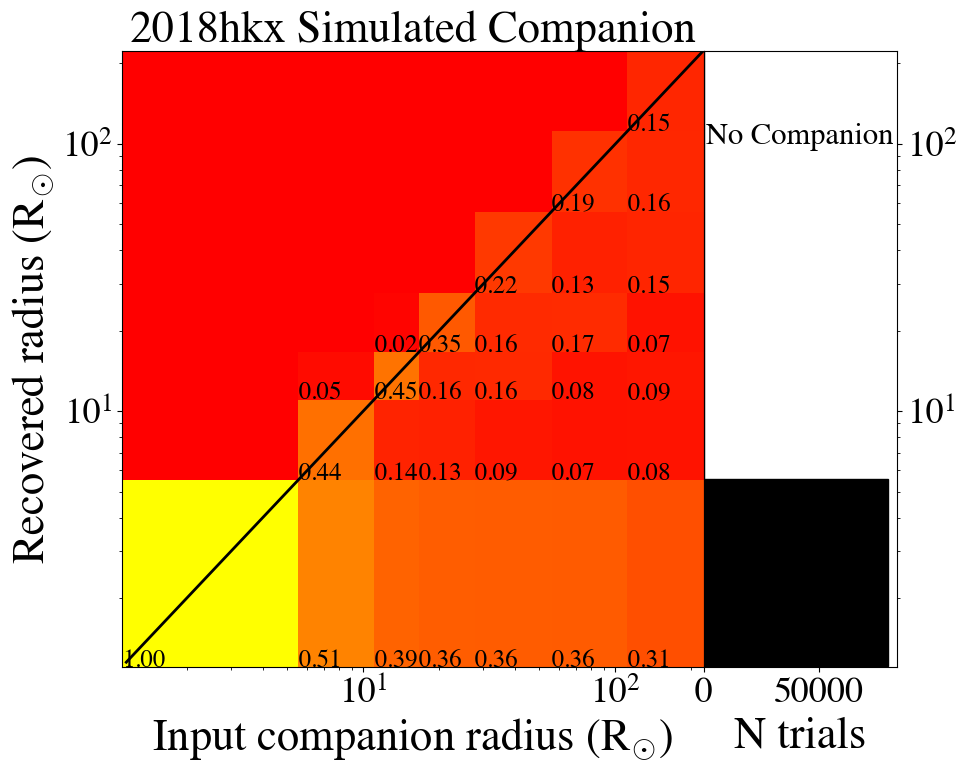}& 
         \includegraphics[width=0.5\textwidth]{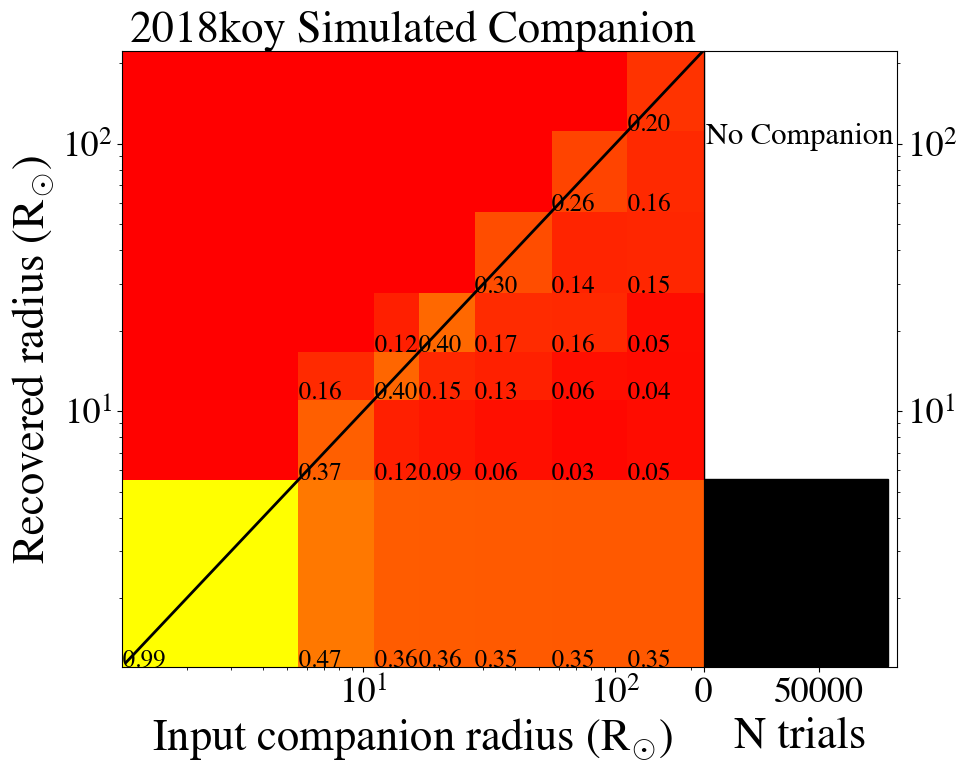} \\
    \end{tabular}
    \caption{Left-hand panels:  Two-dimensional histograms showing the results of the Monte Carlo simulations that test the fitting procedure detailed in \S\ref{sec:companion_search}. For each bin along the x-axis, 10\,000 trials were run at that companion radius while the bins along the y-axis show the distributions of the recovered companion radius.  The fraction of trials in a given bin sum to unity along columns.  Right-hand panels:  One-dimensional histograms that show the recovered companion radius when no companion is actually present.  In all cases, the minimum allowable companion radius is returned. }
    \label{fig:sim_results}
\end{figure*}

\begin{figure*}
    \centering
    \begin{tabular}{c}
         \includegraphics[width=0.8\textwidth]{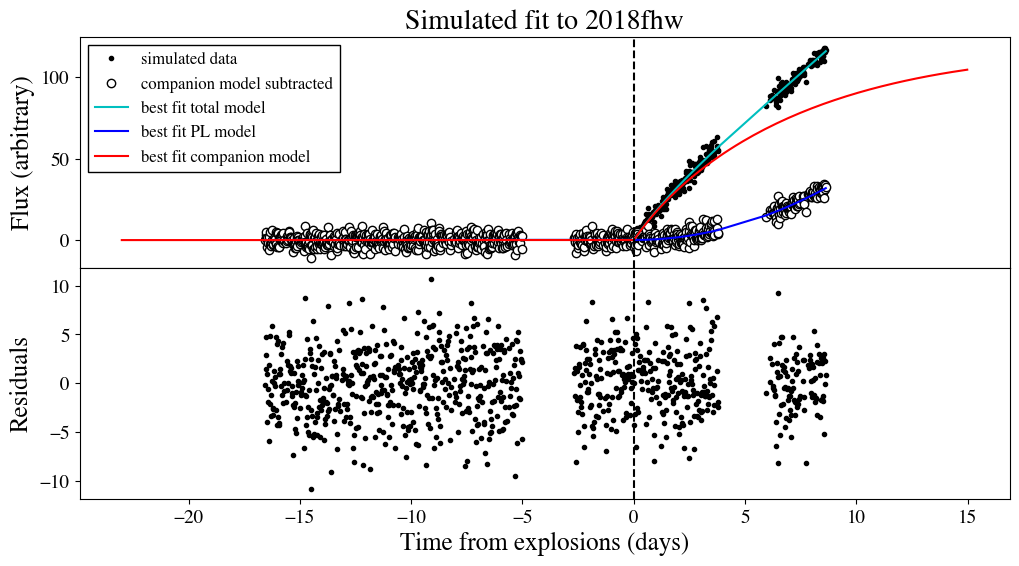}\\
         \includegraphics[width=0.8\textwidth]{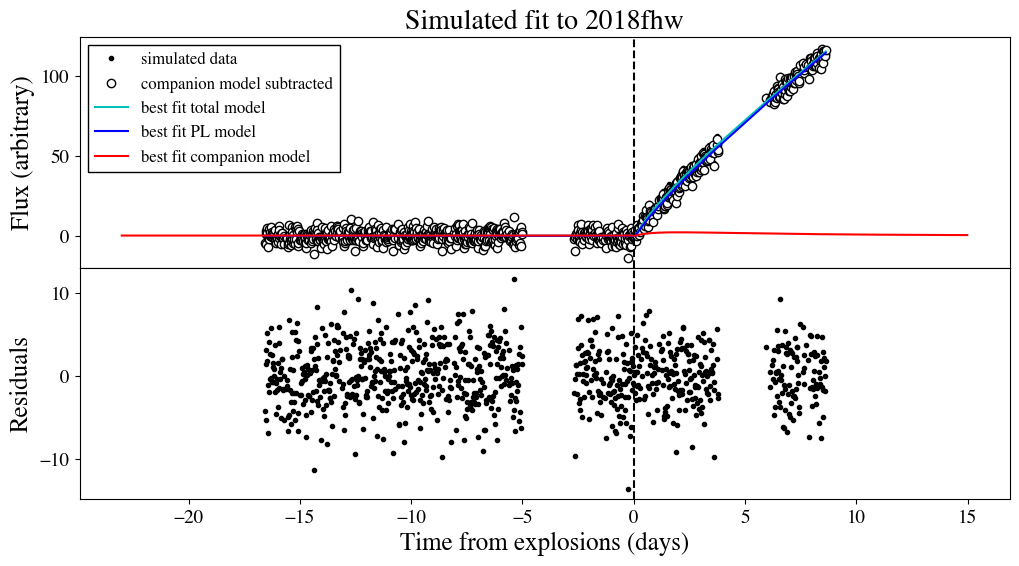}\\ 
    \end{tabular}
    \caption{Examples of simulated data and fits for SN2018fhw.  The top panel shows a simulated companion model with input radius of 223 \rsun.  In this case, the grid search recovers the correct companion model.   The bottom panel shows a different realization using the same input  model, but here the fit recovers a small (1.1 \rsun) companion by compensating for the companion using a power law index $\beta = 0.86$. This latter case accounts for the incorrectly recovered companion parameters in Figure~\ref{fig:sim_results}.  Such solutions are easy to identify, and do not correspond to any results we found when fitting the TESS data in \S\ref{sec:companion_search}.}
    \label{fig:sim_examples}
\end{figure*}

In Figure~\ref{fig:sim_results}, we show the 2D histograms of the recovered companion radius as a function of input companion radius for each SN.  There is good correspondence between the input and recovered radii.  For SN2018exc, SN2018hib, and SN2018fub, the highest probability recovered radius matches the input radius, at least to the closest bin of the grid search.  For SN2018exc, SN2018hib, and SN2018fub there is also a smaller but non-negligible probability that the recovered radius is off by one grid step towards a smaller companion radius than the true value.  For SN2018fhw, SN2018hkx, and SN2018koy (and in some rare cases, SN2018fub), a range of possible companion radii are recovered for the largest simulated companions; in fact, for SN2018hkx and SN2018koy it is more likely to recover a small companion with radius $<$1.1 \rsun\ rather than the true companion radius ($>$50 \rsun).  Inspection of these solutions shows that the fits are driven by a small power law index, with $\beta < 1.0$, which mimics the dominant companion light curve at early times.   Thus, there is a degeneracy in early time Type Ia SN light curves between intrinsically low values of $\beta$ and large companions.  An example of this degeneracy is shown in Figure~\ref{fig:sim_examples}.  No such fits were found for the real TESS data in \S\ref{sec:companion_search}.  It is somewhat unclear why this effect is not seen in SN2018exc, SN2018hib, or (to some extent) SN2018fub, although it is probably related to individual noise properties, sampling patterns, and time ranges over which to fit a particular light curve.  An interesting corollary of this result is that there may be an optimum time range over which to fit early time Type Ia light curves (i.e., up to a larger or smaller fraction of the peak flux than the 40\% adopted here) in order to find signatures of companion stars, but determining if this is the case is beyond the scope of this work.
 
The companion radius recovered in our third set of fits, when no companion model is present in the simulated light curves, is shown by the 1D histograms in the right-hand panels of Figure~\ref{fig:sim_results}.  In all cases, our procedure returns the minimum allowable companion radius of the grid search.  
 
\begin{figure*}
    \centering
    \begin{tabular}{cc}
         \includegraphics[width=0.5\textwidth]{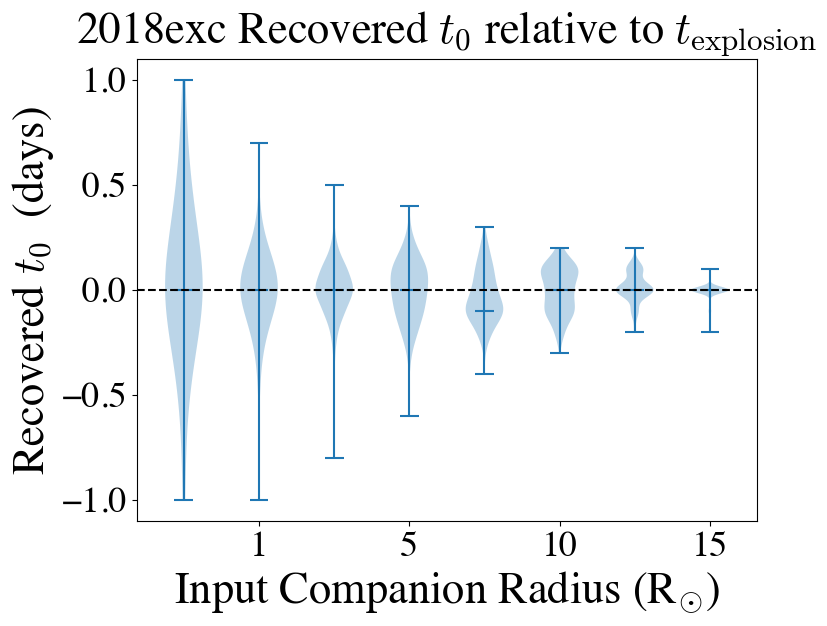}&
         \includegraphics[width=0.5\textwidth]{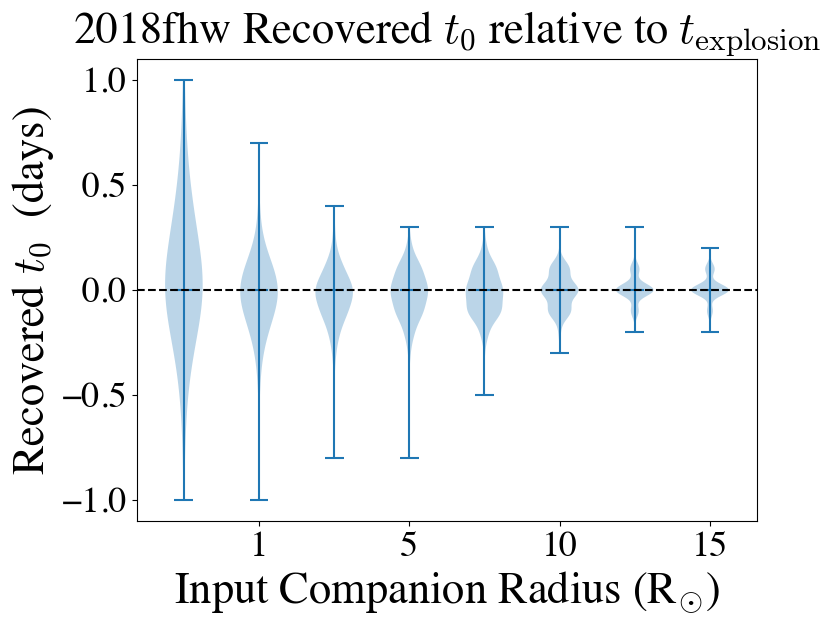}\\ 
         \includegraphics[width=0.5\textwidth]{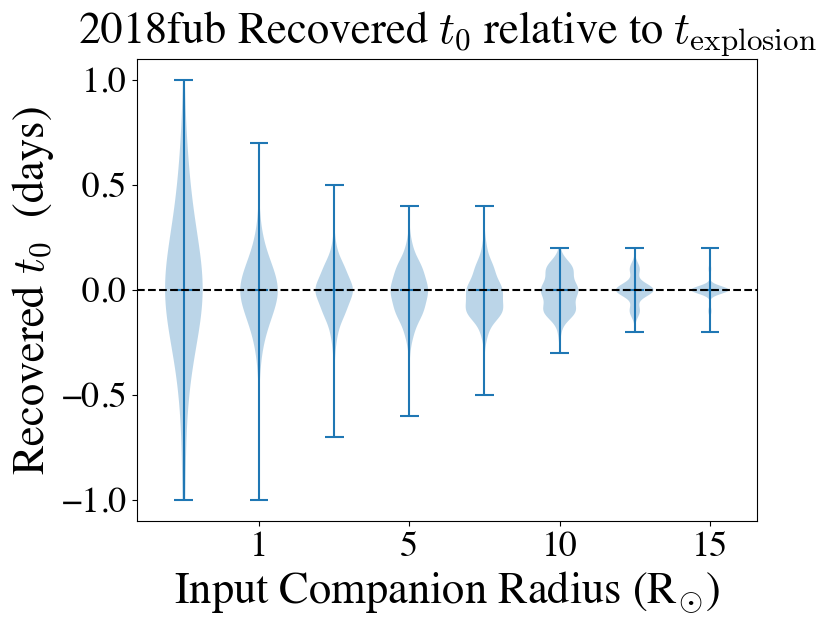}&
         \includegraphics[width=0.5\textwidth]{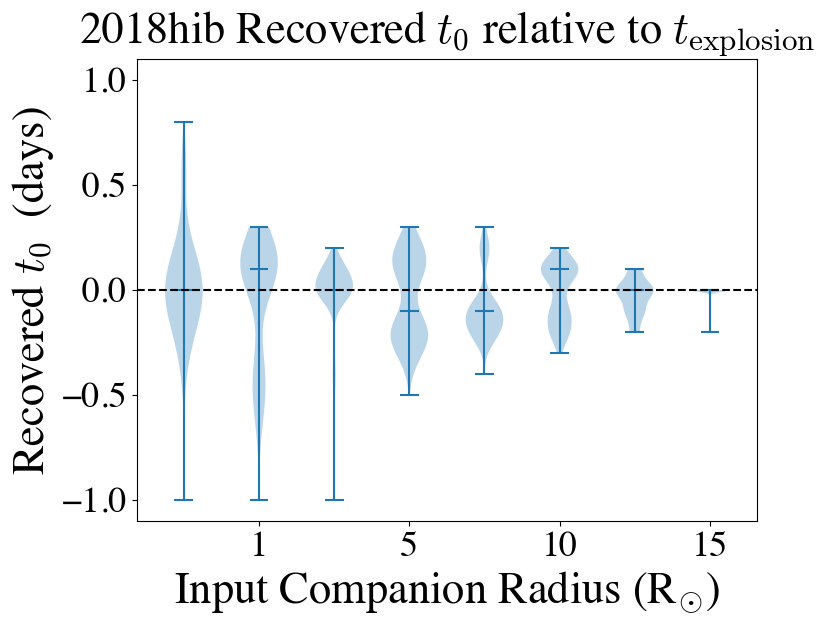}\\
         \includegraphics[width=0.5\textwidth]{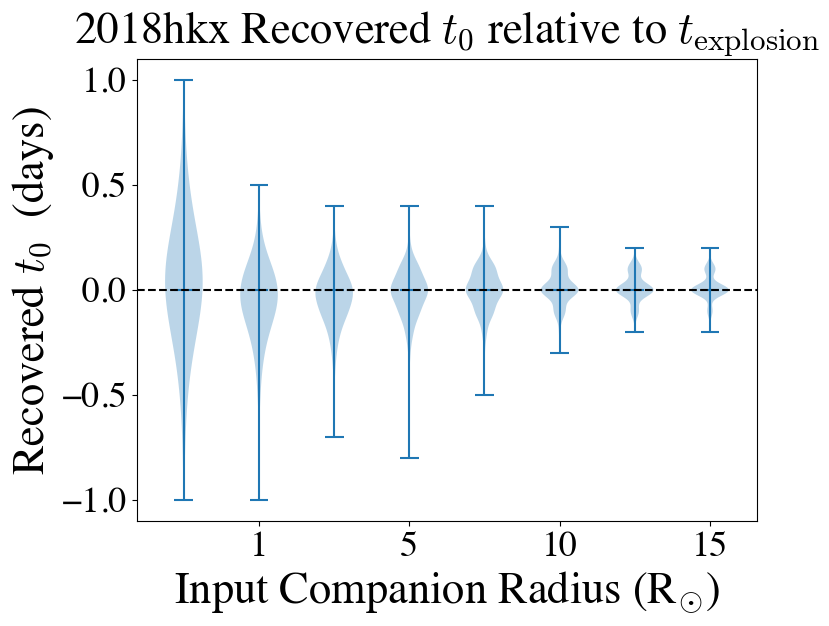}& 
         \includegraphics[width=0.5\textwidth]{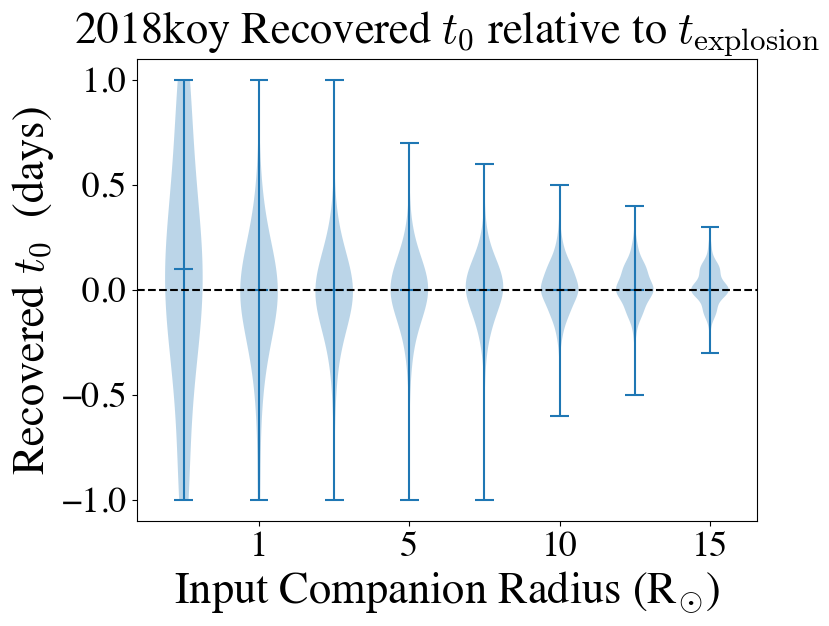} \\
    \end{tabular}
    \caption{Distributions of recovered values of explosion times $t_0$ for the Monte Carlo simulations described in Appendix~C.  The simulated light curves had $t_0$ set to 0.0 days.}
    \label{fig:sim_results_t0}
\end{figure*}

Figure~\ref{fig:sim_results_t0} shows the distributions of the recovered values of $t_0$ for each input companion radius and each SN.  For small values of the input companion radius, $t_0$ is poorly constrained.  This result is expected, because a very small companion signature will be comparable to the measurement noise and have a small effect on the final fit, while the parameters $A$ and $\beta$ can compensate for changes in $t_0$.  However, as the input companion radius increases, the distributions of $t_0$ become narrower and more accurate, showing how the effects of large companions on the light curve are difficult to hide at early times.  For the simulations with no companion signature model, the distributions of $t_0$ are well constrained but biased to $>0.5$\ days.

\begin{figure*}
    \centering
    \begin{tabular}{cc}
         \includegraphics[width=0.5\textwidth]{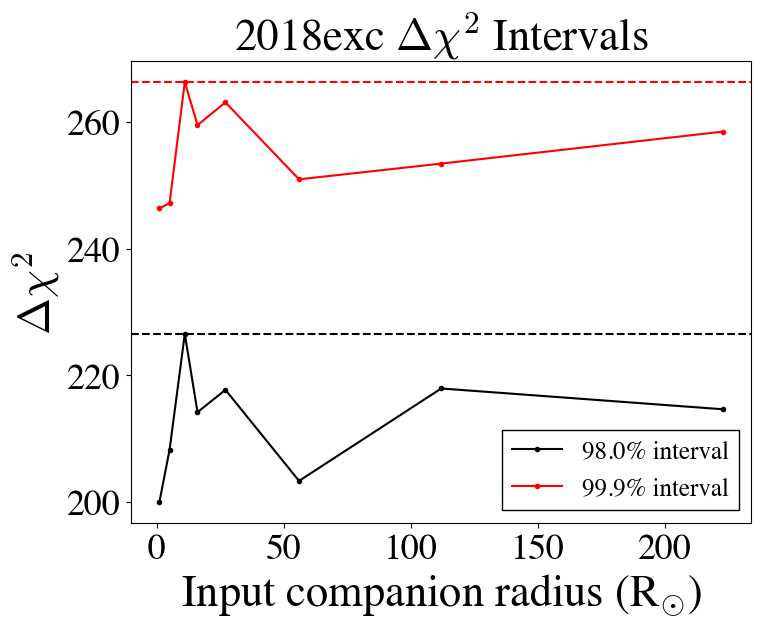}&
         \includegraphics[width=0.5\textwidth]{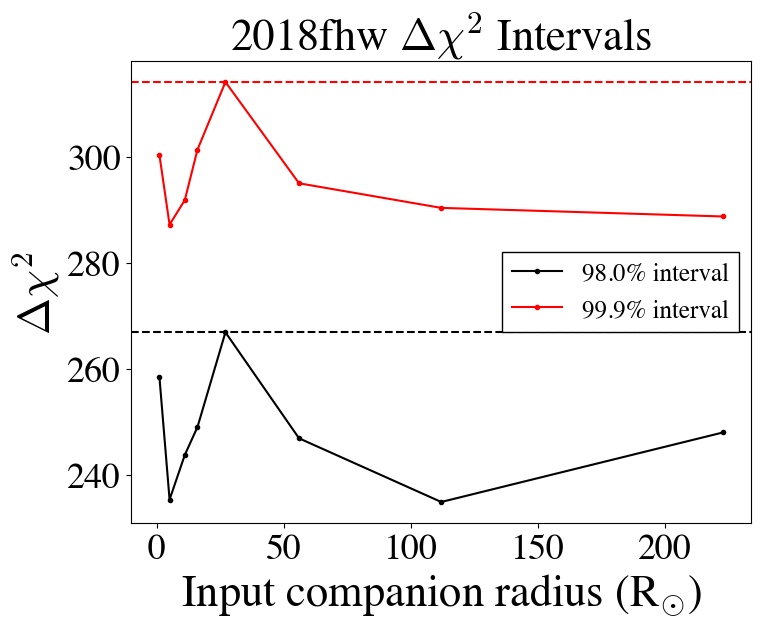}\\ 
         \includegraphics[width=0.5\textwidth]{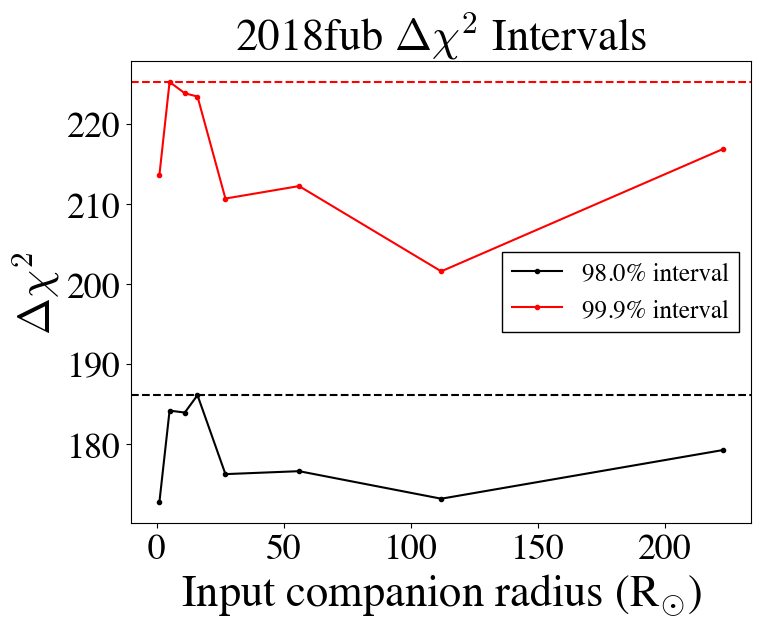}&
         \includegraphics[width=0.5\textwidth]{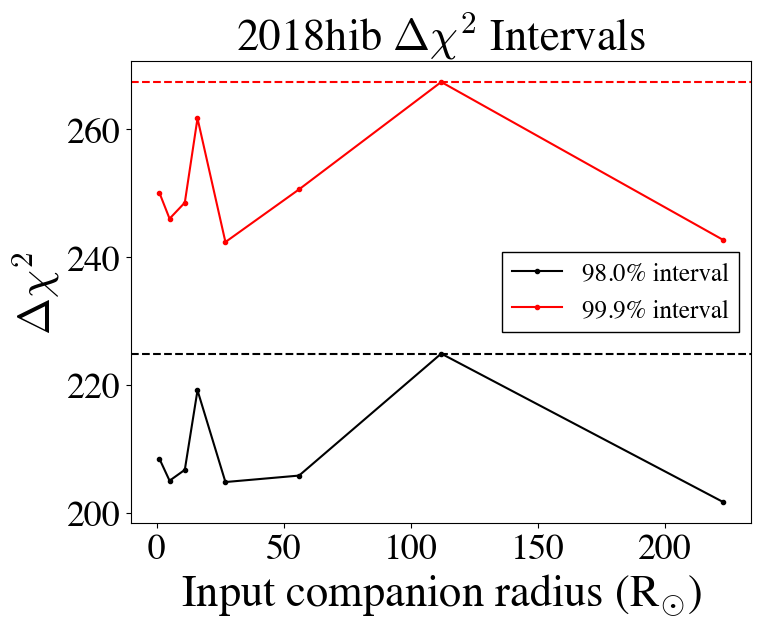}\\
         \includegraphics[width=0.5\textwidth]{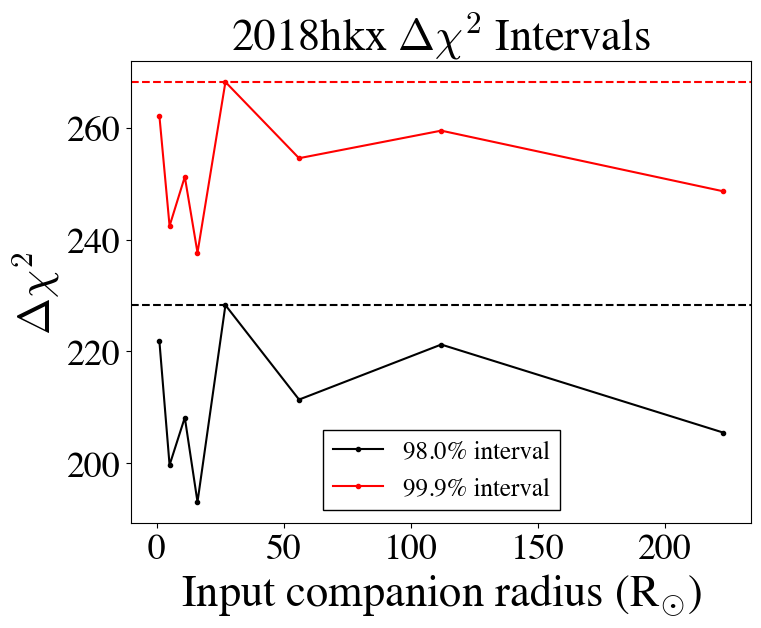}& 
         \includegraphics[width=0.5\textwidth]{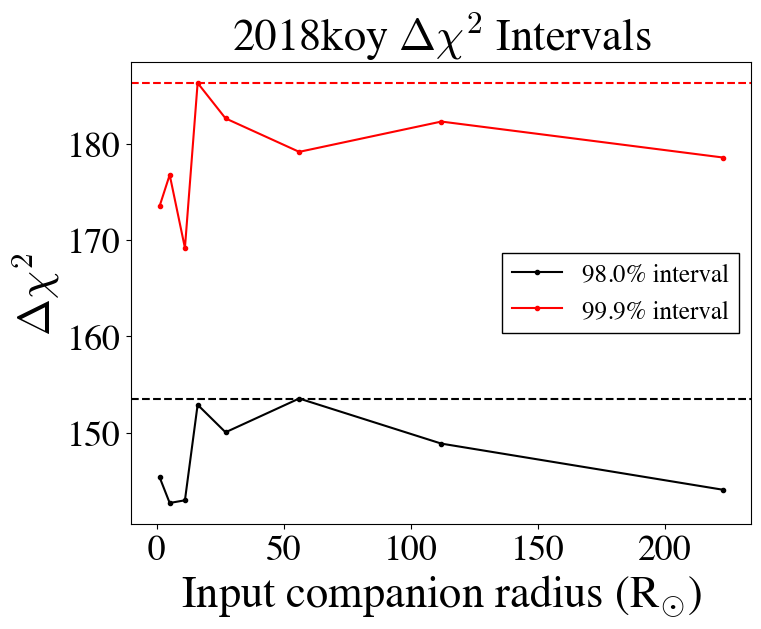} \\
    \end{tabular}
    \caption{98 and 99.9 percentiles of the $\Delta\chi^2$ distributions for the Monte Carlo simulations described in Appendix~C.  The maximum value for each SN is adopted as the 2 and 3$\sigma$ statistical limits, and were used to calculate the blue and red contours in Figure~\ref{fig:companion_contours}.}
    \label{fig:delta_chi2_sims}
\end{figure*}

 Finally, we used the distributions of $\chi^2$ to identify the 2 and 3$\sigma$ statistical limits that can be placed on the model parameters in the grid search.  
  First, we calculate $\Delta\chi^2 = \chi^2 - \min\left( \chi^2\right)$ for all 1000 trials of each simulation.  We then use the 98 and 99.9 percentiles of the distributions of $\Delta\chi^2$ for each simulation to define the statistical 2 and 3$\sigma$ limits.  Figure~\ref{fig:delta_chi2_sims} shows these limits for all SN as a function of input model companion separation. We adopt the largest $\Delta\chi^2$ intervals across input companion separation/radius for the final 2 and 3$\sigma$ statistical limits.  Table~\ref{tab:chi2_confidence} gives the values of $\Delta\chi^2$ used for each SNe.  These intervals are transferred to the TESS data by finding the model with the minimum $\chi^2$ value in the grid search and adding the 2 and 3$\sigma$ $\Delta\chi^2$ values.  Contours of these values of $\chi^2$ are shown in Figure~\ref{fig:companion_contours}.  Any models with $\chi^2$ values outside of these $\chi^2$ contours arise by chance 2\% or 0.1\% of the time.
  
\bibliography{refs}

\end{document}